\documentclass[aps,prl,reprint,nobalancelastpage,nofootinbib]{revtex4-2}

\usepackage[utf8]{inputenc}

\usepackage{amsmath}
\usepackage{amssymb}
\usepackage[all]{xy}
\usepackage{tikz}
\usetikzlibrary{calc}

\allowdisplaybreaks

\usepackage[toc,page]{appendix}
\usepackage{enumerate}
\usepackage{tensor}
\usepackage{booktabs}
\usepackage{bbm}
\usepackage{youngtab}
\usepackage{slashed}
\usepackage{multirow}
\usepackage{makecell}
\usepackage{float}
\usepackage{comment}
\usepackage{verbatim}
\usepackage{xcolor}
\usepackage[framemethod=tikz]{mdframed}
\usepackage{soul}
\usepackage{mathtools}
\usepackage[inline]{enumitem}
\usepackage{accents}
\usepackage{mathtools}
\usepackage{upgreek}

\usepackage{amsthm}

\newcommand{\ab}{|}
\newcommand{\der}{\partial}
\newcommand{\de}{\mathrm{d}}

\newcommand{\e}{{\mathrm{e}}}
\newcommand{\I}{{\mathrm{i}}}

\newcommand{\A}{{A}}
\newcommand{\sigmahat}{{\eta}}
\newcommand{\zetahat}{{\zeta}}
\newcommand{\expansion}{{\theta}}
\newcommand{\h}{{\alpha}}
\newcommand{\B}{{\beta}}
\newcommand{\uniformbound}{{T}}

\newcommand{\ubreve}[1]{{\underaccent{\breve}{#1}}}

\newcommand{\vev}[1]{\langle {#1} \rangle}

\usepackage{hyperref}
\hypersetup{
    colorlinks=true,
    linkcolor=purple,
    citecolor=purple,
    filecolor=purple,      
    urlcolor=purple,
}

\makeatletter
\renewcommand{\thesubsubsection}{\thesection.\arabic{subsection}.\arabic{subsubsection}}
\renewcommand{\subsubsection}{\@startsection{subsubsection}{3}{\z@}{3.25ex \@plus 1ex \@minus .2ex}{1.5ex \@plus .2ex}{\normalfont\small\bfseries\raggedright}}
\makeatother

\renewcommand{\thesubsection}{\Alph{section}.\arabic{subsection}}

\makeatletter
\renewcommand{\p@subsection}{}
\renewcommand{\p@subsubsection}{}
\makeatother
\def\thesection{\arabic{section}}
\def\thesubsection{\arabic{section}.\arabic{subsection}}
\def\thesubsubsection{\arabic{section}.\arabic{subsection}.\arabic{subsubsection}}

\begin{document}

\title{No cosmic horizons from a strengthened higher-dimensional strong energy condition}

\author{Flavio Tonioni}
\email{flavio.tonioni@unipd.it}
\affiliation{Dipartimento di Fisica e Astronomia, Università degli Studi di Padova, and INFN-Padova, via F. Marzolo 8, 35131 Padua, Italy}

\date{\today}

\begin{abstract}
Time-dependent supergravity compactifications lie beyond the setting of the Gibbons/Maldacena-Nuñez no-de Sitter theorem and can even support eternal cosmic acceleration, raising the question of whether they can also produce future cosmic horizons.
For general higher-dimensional matter, the strong energy condition alone does not exclude such horizons.
Here, we demonstrate that a theory fulfilling a strengthened version of the strong energy condition cannot be compactified down to a lower-dimensional theory able to support a future cosmic horizon.
For 10d and 11d supergravities reduced to at least four large spacetime dimensions, this energy condition is fulfilled by the sources that already fulfill the strong energy condition in the higher-dimensional theory.
Our assumption on the dimensional reduction is that it is performed on a smooth compact and boundaryless space.
\end{abstract}

\maketitle

\section{Introduction}

A major question in high-energy physics is the relationship between string-theoretic reductions to Einstein-frame lower-dimensional Friedmann-Lemaître-Robertson-Walker (FLRW) spacetimes and extended epochs of accelerated expansion.
As shown by Gibbons and Maldacena-Nuñez (GMN), de Sitter (dS) solutions cannot exist in the lower-dimensional theory as long as the parent higher-dimensional theory fulfills the strong energy condition (SEC) and the compactification takes place on a static boundaryless space \cite{Gibbons:1984kp, Maldacena:2000mw}.
Because SEC is fulfilled by large classes of higher-dimensional supergravities (SUGRAs), such as those involving field strengths and membranes, this is a far-reaching conclusion, but it is not sufficient to formulate a definitive statement in time-dependent scenarios.

With respect to time-dependent solutions, the situation is significantly more intricate, and the question about long-lived cosmic acceleration conflates several aspects that must be handled with due rigor.
Let us momentarily restrict our attention to higher-dimensional SEC-fulfilling classical Einstein-Hilbert SUGRA.
\begin{itemize}[leftmargin=*]
	\item There is no general obstruction to cosmic acceleration for a finite number of Hubble times, as long as there exists a region where the scalar potential is positive.
    A scalar turnaround after uphill motion provides a familiar mechanism for such transient acceleration \cite{Townsend:2003fx, Emparan:2003gg}.
	\item There is no general obstruction to solutions that accelerate forever, if the large external space is negatively curved \cite{Andersson:2006du}.
	Indeed, negative spatial curvature acts as a perfect fluid saturating the lower-dimensional SEC.
\end{itemize}
Eternal acceleration does not, by itself, imply the existence of a future cosmic horizon \cite{Boya:2002mv}.
Yet, among all the solutions above, none are known that generate a future cosmic horizon.
This means that the maximal comoving distance admitted by the solution is infinite.
In fact, even the eternally accelerating solutions actually asymptote to a state that is exactly neither accelerating nor decelerating.
Even admitting additional elements, such as SEC-violating sources, higher-order gravitational interactions, and quantum-generated potentials, and possibly renouncing to a controlled higher-dimensional embedding, the existence of a single string-theoretic solution exhibiting a future cosmic horizon has not been universally agreed upon \cite{Danielsson:2018ztv}.
Challenges related to realizing cosmic horizons in string compactifications are also evidenced by effective field-theoretic (EFT) considerations \cite{Hellerman:2001yi, Fischler:2001yj, Hebecker:2023qke}.
Additionally, at least the dS horizon underlies longstanding difficulties in defining observables and understanding the microscopic origin of its entropy \cite{Witten:2001kn, Bousso:2004tv, Anninos:2012qw}.
In fact, it has been proposed that the presence of future cosmic horizons is the actual fundamental physical property that is prohibited by quantum gravity \cite{Andriot:2023wvg, Hassfeld:2025hjx}.
This also resonates with conjectures constraining shallow scalar potentials in ultraviolet-complete EFTs \cite{Bedroya:2019snp, Rudelius:2021azq}, for which evidence has been gathered in asymptotic corners of the string landscape \cite{Shiu:2023fhb, Andriot:2026lac, Etheredge:2026cha}.

Let us also comment on two further results.
\begin{itemize}[leftmargin=*]
    \item Although no classical SEC-fulfilling SUGRA reductions are known that accelerate eternally in spatially-flat FLRW spacetime, there exist mathematical constructions that do, including a future cosmic horizon \cite{Russo:2018akp}.
    No microscopic completions of these constructions have been established, but these are counter-examples to the statement that higher-dimensional SEC per se forbids eternal cosmic acceleration.
    \item By definition, the lower-dimensional Einstein frame is the frame where the Einstein-Hilbert coupling is spacetime-independent.
    With a generic warp factor, it is sufficient for an internal-space-averaged condition to hold \cite{Russo:2018akp}.
    The corresponding condition imposed locally in internal space is a sufficient but not necessary requirement.
    Without imposing the Einstein-frame condition locally, constructions with lower-dimensional cosmic acceleration are even less constrained \cite{Koster:2011xg, Russo:2019fnk}.
\end{itemize}
In seeking a universal statement, therefore, we must go beyond the simple assumption of higher-dimensional SEC, and we must construct the lower-dimensional Einstein frame in its general averaged-only sense.

Given the observations above, we should ask the following question.
If we drop the assumption of staticity and suitably modify the requirement of higher-dimensional SEC, what kind of lower-dimensional features are prohibited by the compactification over a boundaryless space?
The answer that we will prove is that a specific strengthened form of SEC (SSEC) in the parent theory is sufficient to rule out cosmic horizons in the lower-dimensional theory.
Explicitly, we can prove the following theorem.
\vspace{-4pt}
\begin{itemize}[wide, labelwidth=!, labelindent=*]
    \item[] \textbf{No-horizon theorem.} Let a $D$-dimensional theory coupled to Einstein-Hilbert gravity fulfill SSEC.
    Then, a reduction over a smooth compact space with no boundary down to a $d$-dimensional Einstein-frame FLRW spacetime, with $d>2$, is necessarily such that
    \begin{equation} \label{no horizon}
        \int_{t_\star}^\infty \dfrac{\de t}{a(t)} = \infty,
    \end{equation}
    where $a$ is the scale factor and $t$ is cosmological time, with $t_\star$ an arbitrary initial time.
\end{itemize}
\vspace{-4pt}
For 10d and 11d SUGRAs reduced to at least four large spacetime dimensions, SSEC is fulfilled by all field-strength tensors and membranes that already fulfill SEC in the higher-dimensional theory.
In other words, for these sources, fulfilling the higher-dimensional SEC is a sufficient condition for fulfilling SSEC, too.
Our argument is entirely carried out at the level of the higher-dimensional field equations.
Nothing in it depends on the formulation of a corresponding lower-dimensional EFT, besides it being an Einstein-frame FLRW reduction.
As a special case, one can see that the GMN no-go theorem, prohibiting dS vacua, is in fact prohibiting the maximally-symmetric example of a cosmic horizon.

We provide two distinct proofs, with different levels of complexity that mirror their applicability.
First, we present a simplified proof that is reminiscent of the derivation of the GMN no-dS theorem as a consequence of the higher-dimensional SEC \cite{Townsend:2001ea, Gibbons:2003gb, Townsend:2003qv}.
Besides providing a simple argument, it motivates the formulation of SSEC.
The tradeoff is that it only works if the Einstein-frame condition is imposed locally.
Second, we present a proof that imposes the Einstein-frame condition in its general averaged-only sense.
This proof involves significantly heavier machinery, and it is reminiscent in spirit of the classic singularity theorems \cite{Penrose_1965, Hawking_1966, Hawking-Penrose_1970} and it largely builds on similar tools.
In both cases, we relegate all technical details to the appendix.

\section{SSEC, and a simplified proof}

Let a $(1,D-1)$-dimensional line element be reduced down to $d>2$ large spacetime dimensions as
\begin{equation} \label{metric}
    d s_{1,D-1}^2 = \e^{2 \tau} \bigl[ \e^{-2 \frac{D-d}{d-2} \omega} d \tilde{s}_{1,d-1}^2 + \e^{2 \omega} d \bar{s}_{D-d}^2 \bigr].
\end{equation}
Here, the external $d$-dimensional spacetime and internal $(D-d)$-dimensional space have line elements $\smash{d \tilde{s}_{1,d-1}^2 = \tilde{g}_{\mu \nu} \de x^\mu \de x^\nu}$ and $\smash{d \bar{s}_{D-d}^2 = \bar{g}_{mn} \de y^m \de y^n}$, respectively, and we have conveniently singled out the universal $\smash{y^m}$-independent volume mode $\smash{\omega}$.
For $\smash{d \tilde{s}_{1,d-1}^2}$ to represent the $d$-dimensional Einstein frame, it suffices to have
\begin{equation} \label{einstein-frame condition}
    \der_\mu \int \de^{D-d} y \, \sqrt{\bar{g}_{D-d}} \; \e^{(D-2) \tau} = 0.
\end{equation}
Our interest is in $d$-dimensional FLRW reductions, i.e.
\begin{equation} \label{d-dimensional FLRW-metric}
    d \tilde{s}_{1,d-1}^2 = - \de t^2 + a^2(t) \, d \mathrm{l}_{d-1}^2,
\end{equation}
where $\smash{d \mathrm{l}_{d-1}^2 = \mathrm{g}_{ij} \de x^i \de x^j}$ is a maximally-symmetric space with constant curvature $k = 0$, $-1/\ell^2$, or $+1/\ell^2$.
We assume the most general spacetime dependencies compatible with the $d$-dimensional FLRW ansatz, only requiring $\smash{\der_i \tau = 0}$ and $\smash{\der_i \bar{g}_{mn} = 0}$, as well as $\smash{\der_i \omega = \der_m \omega = 0}$.
The cosmological observables of our interest are the Hubble parameter $\smash{H = \dot{a}/a}$ and its time derivative, expressed via $\smash{\epsilon = - \dot{H}/H^2}$.
All details on the geometric reduction are in app. \ref{subapp.: higher- and lower-dimensional geometry}.

We can arrange the $D$-dimensional Einstein equations as $d$-dimensional Friedmann equations, including the volume-mode dynamics, through linear combinations of the $(0,0)$-component and $d$- and $(D-d)$-dimensional component traces; additionally, the traceless $(D-d)$-dimensional components control the dynamics of the volume-preserving modes.
All details are in app. \ref{subapp.: Einstein/Friedmann and continuity equations}.
As our interest is in the $d$-dimensional physics, we average out all the equations via a suitable integration over the internal coordinates.
Our assumption is that the smooth $(D-d)$-dimensional space is compact and boundaryless.
Let the weighted internal-space volume element be $\smash{\de \mu = \e^{(D-2) \tau} \sqrt{\bar{g}_{D-d}} \; \de^{D-d} y}$.
Then, given the positive integral $\smash{K = \int \de \mu}$, which is time-independent by the Einstein-frame condition in eq. (\ref{einstein-frame condition}), we define the weighted average of a function $f$ as $\smash{\vev{f} = \int \de \mu \, f/K}$.
The Einstein-frame condition in eq. (\ref{einstein-frame condition}) can be expressed as $\smash{\vev{\varphi} = 0}$, where $\smash{\varphi = (1/2) \, \bar{g}^{mn} \dot{\bar{g}}_{mn} + (D-2) \dot{\tau}}$.
We also define the matter integrals
\begin{subequations}
\begin{align}
    I_{0} & = \dfrac{\kappa_D^2}{d-2} \, \vev{T_{00}}, \label{I(0)} \\
    I_{d-1} & = \dfrac{\kappa_D^2}{d-2} \vev{\tilde{g}^{ij} T_{ij}}, \label{I(d-1)} \\
    I_{D-d} & = \dfrac{\kappa_D^2}{d-2} \, \e^{-2 \frac{D-2}{d-2} \omega} \vev{\bar{g}^{mn} T_{mn}}, \label{I(D-d)}
\end{align}
\end{subequations}
where $\kappa_D$ is the $D$-dimensional gravitational coupling, as well as the shifted and rescaled volume velocity
\begin{equation} \label{dotomegatau}
    \dot{\chi} = \dfrac{D-d}{d-2} \, \dot{\omega} - \vev{\dot{\tau}}.
\end{equation}
After defining the additional integrals $\smash{I_{\mathrm{cw}}}$ and $\smash{I_{\theta \delta} \geq 0}$, capturing the internal-space curvature, the warp factor, and the volume-preserving modes, we can express the $d$-dimensional Friedmann equations and the volume-mode field equation, respectively, as
\begin{subequations}
\begin{align}
    & (d-1) \Bigl[ H^2 + \dfrac{k}{a^2} \Bigr] = \dfrac{D-2}{D-d} \, \dot{\chi}^2 + \Delta + 2 I_0 - I_{\mathrm{cw}}, \label{integrated master F1-eq.} \\
    & \epsilon H^2 + \dfrac{k}{a^2} = \dfrac{D-2}{D-d} \, \dot{\chi}^2 + \Delta + \Bigl( I_0 + \dfrac{I_{d-1}}{d-1} \Bigr), \label{integrated master F2-eq.}
\end{align}
\end{subequations}
and
\begin{align} \label{integrated master Eomega-eq.}
    \ddot{\chi} + (d-1) H \dot{\chi} = \dfrac{D-d}{D-2} (I_0 - I_{d-1}) + \dfrac{d-2}{D-2} I_{D-d} - I_{\mathrm{cw}},
\end{align}
where we defined $\smash{\Delta = I_{\theta \delta} - [(d-1)/(d-2)^2] \vev{\varphi^2}}$ for brevity.
All details are in app. \ref{subapp.: averaged-out equations}.

The $D$-dimensional SEC is a condition on the energy-momentum tensor of a theory that, intuitively, makes gravity attractive \cite{Wald, Carroll}.
As reviewed in app. \ref{subapp.: higher-dimensional SEC and lower-dimensional physics}, on the reduced theory, it can be shown to imply, among others, the non-negativity conditions
\begin{subequations}
\begin{align}
    (D-3) I_0 + I_{d-1} + I_{D-d} & \geq 0, \label{integrated basic SEC} \\
    (d-1) I_0 + I_{d-1} & \geq 0. \label{integrated d-dim. EC}
\end{align}
\end{subequations}
It turns out these are not enough to rule out $d$-dimensional solutions exhibiting cosmological horizons.
It is a different energy condition that does.
To introduce it, given the external spatial vielbeins $\smash{\tilde{e}_a{}^i}$, we define $\smash{e_a{}^M = \e^{- \tau} \, \e^{[(D-d)/(d-2)] \omega} \, \tilde{e}_a{}^i \delta^M_i}$ with respect to the $D$-dimensional metric.
Then, we define the projector on the external space $\smash{P^{MN} = \delta^{ab} e_a{}^M e_b{}^N }$, i.e.
\begin{equation} \label{(d-1)-space projector}
    P^{MN} = \e^{-2\tau} \, \e^{2 \frac{D-d}{d-2} \omega} \tilde{g}^{ij} \delta_i^M \delta_j^N.
\end{equation}
We postulate a strengthened version of the SEC, as follows.
Let us consider any future-directed unit time-like vector $v^M$ such that $\smash{P_{MN} v^N = 0}$.
Then, we say that the theory fulfills the SSEC if
\begin{equation} \label{ricci-tensor SSEC}
    R_{MN} v^M v^N \geq \dfrac{1-\lambda}{\lambda} \, \dfrac{1}{d-1} \, R_{MN} P^{MN},
\end{equation}
where $\lambda$ is a constant in the interval
\begin{equation} \label{lambda proof interval}
    \dfrac{1}{2} \dfrac{D-d-1+\sqrt{D-d+1}}{D-2} \! < \lambda \leq \! \dfrac{1}{2} \Bigl( 1 + \dfrac{\sqrt{d-2}}{\sqrt{D-2}} \Bigr).
\end{equation}
Throughout this work, whenever we mention SSEC, it is understood that we mean it with a $\lambda$-value belonging to the interval in eq. (\ref{lambda proof interval}).
We refer to this as a strengthened version of the $D$-dimensional SEC for two reasons.
\begin{itemize}[leftmargin=*]
    \item While the SEC statement holds for arbitrary time-like vectors, SSEC is formulated for time-like vectors orthogonal to the external spatial directions.
    \item By expressing the SSEC statement as $\smash{R_{MN} v^M v^N} \geq \smash{(1-\lambda)} \smash{[R_{MN} v^M v^N + R_{MN} P^{MN}/(d-1)}]$, we see that SEC would separately imply the non-negativity of the SSEC left- and right-hand sides, in FLRW reductions, but not establish any general hierarchy among these.
\end{itemize}
Unfortunately, we are unaware of a different motivation of SSEC from first principles, or of a universal argument to fix the constant $\lambda$.
Still, for the tree-level 2-derivative higher-dimensional classical actions from the weak-coupling limit of critical string theories, i.e. the 10d type-I, type-II and heterotic SUGRAs, SEC implies SSEC if $d \geq 4$.
(Here, we are considering field strengths with either no or all the external spatial legs, and smeared membranes filling the external spacetime.)
The implication is true more generally for field strengths and membranes if $\smash{\lambda \geq (D-d)/(D-2)}$, for $D \leq 4d-6$.
In the special case of 11d SUGRA, though, we find SSEC even for $d=3,4$, with different $\lambda$s.
All details are in app. \ref{app.: SEC and SSEC for higher-dimensional SUGRA-like sources}.

Below, we illustrate a simplified proof of the no-horizon theorem.
For here only, we assume that the Einstein-frame condition holds through a point-wise identity in the internal space, rather than merely eq. (\ref{einstein-frame condition}), i.e.
\begin{equation} \label{point-wise einstein frame condition}
    \varphi = \dfrac{1}{2} \, \bar{g}^{mn} \dot{\bar{g}}_{mn} + (D-2) \dot{\tau} = 0.
\end{equation}
Albeit restrictive, this holds for instance whenever the time dependence of the warp factor can be neglected, i.e. $\smash{\dot{\tau} = 0}$, and the remaining internal-metric deformations are locally volume-preserving, i.e. $\smash{\bar{g}^{mn} \dot{\bar{g}}_{mn} = 0}$ \cite{Giddings:2005ff, Grana:2005jc}.
If eq. (\ref{point-wise einstein frame condition}) is in place, we do not only have $\smash{\vev{\varphi} = 0}$, but also $\smash{\vev{\varphi^2} = 0}$.
This ensures the non-negativity of $\smash{\Delta \geq 0}$, which offers a crucial simplification.
For this simplified proof only, we also make additional assumptions: we assume the $D$-dimensional SEC to hold, with its implication in ineq. (\ref{integrated d-dim. EC}), and we consider non-positive FLRW spatial curvature $k=0,-1/\ell^2$.

To start, we observe an obvious implication of the $D$-dimensional SEC.
By ineq. (\ref{integrated d-dim. EC}), and thanks to the non-negativity of $\smash{\dot{\chi}^2, \Delta \geq 0}$, eq. (\ref{integrated master F2-eq.}) implies that $\smash{\epsilon H^2 + k/a^2 \geq 0}$.
This also means that
\begin{equation} \label{point-wise SEC 0-implication}
    \epsilon \geq 0.
\end{equation}
This would not be implied if eq. (\ref{point-wise einstein frame condition}) were not holding.

By subtracting out eqs. (\ref{integrated master F1-eq.}, \ref{integrated master F2-eq.}), and plugging in eq. (\ref{integrated master Eomega-eq.}), we find the crucial equation to motivate SSEC, i.e.
\begin{equation} \label{integrated Einstein-equation master combination}
\begin{split}
    & \dfrac{1}{d-2} \dfrac{1}{a} \dfrac{\de}{\de t} [a(H - \dot{\chi})] + H (H - \dot{\chi}) + \dfrac{k}{a^2} \\
    = \, & \Bigl( I_0  + \dfrac{I_{d-1}}{d-1} \Bigr) - \Bigl[\dfrac{D-3}{D-2} I_0 + \dfrac{I_{d-1} + I_{D-d}}{D-2} \Bigr].
\end{split}
\end{equation}
On the left hand-side, we recognize purely geometric terms.
On the right hand-side, we recognize the difference of two terms that are non-negative in view of the $D$-dimensional SEC, i.e. ineq. (\ref{integrated basic SEC}) and ineq. (\ref{integrated d-dim. EC}).
Unfortunately, however, the sign of their combination is not also fixed.
This is precisely the place where SSEC enters into play.
In its integrated form, it reads
\begin{equation} \label{integrated SSEC}
    \Bigl[\dfrac{D-3}{D-2} I_0 + \dfrac{I_{d-1} + I_{D-d}}{D-2} \Bigr] \geq (1-\lambda) \, \Bigl( I_0  + \dfrac{I_{d-1}}{d-1} \Bigr).
\end{equation}
If ineq. (\ref{integrated SSEC}) is in place, then eq. (\ref{integrated Einstein-equation master combination}) provides a sharp bound, as follows.
We can replace the $\lambda$-dependent source term via eq. (\ref{integrated master F2-eq.}), and, thanks to the non-negativity of $\smash{\Delta \geq 0}$, a completion of squares gives
\begin{equation} \label{optimized differential proof ineq.}
    \dfrac{1}{d-2} \dfrac{\de}{\de t} [a (H - \dot{\chi})] \leq - \lambda \, \dfrac{\de}{\de t} (a H) - (1- \lambda) \dfrac{k}{a}.
\end{equation}
for $\lambda$ in the interval in eq. (\ref{lambda proof interval}).
(Actually, the current step requires a weaker condition.
The shorter interval in eq. (\ref{lambda proof interval}) comes from the general proof in sec. \ref{sec.: a general no-horizon proof}.)

Now, by contradiction, let us assume that
\begin{equation} \label{point-wise would-be horizon condition}
    \int_{t_\star}^\infty \dfrac{\de t}{a(t)} < \infty,
\end{equation}
for any initial time $t_\star$.
By integrating ineq. (\ref{optimized differential proof ineq.}), we find
\begin{equation}
    \Bigl[ a \Bigl(\dfrac{H - \dot{\chi}}{d-2} + \lambda H \Bigr) \Bigr] \Bigr\ab_{t_\star}^t \leq - (1-\lambda) \int_{t_\star}^t \de s \, \dfrac{k}{a(s)}.
\end{equation}
In view of ineq. (\ref{point-wise would-be horizon condition}), for any finite initial time $\smash{t_\star}$, we may always find a finite positive constant $\Gamma > 0$ such that
\begin{equation} \label{optimized integrated proof ineq.}
    \dfrac{H - \dot{\chi}}{d-2} + \lambda H \leq \dfrac{1}{d-2} \, \dfrac{\Gamma}{a}.
\end{equation}
By ineq. (\ref{point-wise SEC 0-implication}), we have $\smash{a_2 \leq a_1 \, \e^{H_1 (t_2-t_1)}}$ for times $t_2 \geq t_1$, whence $\smash{\int_t^{\infty} \de s / a(s) \geq 1/(a H)}$.
Thus, ineq. (\ref{point-wise would-be horizon condition}) requires $\smash{\lim_{t \to \infty} a(t)H(t) = \infty}$, and therefore
\begin{equation} \label{dotomegatau asymptotic proof bound}
    \liminf_{t \to \infty} \, \dfrac{\dot{\chi}(t)}{H(t)} > 1.
\end{equation}
At the same time, eq. (\ref{integrated master F2-eq.}), via ineq. (\ref{integrated d-dim. EC}) and the non-negativity of $\smash{\dot{\chi}^2}, \smash{\Delta}, \smash{-k \geq 0}$, implies $\smash{\epsilon H^2} \geq \smash{[(D-2)/(D-d)] \dot{\chi}^2}$.
In view of ineq. (\ref{dotomegatau asymptotic proof bound}), there must exist a finite time $t_0$ such that
\begin{equation}
    - \dfrac{\dot{H}}{H^2} \geq \dfrac{D-2}{D-d}
\end{equation}
at all later times $t \geq t_0$.
In the convenient time parameterization where $\smash{t_0 H_0 = (D-d)/(D-2)}$, a double time integration immediately implies
\begin{equation}
    \int_{t_0}^\infty \dfrac{\de t}{a(t)} \geq \dfrac{1}{a_0} \int_{t_0}^\infty \de t \, \Bigl( \dfrac{t_0}{t} \Bigr)^{\frac{D-d}{D-2}} = \infty,
\end{equation}
which contradicts ineq. (\ref{point-wise would-be horizon condition}), and therefore proves the absence of a future cosmic horizon.

\section{A general no-horizon proof} \label{sec.: a general no-horizon proof}

Here, we discuss the key elements towards the general no-horizon theorem, with the integral Einstein-frame condition.
All details are delegated to app. \ref{app.: no-horizon proof}.

Our assumptions are that the $(D-d)$-dimensional reduction space is Riemannian, compact, and boundaryless, and that the $D$-dimensional theory fulfills SSEC.
Our strategy consists in showing that, if we assume by contradiction the finiteness of the integral
\begin{equation} \label{would-be horizon condition}
    L_\star = \int_{t_\star}^\infty \dfrac{\de t}{a(t)} < \infty,
\end{equation}
where $t_\star$ is an initial time, then we necessarily run into an inconsistency.
In particular, by the integral Einstein-frame condition in eq. (\ref{einstein-frame condition}), we have $\smash{\dot{K} = 0}$.
Therefore, for any time $t \geq t_\star$, we can write
\begin{equation} \label{proof-solving integral}
    \int_{t_\star}^t \de s \int \de \mu \, a^{d-3} = K \int_{t_\star}^t \de s \, a^{d-3}(s).
\end{equation}
Intuitively, with $d>2$, the existence of a horizon requires a scale factor that does not vanish at infinity, forcing this quantity to diverge as $t \to \infty$.
What we will show is that SSEC, instead, would bound it from above by a constant if the remaining conformal time were finite, as would be implied by eq. (\ref{would-be horizon condition}).

For the proof, our idea consists in defining coordinates following a family of geodesics and bounding the global-coordinate integrals through these.
At fixed time and internal position, the line element associated to the metric in eq. (\ref{metric}) is $\smash{d l_{d-1}^2 = a^2 \, \e^{2 \tau} \, \e^{- 2 [(D-d)/(d-2)] \omega} d \mathrm{l}_{d-1}^2}$.
It is therefore convenient to define
\begin{equation} \label{alpha}
    \e^\A = \e^{\tau - \frac{D-d}{d-2} \omega} \dfrac{a}{a_\star}.
\end{equation}
From a physical perspective, with respect to cosmological time, we build our argument around the worldline of a probe particle moving in the reduced $(1,D-d)$-dimensional internal space with a spacetime-dependent mass $\smash{\mu = \mu_0 \, (a_\star/a) \, \e^{\A/\lambda}}$, where $\mu_0>0$ is a positive constant.
This means that, at fixed external position, we consider the reduced Weyl-rescaled metric $\smash{d \breve{s}_{1,D-d}^2} = \smash{-\de t^2 + \e^{2 [(D-2)/(d-2)] \omega} d \bar{s}_{D-d}^2}$, and the action
\begin{equation} \label{cosmological-time action}
    S = - \mu_0 \int \de t \, \sqrt{1 - \e^{2 \frac{D-2}{d-2} \omega} \bar{g}_{mn} \dot{y}^m \dot{y}^n} \, \dfrac{a_\star}{a} \, \e^{\frac{\A}{\lambda}}.
\end{equation}
We build a family of time-like curves whose unit velocities $v^M$ solve the equations of motion for eq. (\ref{cosmological-time action}) that are initially orthogonal to all internal spatial directions, i.e.
\begin{subequations}
\begin{align}
    v^N \nabla_N v^M & = - \dfrac{1-\lambda}{\lambda}\,(g^{MN} + v^M v^N) \der_N \A, \label{v dynamics} \\
    v^M(t_\star,y_\star) & = \e^{-\tau(t_\star,y_\star)} \, \e^{\frac{D-d}{d-2}\omega(t_\star)} \, \delta^M_0. \label{v initial conditions}
\end{align}
\end{subequations}
We can build a family of curves sweeping the internal directions by labelling each element of the family by its initial conditions $\smash{y_\star^m}$.
Seen as curves in the full $(1,D-1)$-dimensional spacetime, these curves do not move in the external directions.
If we define an auxiliary metric $\smash{d \ubreve{s}_{1,D-d}^2} = \smash{(\mu/\mu_0)^2 d \breve{s}_{1,D-d}^2}$, then proper geodesic curves orthogonal to the initial hypersurface are identified by the vectors $\smash{\ubreve{v}^M = \e^{- [(1-\lambda)/\lambda] \A} v^M}$, which fulfill $\smash{\ubreve{v}^N \ubreve{\nabla}{}_N \ubreve{v}^M = 0}$ by eq. (\ref{v dynamics}).
Let
\begin{equation} \label{dsigma}
    \de \sigma = \sqrt{1 - \e^{2 \frac{D-2}{d-2} \omega} \bar{g}_{mn} \dot{y}^m \dot{y}^n} \, \dfrac{a_\star}{a} \, \de t.
\end{equation}
A consequence of the horizon hypothesis, in eq. (\ref{would-be horizon condition}), is that irrespective of the initial conditions $y_\star^m$ the maximal extent of the $\sigma$-coordinate is bounded as
\begin{equation} \label{sigma-horizon inequality}
    0 \leq \sigma(t;y_\star) \leq \int_{t_\star}^t \de s \, \dfrac{a_\star}{a(s)} \leq a_\star L_\star.
\end{equation}
Our idea is that of working with the local $\smash{(\sigma,x^i,y_\star^m)}$-coordinates, where each initial condition $\smash{y_\star^m}$ identifies a different curve through eqs. (\ref{v dynamics}, \ref{v initial conditions}), to bound integrals in the globally-defined $\smash{(t,x^i,y^m)}$-coordinates.
Given the infinitesimal volume-element transformation
\begin{equation} \label{jacobian volume-element transformation}
    \Bigl( \dfrac{a}{a_\star} \Bigr)^{d-3} \, \de t \, \de \mu = J \, \de \sigma \, \de \mu_\star,
\end{equation}
through SSEC, compactness and boundarylessness of the internal space, we can bound the time integral of the Jacobian factor $J$, irrespective of the $\smash{y_\star^m}$-fixed curve being selected.
In particular, the key to the reasoning is that $\lambda$-dependent terms emerging above allow us to single out a differential inequality along the flow in eq. (\ref{v dynamics}), with a peculiar positivity condition implied by SSEC.
We simply expand the identity $\smash{v^M \nabla_M (\nabla_N v^N)} = \smash{v^M \nabla_N \nabla_M v^N - R_{MN} v^M v^N}$.
Whenever $\lambda$ is in the interval of eq. (\ref{lambda proof interval}), we have the fundamental inequality
\begin{equation} \label{fundamental ineq.}
\begin{split}
    \e^{2 \A} \biggl[ R_{MN} v^M v^N - \dfrac{1-\lambda}{\lambda} \dfrac{1}{d-1} \, R_{MN} P^{MN} \biggr] \leq \\
    - \dfrac{1-\lambda}{\lambda} \, \dfrac{(d-2) k}{a_\star^2} - \dfrac{\der^2}{\der \sigma^2} \Bigl[ \ln J + \dfrac{\A}{\lambda} & \Bigr] \\
    - \dfrac{1}{D-d} \, \biggl[ \dfrac{\der}{\der \sigma} \Bigl[ \ln J - \Bigl( D - 2 - \dfrac{D-d}{2 \lambda} \Bigr) \A \Bigr] \biggr]^2 &,
\end{split}
\end{equation}
which we can also see as following from the auxiliary-frame Raychaudhuri equation.
Through SSEC, in eq. (\ref{ricci-tensor SSEC}), this reduces to an inequality involving just the geometric terms $J$ and $\A$.
In particular, we can prove that there exists a finite positive constant $\uniformbound > 0$ such that
\begin{equation} \label{J-integral key inequality}
    \int_{0}^{\sigma} \de \sigma' \, J(\sigma', y_\star) \leq \uniformbound.
\end{equation}
Crucially, $\uniformbound$ is independent of the initial position $y_\star$ and of the upper integration limit $\sigma$.
Its finiteness follows from ineq. (\ref{fundamental ineq.}), jointly with the uniform boundedness of smooth initial data ensured by compactness, and the finite $\sigma$-range in ineq. (\ref{sigma-horizon inequality}).
To extend the local bound to the global integration region, we exploit the global hyperbolicity of the reduced auxiliary spacetime: every point in the initial-slice future is reached by a time-like geodesic maximizing auxiliary proper time, and in particular such geodesics start orthogonally and have no focal point before their endpoints \cite{Wald, Witten:2019qhl}.
This requires boundarylessness.
The singular locus of focal points occupies a measure-zero volume, which allows us to bound the global-coordinate integral as $\smash{\int_{t_\star}^t \de s \int \de \mu \, (a/a_\star)^{d-3}} \leq \smash{\int_0^{a_\star L_\star} \de \sigma \int \de \mu_\star \, J}$, by eqs. (\ref{sigma-horizon inequality}, \ref{jacobian volume-element transformation}).
Through eq. (\ref{proof-solving integral}) and ineq. (\ref{J-integral key inequality}), we thus get
\begin{equation} \label{global integral bound}
    \int_{t_\star}^t \de s \, a^{d-3}(s) \leq a_\star^{d-3} \uniformbound.
\end{equation}
Because $\smash{d \geq 3}$, we always have $\smash{1 \leq (a/a_\star)^{d-3} + a_\star/a}$.
Then, as a consequence of the horizon hypothesis in eq. (\ref{would-be horizon condition}), a time integration provides the bound
\begin{equation}
    t - t_\star \leq \uniformbound + a_\star L_\star.
\end{equation}
By finiteness of $\smash{L_\star < \infty}$, this implies that the constant term $\uniformbound$ must be arbitrarily large, which contradicts its finiteness.
Therefore, the horizon assumption in eq. (\ref{would-be horizon condition}) is inherently inconsistent.
It is necessary to conclude, instead, that there is no horizon, i.e. eq. (\ref{no horizon}) is in place.

\section{Discussion} \label{sec.: discussion}

The absence of future cosmic horizons due to SSEC can be interpreted as a minimal remnant, in the lower-dimensional theory, of the attractive character of gravity in the parent theory.
Despite being able to be repulsive, the reduced lower-dimensional gravity cannot repel efficiently enough to form a horizon.
Different kinds of energy-condition constraints on compactifications have appeared before \cite{Das:2019vnx, Bernardo:2022ony, Faruk:2024usu}.

There are several clear future targets.
If we believe that a lack of string-theoretic constructions with future cosmic horizons is evidence for a deeper-level fundamental obstruction, then we may try and reverse the logic of the letter.
This motivates seeking a fundamental origin for a condition like SSEC, including a definite interpretation of $\lambda$.
In that case, an additional task is that of understanding the effective impact of SEC- and/or SSEC-violating sources.
If they cannot provide genuine ways out, then there should be a dynamical reason why their presence does not contradict the absence of horizons.
A further direction is to clarify the role of the Kaluza-Klein vector sector in time-dependent compactifications \cite{Douglas:2008jx, Frey:2008xw}, focusing on mixed metric components of the form $\smash{A^m{}_\mu = A^m \delta^0_\mu}$, with $\smash{\der_i A^m = 0}$.

As an application, our no-horizon theorem has an obvious relationship with the dS and transplanckian censorship conjectures \cite{Obied:2018sgi, Ooguri:2018wrx, Bedroya:2019snp, Rudelius:2021azq}.
For scalar EFTs whose microscopic uplift satisfies the hypotheses of our theorem, and whose potential remains positive throughout the future evolution, the norm of its logarithmic gradient in Planck units along an on-shell trajectory cannot remain uniformly below the critical value $\gamma_d = 2/\sqrt{d-2}$, if $\smash{k=0,-1/\ell^2}$.
This provides a dynamical bound on the supremum of the slope along the trajectory, rather than a point-wise strong-dS bound, in line with previous arguments \cite{Shiu:2023fhb}.
A detailed account is in app. \ref{app.: no horizons and de Sitter and transplanckian censorship conjectures}.

\begin{acknowledgments}
\subsection*{Acknowledgments}

I am deeply indebted to Thomas Van Riet, who pushed the fundamental research questions explored in this letter and who has been providing feedback and insights on this project throughout its development, including comments on the manuscript draft.
Also, I acknowledge the use of OpenAI's GPT-5.6 and GPT-6 Astra for assistance in tightening formal proofs, speeding up calculations, and formulating challenging test cases.
I have independently verified all analytic steps and all calculations, and I take full scientific responsibility for all the contents of this letter.
My work is funded by the European Union-NextGenerationEU/PNRR (mission 4, component 2, CUP C93C24004950006).

\end{acknowledgments}

\bibliographystyle{apsrev4-1}
\bibliography{refs.bib}

\newpage

\appendix
\onecolumngrid

\numberwithin{equation}{section}

\renewcommand{\thesubsection}{\Alph{section}.\arabic{subsection}}

\makeatletter
\renewcommand{\thesubsubsection}{\thesection.\arabic{subsection}.\arabic{subsubsection}}
\makeatother

\section{Dimensional reductions and a simplified no-horizon proof} \label{app. eq.: dimensional reductions and a simplified no-horizon proof}
Here we overview the mathematical derivation of the dimensional reductions of interest, including the energy conditions, and illustrate in detail a simplified version of the no-horizon proof.

\subsection{Higher- and lower-dimensional geometry} \label{subapp.: higher- and lower-dimensional geometry}

Let a $(1,D-1)$-dimensional line element be reduced down to $d>2$ large spacetime dimensions as
\begin{equation} \label{app. eq.: metric}
    d s_{1,D-1}^2 = \e^{2 \tau(x,y)} \bigl[ \e^{-2 \frac{D-d}{d-2} \omega(x)} d \tilde{s}_{1,d-1}^2(x) + \e^{2 \omega(x)} d \bar{s}_{D-d}^2(x,y) \bigr].
\end{equation}
Here, the line element $\smash{d \tilde{s}_{1,d-1}^2 = \tilde{g}_{\mu \nu} \de x^\mu \de x^\nu}$, and the small compact $(D-d)$-dimensional space, with line element $\smash{d \bar{s}_{D-d}^2 = \bar{g}_{mn} \de y^m \de y^n}$, have coordinates $x^\mu$ and $y^m$, with $\mu = 0,1, \dots, d-1$ and $m=1\dots,D-d$, respectively.

Several considerations are in order.
\begin{itemize}
    \item Our interest is in working in the $d$-dimensional Einstein frame.
    A simple scaling argument for the the $D$- and $d$-dimensional Einstein-Hilbert actions reveals that the $d$-dimensional gravitational coupling $\kappa_d$ is such that
    \begin{equation}
        \dfrac{1}{\kappa_d^2} = \dfrac{1}{\kappa_D^2} \int \de^{D-d} y \, \sqrt{\bar{g}_{D-d}} \; \e^{(D-2) \tau},
    \end{equation}
    where $\kappa_D$ is the $D$-dimensional gravitational coupling.
    The $d$-dimensional Einstein frame is defined by the requirement that $\smash{\der_\mu \kappa_d = 0}$, which is equivalent to the condition $\smash{\der_\mu \int \de^{D-d} y \, \sqrt{\bar{g}_{D-d}} \; \e^{(D-2) \tau} = 0}$, i.e.
    \begin{equation} \label{app. eq.: einstein-frame condition}
        \int \de^{D-d} y \, \sqrt{\bar{g}_{D-d}} \; \e^{(D-2) \tau} \Bigl[ \dfrac{1}{2} \, (\bar{g}^{mn} \der_\mu \bar{g}_{mn}) + (D-2) \der_\mu \tau \Bigr] = 0.
    \end{equation}
    \item For convenience, we wish to single out the unique scalar mode $\omega$ that controls the internal volume
    \begin{equation}
        \mathrm{vol}_{D-d} = \int \de^{D-d} y \, \sqrt{g_{D-d}} = \e^{(D-d) \omega} \int \de^{D-d} y \, \sqrt{\bar{g}_{D-d}} \; \e^{(D-d) \tau}.
    \end{equation}
    To enforce this, we require the condition $\smash{\der_\mu \int \de^{D-d} y \, \sqrt{\bar{g}_{D-d}} \; \e^{(D-d) \tau} = 0}$, i.e.
    \begin{equation} \label{app. eq.: omega-volume condition}
        \int \de^{D-d} y \, \sqrt{\bar{g}_{D-d}} \; \e^{(D-d) \tau} \Bigl[ \dfrac{1}{2} \, (\bar{g}^{mn} \der_\mu \bar{g}_{mn}) + (D-d) \der_\mu \tau \Bigr] = 0.
    \end{equation}
    Equivalently, this condition can be understood as the fixing of a gauge redundancy, as follows.
    For any function $\lambda$ such that $\smash{\der_m \lambda = 0}$, the metric in eq. (\ref{app. eq.: metric}) is invariant under the transformations $\smash{\omega \to \omega + \lambda}$, $\smash{\tau \to \tau + [(D-d)/(d-2)] \lambda}$, and $\smash{\bar{g}_{mn} \to \bar{g}_{mn} \, \e^{-2 [(D-2)/(d-2)] \lambda}}$.
    Then, clearly eq. (\ref{app. eq.: omega-volume condition}) requires $\smash{\der_\mu \lambda = 0}$.
\end{itemize}
These conditions are integrated over the internal coordinates and imposed point-wise in the external coordinates.
Their simultaneous fulfilling still allows for non-trivial spacetime-dependent internal profiles satisfying the two integral constraints.

Both eqs. (\ref{app. eq.: einstein-frame condition}, \ref{app. eq.: omega-volume condition}) are off-shell identities, and therefore they can be differentiated.
In particular, we require the vanishing in eq. (\ref{app. eq.: einstein-frame condition}) as an identity for the $d$-dimensional theory, i.e. $\smash{\tilde{\nabla}_\mu \der_\nu \int \de^{D-d} y \, \sqrt{\bar{g}_{D-d}} \; \e^{(D-2) \tau} = 0}$, which explicitly reads
\begin{equation} \label{app. eq.: differentiated einstein-frame condition}
\begin{split}
    \int \de^{D-d} y \, \sqrt{\bar{g}_{D-d}} \; \e^{(D-2) \tau} \Bigl[ & - \dfrac{1}{2} \, (\bar{g}^{mn} \der_\mu \bar{g}_{np}) (\bar{g}^{pq} \der_\nu \bar{g}_{qm}) + \dfrac{1}{2} \, (\bar{g}^{mn} \tilde{\nabla}_\mu \der_\nu \bar{g}_{mn}) + (D-2) \tilde{\nabla}_\mu \der_\nu \tau \\
        & + \dfrac{1}{4} \, (\bar{g}^{mn} \der_\mu \bar{g}_{mn}) (\bar{g}^{pq} \der_\nu \bar{g}_{pq}) \\
        & + (D-2)^2 \der_\mu \tau \der_\nu \tau + \dfrac{D-2}{2} \, (\bar{g}^{mn} \der_\mu \bar{g}_{mn}) \der_\nu \tau + \dfrac{D-2}{2} \, (\bar{g}^{mn} \der_\nu \bar{g}_{mn}) \der_\mu \tau \Bigr] = 0,
\end{split}
\end{equation}
and similarly for eq. (\ref{app. eq.: omega-volume condition}).

\subsubsection{Higher- and lower-dimensional geometric mapping}
Below, we work out the fundamental geometric properties of the spacetime of interest, mapping the $D$-dimensional description into $d$-dimensional terms.

A straightforward calculation allows one to see that the Christoffel symbols are
\begin{subequations}
\begin{align}
    \Gamma^\mu_{\rho \sigma} & = \tilde{\Gamma}^\mu_{\rho \sigma} - \dfrac{D-d}{d-2} \Bigl[ \delta^\mu_\rho \der_\sigma \omega + \delta^\mu_\sigma \der_\rho \omega - \tilde{g}_{\rho \sigma} \tilde{g}^{\mu \nu} \der_\nu \omega \Bigr] + \delta^\mu_\rho \der_\sigma \tau + \delta^\mu_\sigma \der_\rho \tau - \tilde{g}_{\rho \sigma} \tilde{g}^{\mu \nu} \der_\nu \tau, \\[0.50ex]
    \Gamma^\mu_{\rho q} & = \delta^{\mu}_\rho \der_q \tau, \\[0.65ex]
    \Gamma^m_{\rho \sigma} & = - \e^{-2 \frac{D-2}{d-2} \omega} \tilde{g}_{\rho \sigma} \bar{g}^{mn} \der_n \tau, \\[0.35ex]
    \Gamma^m_{p \sigma} & = \delta^m_p \der_\sigma \omega + \dfrac{1}{2} \, \bar{g}^{mn} \der_\sigma \bar{g}_{np} + \delta^m_p \der_\sigma \tau, \\[-0.35ex]
    \Gamma^\mu_{p q} & = - \e^{2 \frac{D-2}{d-2} \omega} \bar{g}_{p q} \tilde{g}^{\mu \nu} \der_\nu \omega - \dfrac{1}{2} \, \e^{2 \frac{D-2}{d-2} \omega} \tilde{g}^{\mu \nu} \der_\nu \bar{g}_{pq} - \e^{2 \frac{D-2}{d-2} \omega} \bar{g}_{p q} \tilde{g}^{\mu \nu} \der_\nu \tau, \\[0.65ex]
    \Gamma^m_{p q} & = \bar{\Gamma}^m_{p q} + \delta^m_p \der_q \tau + \delta^m_q \der_p \tau - \bar{g}_{p q} \bar{g}^{m n} \der_n \tau.
\end{align}
\end{subequations}
Then, the Ricci-tensor components are
\begin{subequations}
\begin{align}
    R_{\mu \nu} & =
    \begin{aligned}[t]
        \tilde{R}_{\mu \nu} & - \biggl[ \dfrac{\e^{- 2 \frac{D-2}{d-2} \omega}}{D-2} \, \dfrac{\bar{\nabla}^2 \e^{(D-2) \tau}}{\e^{(D-2) \tau}} - \dfrac{D-d}{d-2} \tilde{\nabla}^2 \omega \biggr] \, \tilde{g}_{\mu \nu} - (D-d) \, \dfrac{D-2}{d-2} \, \der_\mu \omega \der_\nu \omega \\
        & - \dfrac{1}{4} \, (\bar{g}^{mn} \der_\mu \bar{g}_{np}) (\bar{g}^{pq} \der_\nu \bar{g}_{qm}) - \dfrac{1}{4} \, \tilde{\nabla}_\mu (\bar{g}^{pq} \der_\nu \bar{g}_{pq}) - \dfrac{1}{4} \, \tilde{\nabla}_\nu (\bar{g}^{pq} \der_\mu \bar{g}_{pq}) \\
        & - \dfrac{1}{2} \, \dfrac{D-d}{d-2} \, (\bar{g}^{pq} \der_\rho \bar{g}_{pq}) \, [\delta^\rho_\mu \der_\nu \omega + \delta^\rho_\nu \der_\mu \omega - \tilde{g}_{\mu \nu} \tilde{g}^{\rho \sigma} \der_\sigma \omega] - \dfrac{1}{2} \, (\bar{g}^{pq} \der_\mu \bar{g}_{pq}) \der_\nu \omega - \dfrac{1}{2} \, (\bar{g}^{pq} \der_\nu \bar{g}_{pq}) \der_\mu \omega \\
        & - \biggl[ \dfrac{1}{D-2} \, \dfrac{\tilde{\nabla}^{2} \e^{(D-2) \tau}}{\e^{(D-2) \tau}} + \dfrac{1}{2} \, \tilde{g}^{\rho \sigma} \, (\bar{g}^{m n} \der_\rho \bar{g}_{m n}) \der_\sigma \tau \biggr] \tilde{g}_{\mu \nu} \\
        & - (D-2) \, \dfrac{D-d}{d-2} \, [\der_\mu \omega \der_\nu \tau + \der_\nu \omega \der_\mu \tau - \tilde{g}^{\rho \sigma} \der_\rho \omega \, \der_\sigma \tau \, \tilde{g}_{\mu \nu}] + (D-2) \, \e^{\tau} \, \tilde{\nabla}_\mu \der_\nu \, \e^{- \tau},
    \end{aligned} \label{app. eq.: Rmunu} \\[1.4ex]
    R_{\mu n} & =
    \begin{aligned}[t]
        (D-2) \, \der_\mu \omega \der_n \tau & + \dfrac{1}{2} \, \bar{\nabla}_p (\bar{g}^{pq} \der_\mu \bar{g}_{nq}) + \dfrac{D-2}{2} \, \der_p \tau \, (\bar{g}^{pq} \der_\mu \bar{g}_{qn}) - \dfrac{1}{2} \, \der_n (\bar{g}^{pq} \der_\mu \bar{g}_{pq}) \\[0.85ex]
        & - (D-2) \, \der_\mu \der_n \tau + (D-2) \, \der_\mu \tau \der_n \tau,
    \end{aligned} \label{app. eq.: Rmun} \\[1.4ex]
    R_{m n} & =
    \begin{aligned}[t]
        \bar{R}_{m n} & + (D-2) \, \e^{\tau} \, \bar{\nabla}_m \der_n \, \e^{- \tau} - \biggl[\dfrac{1}{D-2} \, \dfrac{\bar{\nabla}^2 \e^{(D-2) \tau}}{\e^{(D-2) \tau}} + \e^{2 \frac{D-2}{d-2} \omega} \tilde{\nabla}^2 \omega \biggr] \bar{g}_{m n} \\[1.0ex]
        & + \e^{2 \frac{D-2}{d-2} \omega} \Bigl[ - \dfrac{1}{2} \, \tilde{\nabla}^2 \bar{g}_{mn} + \dfrac{1}{2} \, \tilde{g}^{\rho \sigma} \bar{g}^{pq} \der_\rho \bar{g}_{mp} \der_\sigma \bar{g}_{nq} - \dfrac{1}{2} \, \tilde{g}^{\rho \sigma} (\bar{g}^{pq} \der_\rho \bar{g}_{pq}) \Bigl( \bar{g}_{mn} \der_\sigma \omega + \dfrac{1}{2} \, \der_\sigma \bar{g}_{mn} \Bigr) \Bigr] \\[0.4ex]
        & - \e^{2 \frac{D-2}{d-2} \omega} \Bigl[ \dfrac{1}{D-2} \, \dfrac{\tilde{\nabla}^2 \e^{(D-2) \tau}}{\e^{(D-2) \tau}} + (D-2) \, \tilde{g}^{\rho \sigma} \der_\rho \omega \, \der_\sigma \tau + \dfrac{1}{2} \, \tilde{g}^{\rho \sigma} (\bar{g}^{p q} \der_\rho \bar{g}_{p q}) \der_\sigma \tau \Bigr] \bar{g}_{m n} \\
        & - \dfrac{D-2}{2} \, \e^{2 \frac{D-2}{d-2} \omega} \tilde{g}^{\rho \sigma} \der_\rho \bar{g}_{m n} \der_\sigma \tau,
    \end{aligned} \label{app. eq.: Rmn}
\end{align}
\end{subequations}
and the Ricci scalar is
\begin{equation} \label{app. eq.: R}
\begin{split}
    R = \e^{-2 \tau} \e^{2 \frac{D-d}{d-2} \omega} \biggl[ \tilde{R} & + 2 \, \dfrac{D-d}{d-2} \, \tilde{\nabla}^2 \omega - (D-d) \, \dfrac{D-2}{d-2} \, \der \omega \, \tilde{\cdot} \, \der \omega \\[-0.65ex]
    & + \e^{-2 \frac{D-2}{d-2} \omega} \Bigl[ \bar{R} - 2 \, \dfrac{D-1}{D-2} \dfrac{\bar{\nabla}^2 \e^{(D-2) \tau}}{\e^{(D-2) \tau}} + (D-1)(D-2) \, \der \tau \, \bar{\cdot} \, \der \tau \Bigr] \\
    & - \dfrac{1}{2} \, \bar{g}^{pq} \tilde{\nabla}^2 \bar{g}_{pq} + \dfrac{1}{4} \, \tilde{g}^{\rho \sigma} \bigl[ (\bar{g}^{mn} \der_\rho \bar{g}_{np}) (\bar{g}^{pq} \der_\sigma \bar{g}_{qm}) - (\bar{g}^{pq} \der_\rho \bar{g}_{pq}) (\bar{g}^{rs} \der_\sigma \bar{g}_{rs}) \bigr] \\
    & - \dfrac{1}{2} \, \tilde{g}^{\rho \sigma} \tilde{\nabla}_\rho (\bar{g}^{rs} \der_\sigma \bar{g}_{rs}) - \tilde{g}^{\rho \sigma} \der_\rho \omega \, (\bar{g}^{pq} \der_\sigma \bar{g}_{pq}) \\
    & - 2 \, \dfrac{D-1}{D-2} \, \dfrac{\tilde{\nabla}^2 \e^{(D-2) \tau}}{\e^{(D-2) \tau}} + (D-1) (D-2) \, \der \tau \, \tilde{\cdot} \, \der \tau - (D-1) \, \tilde{g}^{\rho \sigma} (\bar{g}^{mn} \der_\rho \bar{g}_{mn}) \, \der_\sigma \tau \biggr].
\end{split}
\end{equation}

One can also verify that the $D$-dimensional Laplace operator on a scalar function $f=f(x,y)$ as
\begin{equation} \label{app. eq.: nabla2}
    \nabla^2 f = \e^{-2 \tau} \, \e^{2 \frac{D-d}{d-2} \omega} \biggl[ \tilde{\nabla}^2 f + \Bigl[ \dfrac{1}{2} \, (\bar{g}^{mn} \der \bar{g}_{mn}) + (D-2) \der \tau \Bigr] \, \tilde{\cdot} \, \der f + \e^{-2 \frac{D-2}{d-2} \omega} \bigl[ \bar{\nabla}^2 f + (D-2) \, \der \tau \, \bar{\cdot} \, \der f \bigr] \biggr].
\end{equation}

\subsubsection{FLRW-spacetime}
Let the $d$-dimensional metric be
\begin{equation} \label{app. eq.: d-dimensional FLRW-metric}
    d \tilde{s}_{1,d-1}^2 = - \de t^2 + a^2(t) \, d \mathrm{l}_{d-1}^2,
\end{equation}
where the scale factor is only a function of cosmological time, i.e. $a=a(t)$, and $\smash{d \mathrm{l}_{d-1}^2 = \mathrm{g}_{ij} \de x^i \de x^j}$ represents one of the three maximally-symmetric spaces: the Euclidean space $\mathbb{E}^{d-1}$, the hyperboloid $\mathbb{H}^{d-1}$, or the sphere $\mathrm{S}^{d-1}$, with cosmological curvature $k = 0$, $-1/\ell^2$, or $+1/\ell^2$, respectively.
Their Riemann tensor reads $\smash{\mathrm{R}_{i j k l} = k (\mathrm{g}_{ik} \mathrm{g}_{jl} - \mathrm{g}_{il} \mathrm{g}_{jk})}$, implying Ricci tensor $\smash{\mathrm{R}_{ij} = k (d-2) \, \mathrm{g}_{ij}}$ and Ricci scalar $\smash{\mathrm{R} = (d-1)(d-2) k}$.
Given the Hubble parameter $H = \dot{a}/a$, we define the acceleration parameter $\smash{\epsilon = - \dot{H}/H^2 = 1 - \ddot{a} a / \dot{a}^2}$.
Then, one can write
\begin{subequations}
    \begin{align}
        \tilde{R}_{00} & = - (d-1) \, \dfrac{\ddot{a}}{a} = - (d-1) (1 - \epsilon) H^2, \\
        \tilde{R}_{ij} & = \bigl[ a \ddot{a} + (d-2) \, (\dot{a}^2 + k) \bigr] \, \mathrm{g}_{ij} = \Bigl[ (d - 1 - \epsilon) \, H^2 + (d-2) \dfrac{k}{a^2} \Bigr] \, \tilde{g}_{ij}, \label{app. eq.: FLRW Rij}
    \end{align}
\end{subequations}
and
\begin{equation}
    \tilde{R} = (d-1) \Bigl[ 2 \, \dfrac{\ddot{a}}{a} + (d-2) \, \dfrac{\dot{a}^2 + k}{a^2} \Bigr] = (d-1) \Bigl[ (d - 2 \epsilon) H^2 + (d-2) \dfrac{k}{a^2} \Bigr].
\end{equation}
Another useful identity is $\smash{\tilde{\nabla}^2 f = - \ddot{f} - (d-1) H \dot{f}}$ for every function $f=f(t)$ of just cosmological time.
Also, the Christoffel symbols are
\begin{subequations}
\begin{align}
    \tilde{\Gamma}^0_{j k} & = a \dot{a} \, \mathrm{g}_{j k} = H \tilde{g}_{jk}, \label{app. eq.: FLRW Gamma0jk} \\
    \tilde{\Gamma}^i_{j 0} & = \dfrac{\dot{a}}{a} \, \delta^i_j = H \delta^i_j, \label{app. eq.: FLRW Gammaij0} \\
    \tilde{\Gamma}^i_{j k} & = \Upgamma^i_{j k}. \label{app. eq.: FLRW Gammaijk}
\end{align}
\end{subequations}
For later convenience, we rewrite the metric in spherical coordinates as
\begin{equation} \label{app. eq.: d-dimensional FLRW-metric, spherical coordinates}
    d \tilde{s}_{1,d-1}^2 = - \de t^2 + a^2(t) \, \Bigl[ \dfrac{\de r^2}{1 - k r^2} + r^2 \, \de \Omega_{d-2}^2 \Bigr].
\end{equation}
After writing the angular components for the unit $(d-2)$-sphere via $\smash{\de \Omega_{d-2}^2 = \mathrm{h}_{ab} \de \phi^a \de \phi^b}$, the only non-zero components of the purely-spatial Christoffel symbols are $\smash{\Upgamma^r_{rr} = kr/(1-kr^2)}$, $\smash{\Upgamma^r_{ab} = - r (1-kr^2) \mathrm{h}_{ab}}$, $\smash{\Upgamma^a_{rb} = \Upgamma^a_{br} = \delta^a_b/r}$ and $\smash{\Upgamma^{a}_{bc} = \upgamma^a_{bc}}$, where $\smash{\upgamma^a_{bc}}$ are the Christoffel symbols for such unit $(d-2)$-sphere.

To preserve the FLRW isometries, we also need to restrict higher-dimensional metric functions to be such that $\smash{\der_i \omega = 0}$, $\smash{\der_i \tau = 0}$, and $\smash{\der_i \bar{g}_{mn} = 0}$.

\subsection{Einstein/Friedmann and continuity equations} \label{subapp.: Einstein/Friedmann and continuity equations}
Now, we illustrate the manipulations that allow one to express the $D$-dimensional Einstein equations as $d$-dimensional Friedmann equations.

\subsubsection{Einstein/Friedmann equations}

The trace-reversed $D$-dimensional Einstein equations read
\begin{equation}
    R_{MN} = \kappa_D^2 \Bigl[ T_{MN} - \dfrac{1}{D-2} \, g_{MN} (g^{PQ} T_{PQ}) \Bigr],
\end{equation}
where $T_{MN}$ is the $D$-dimensional energy-momentum tensor and $\kappa_D$ is the $D$-dimensional gravitational coupling.

To recover the $d$-dimensional Friedmann equations, the relevant components of the trace-reversed Einstein equations are the $(0,0)$-component, the $d$-dimensional trace, and the $D$-dimensional trace, or equivalently the $(D-d)$-dimensional trace, for which we exploit the reductions of the $D$-dimensional Ricci tensor block-diagonal terms and scalar in eqs. (\ref{app. eq.: Rmunu}, \ref{app. eq.: Rmn}) and eq. (\ref{app. eq.: R}), respectively, and which we express via the identities $\smash{E_\alpha = 0}$, with $\alpha = 0, d, D, D-d$ in a self-explanatory notation.
Explicitly, we have\footnote{It is useful to observe the identity
\begin{align*}
    \dfrac{1}{D-2} \, \dfrac{\tilde{\nabla}^2 \e^{(D-2) \tau}}{\e^{(D-2) \tau}} & = - \ddot{\tau}  - (d-1) H \dot{\tau} - (D-2) \dot{\tau}^2,
\end{align*}
as well as
\begin{align*}
    \tilde{\nabla}_0 [\bar{g}^{mn} \dot{\bar{g}}_{mn}] & =
    - (\bar{g}^{mn} \dot{\bar{g}}_{np})
    (\bar{g}^{pq} \dot{\bar{g}}_{qm})
    + \bar{g}^{mn} \ddot{\bar{g}}_{mn}, \\
    \e^{\tau} \, \tilde{\nabla}_0 [- \e^{- \tau} \dot{\tau}] & = - \ddot{\tau} + \dot{\tau}^2,
\end{align*}
and
\begin{align*}
    \tilde{g}^{\mu \nu} \tilde{\nabla}_\mu [\bar{g}^{mn} \der_\nu \bar{g}_{mn}]
    & =
    \begin{aligned}[t]
        & - \bar{g}^{mn} \ddot{\bar{g}}_{mn} - (d-1) H (\bar{g}^{mn} \dot{\bar{g}}_{mn}) \\
        & + (\bar{g}^{mn} \dot{\bar{g}}_{np}) (\bar{g}^{pq} \dot{\bar{g}}_{qm}),
    \end{aligned} \\
    \e^{\tau} \, \tilde{g}^{\mu \nu} \tilde{\nabla}_\mu \der_\nu \, \e^{- \tau} & = \ddot{\tau} + (d-1) H \dot{\tau} -\dot{\tau}^2.
\end{align*}
}
\begin{subequations}
\begin{align}
    E_0 & =
    \begin{aligned}[t]
        - R_{00} & - (d-1) (1 - \epsilon) H^2 - (D-d) \, \dfrac{D-2}{d-2} \, \dot{\omega}^2 + \dfrac{D-d}{d-2} \, [\ddot{\omega} + (d-1) H \dot{\omega}] + \dfrac{\e^{-2 \frac{D-2}{d-2} \omega}}{D-2} \, \dfrac{\bar{\nabla}^2 \e^{(D-2) \tau}}{\e^{(D-2) \tau}} \\[0.4ex]
        & + \dfrac{1}{4} \, (\bar{g}^{mn} \dot{\bar{g}}_{np}) (\bar{g}^{pq} \dot{\bar{g}}_{qm}) - \dfrac{1}{2} \, (\bar{g}^{mn} \ddot{\bar{g}}_{mn}) - (D-1) \, \ddot{\tau} - (d-1) H \dot{\tau} \\
        & - \dfrac{1}{2} \, \dfrac{(D-2) + (d-2)}{d-2} \, (\bar{g}^{mn} \dot{\bar{g}}_{mn}) \dot{\omega} - (D-2) \, \dfrac{D-d}{d-2} \, \dot{\tau} \dot{\omega} - \dfrac{1}{2} \, (\bar{g}^{mn} \dot{\bar{g}}_{mn}) \dot{\tau},
    \end{aligned} \label{app. eq.: E0-eq.} \\
    E_d & =
    \begin{aligned}[t]
         - \tilde{g}^{\mu \nu} R_{\mu \nu} & + (d-1) \Bigl[ (d - 2 \epsilon) H^2 + (d-2) \dfrac{k}{a^2} \Bigr] + (D-d) \, \dfrac{D-2}{d-2} \, \dot{\omega}^2 - d \, \dfrac{D-d}{d-2} \, [\ddot{\omega} + (d-1) H \dot{\omega}] \\[-0.65ex]
         & - \dfrac{d}{D-2} \, \e^{-2 \frac{D-2}{d-2} \omega} \dfrac{\bar{\nabla}^2 \e^{(D-2) \tau}}{\e^{(D-2) \tau}} - \dfrac{1}{4} \, (\bar{g}^{mn} \dot{\bar{g}}_{np}) (\bar{g}^{pq} \dot{\bar{g}}_{qm}) + \dfrac{1}{2} \, \bar{g}^{mn} [\ddot{\bar{g}}_{mn} + (d-1) H \dot{\bar{g}}_{mn}] \\[0.85ex]
         & + [(D-2)+d] \, [\ddot{\tau} + (d-1) H \dot{\tau}] + (d-1) (D-2) \dot{\tau}^2 \\[0.65ex]
         & - \dfrac{1}{2} \, [(D-2) - d] \, (\bar{g}^{mn} \dot{\bar{g}}_{mn}) \dot{\omega} - (D-d)(D-2) \dot{\tau} \dot{\omega} + \dfrac{d}{2} \, (\bar{g}^{mn} \dot{\bar{g}}_{mn}) \dot{\tau},
    \end{aligned} \label{app. eq.: E_d-eq.} \\
    E_D & =
    \begin{aligned}[t]
        - \e^{2 \tau} \, \e^{-2 \frac{D-d}{d-2} \omega} R & + (d-1) \Bigl[ (d - 2 \epsilon) H^2 + (d-2) \dfrac{k}{a^2} \Bigr] + (D-d) \, \dfrac{D-2}{d-2} \, \dot{\omega}^2 - 2 \, \dfrac{D-d}{d-2} \, [\ddot{\omega} + (d-1) H \dot{\omega}] \\
        & + \e^{-2 \frac{D-2}{d-2} \omega} \biggl[ \bar{R} - 2 \, \dfrac{D-1}{D-2} \dfrac{\bar{\nabla}^2 \e^{(D-2) \tau}}{\e^{(D-2) \tau}} + (D-1)(D-2) \, \der \tau \, \bar{\cdot} \, \der \tau \biggr] \\[-0.45ex]
        & + \bar{g}^{mn} [\ddot{\bar{g}}_{mn} + (d-1) H \, \dot{\bar{g}}_{mn}] + \dfrac{1}{4} \, (\bar{g}^{mn} \dot{\bar{g}}_{mn}) (\bar{g}^{pq} \dot{\bar{g}}_{pq}) - \dfrac{3}{4} \, (\bar{g}^{mn} \dot{\bar{g}}_{np}) (\bar{g}^{pq} \dot{\bar{g}}_{qm}) \\[1.55ex]
        & + 2 (D-1) [\ddot{\tau} + (d-1) H \dot{\tau}] + (D-1)(D-2) \dot{\tau}^2 + (\bar{g}^{mn} \dot{\bar{g}}_{mn}) \dot{\omega} + (D-1) \, (\bar{g}^{mn} \dot{\bar{g}}_{mn}) \dot{\tau},
    \end{aligned} \label{app. eq.: E_D-eq.}
\end{align}
\end{subequations}
as well as
\begin{equation} \label{app. eq.: Eomega-eq.}
\begin{split}
    E_{D-d} = & - \e^{-2 \frac{D-2}{d-2} \omega} \bar{g}^{mn} R_{mn} + (D-d) \, [\ddot{\omega} + (d-1) H \dot{\omega}] \\[0.65ex]
    & + \e^{-2 \frac{D-2}{d-2} \omega} \biggl[ \bar{R} - \dfrac{2(D-2)-(d-2)}{D-2} \dfrac{\bar{\nabla}^2 \e^{(D-2) \tau}}{\e^{(D-2) \tau}} + (D-1)(D-2) \, \der \tau \, \bar{\cdot} \, \der \tau \biggr] \\
    & + \dfrac{1}{2} \, \bar{g}^{mn} [\ddot{\bar{g}}_{mn} + (d-1) H \dot{\bar{g}}_{mn}] - \dfrac{1}{2} \, (\bar{g}^{mn} \dot{\bar{g}}_{np}) (\bar{g}^{pq} \dot{\bar{g}}_{qm}) + \dfrac{1}{4} \, (\bar{g}^{mn} \dot{\bar{g}}_{mn}) (\bar{g}^{pq} \dot{\bar{g}}_{pq}) \\[1.25ex]
    & + (D-d) \, [\ddot{\tau} + (d-1) H \dot{\tau}] + (D-d) (D-2) \dot{\tau}^2 \\[0.65ex]
    & + \dfrac{1}{2} \, (D-d) \, (\bar{g}^{mn} \dot{\bar{g}}_{mn}) \dot{\omega} + (D-d) (D-2) \dot{\tau} \dot{\omega} + \dfrac{(D-d)+(D-2)}{2} \, (\bar{g}^{mn} \dot{\bar{g}}_{mn}) \dot{\tau},
\end{split}
\end{equation}
where one can verify the identity $\smash{E_d + E_{D-d} = E_D}$.

As far as the matter sector is concerned, we notice the identities
\begin{subequations}
\begin{align}
    R_{00} - \dfrac{1}{d-2} \, \e^{-2 \frac{D-2}{d-2} \omega} \bar{g}^{mn} R_{mn} & = \dfrac{\kappa_D^2}{d-2} \, [(d-3) \, T_{00} + \tilde{g}^{ij} T_{ij}], \label{app. eq.: master E0-eq. lemma} \\
    - \tilde{g}^{\mu \nu} R_{\mu \nu} - \dfrac{d}{d-2} \, \e^{-2 \frac{D-2}{d-2} \omega} \bar{g}^{mn} R_{mn} & = \dfrac{2 \kappa_D^2}{d-2} \, [- T_{00} + \tilde{g}^{ij} T_{ij}], \label{app. eq.: master E_d-eq. lemma}
\end{align}
\end{subequations}
as well as
\begin{equation} \label{app. eq.: master Eomega-eq. lemma}
    - \dfrac{1}{d-2} \, \e^{-2 \frac{D-2}{d-2} \omega} \bar{g}^{mn} R_{mn} = \dfrac{D-d}{D-2} \dfrac{\kappa_D^2}{d-2} \, \Bigl[ - T_{00} + \tilde{g}^{ij} T_{ij} - \dfrac{d-2}{D-d} \, \e^{-2 \frac{D-2}{d-2} \omega} \bar{g}^{mn} T_{mn} \Bigr].
\end{equation}

Now, by plugging eq. (\ref{app. eq.: Eomega-eq.}) into eqs. (\ref{app. eq.: E0-eq.}, \ref{app. eq.: E_d-eq.}), and by exploiting eqs. (\ref{app. eq.: master E0-eq. lemma}, \ref{app. eq.: master E_d-eq. lemma}) and eq. (\ref{app. eq.: master Eomega-eq. lemma}), we can simplify $E_0$ and $E_d$ after substituting the $\smash{[\ddot{\omega} + (d-1) H \dot{\omega}]}$-terms from them, as well as simplify $E_{D-d}$.
To determine the Friedmann equations, we then perform the linear combinations $\smash{F_1 = (2 E_0 + E_d)/(d-2)}$ and $\smash{F_2 = (d E_0 + E_d)/[(d-1)(d-2)]}$.
In both cases, it is convenient to group together combinations that vanish upon an appropriately weighted internal-space integral, according to eqs. (\ref{app. eq.: einstein-frame condition}, \ref{app. eq.: differentiated einstein-frame condition}).
In particular, in the $d$-dimensional FLRW geometry of interest, these read
\begin{align*}
    & \int \de^{D-d} y \, \sqrt{\bar{g}_{D-d}} \; \e^{(D-2) \tau} \Bigl[ \dfrac{1}{2} \, (\bar{g}^{mn} \dot{\bar{g}}_{mn}) + (D-2) \dot{\tau} \Bigr] = 0, \\
    & \int \de^{D-d} y \, \sqrt{\bar{g}_{D-d}} \; \e^{(D-2) \tau} \biggl[ \dfrac{1}{2} \, (\bar{g}^{mn} \ddot{\bar{g}}_{mn}) - \dfrac{1}{2} \, (\bar{g}^{mn} \dot{\bar{g}}_{np}) (\bar{g}^{pq} \dot{\bar{g}}_{qm}) + (D-2) \, \ddot{\tau} + \Bigl[ \dfrac{1}{2} \, (\bar{g}^{mn} \dot{\bar{g}}_{mn}) + (D-2) \dot{\tau} \Bigr]^2 \biggr] = 0.
\end{align*}
With this in mind, we obtain
\begin{subequations}
\begin{align}
    F_1 & =
    \begin{aligned}[t]
        & - \dfrac{E_{D-d}}{d-2} - \dfrac{2 \kappa_D^2}{d-2} \, T_{00} + (d-1) \Bigl[ H^2 + \dfrac{k}{a^2} \Bigr] + \dfrac{\e^{-2 \frac{D-2}{d-2} \omega}}{d-2} \biggl[ \bar{R} - 2 \, \dfrac{\bar{\nabla}^2 \e^{(D-2) \tau}}{\e^{(D-2) \tau}} + (D-1)(D-2) \, \der \tau \, \bar{\cdot} \, \der \tau \biggr] \\
        & - \dfrac{D-d}{d-2} \, \dfrac{D-2}{d-2} \, \dot{\omega}^2 + \dfrac{D-2}{d-2} \, \dot{\tau} \, [\dot{\tau} + 2 \dot{\omega}] - \dfrac{1}{4} \, \dfrac{1}{d-2} \, (\bar{g}^{mn} \dot{\bar{g}}_{np}) (\bar{g}^{pq} \dot{\bar{g}}_{qm}) \\
        & + \dfrac{1}{d-2} \, \Bigl[ \dfrac{1}{2} \, (\bar{g}^{mn} \dot{\bar{g}}_{mn}) + (D-2) \dot{\tau} \Bigr]^2 + \dfrac{2}{d-2} \, \Bigl[ (d-1) H - \dfrac{D-2}{d-2} \, \dot{\omega} \Bigr] \Bigl[ \dfrac{1}{2} \, (\bar{g}^{mn} \dot{\bar{g}}_{mn}) + (D-2) \dot{\tau} \Bigr],
    \end{aligned} \label{app. eq.: preliminary master F1-eq.} \\
    F_2 & =
    \begin{aligned}[t]
        & - \dfrac{\kappa_D^2}{d-2} \, \Bigl[ T_{00} + \dfrac{\tilde{g}^{ij} T_{ij}}{d-1} \Bigr] + \Bigl[ \epsilon H^2 + \dfrac{k}{a^2} \Bigr] - \dfrac{D-d}{d-2} \, \dfrac{D-2}{d-2} \, \dot{\omega}^2 + \dfrac{D-2}{d-2} \, \dot{\tau} \, [\dot{\tau} + 2 \dot{\omega}] \\
        & - \dfrac{1}{4} \, \dfrac{1}{d-2} \, (\bar{g}^{mn} \dot{\bar{g}}_{np}) (\bar{g}^{pq} \dot{\bar{g}}_{qm}) + \dfrac{1}{d-2} \, \Bigl[ H - 2 \, \dfrac{D-2}{d-2} \, \dot{\omega} \Bigr] \Bigl[ \dfrac{1}{2} \, (\bar{g}^{mn} \dot{\bar{g}}_{mn}) + (D-2) \dot{\tau} \Bigr] \\
        & - \dfrac{1}{d-2} \, \biggl[ \dfrac{1}{2} \, (\bar{g}^{mn} \ddot{\bar{g}}_{mn}) - \dfrac{1}{2} \, (\bar{g}^{mn} \dot{\bar{g}}_{np}) (\bar{g}^{pq} \dot{\bar{g}}_{qm}) + (D-2) \ddot{\tau} \biggr],
    \end{aligned} \label{app. eq.: preliminary master F2-eq.}
\end{align}
\end{subequations}
and we still have
\begin{equation} \label{app. eq.: preliminary master Eomega-eq.}
\begin{split}
    \dfrac{E_{D-d}}{d-2} = & - \dfrac{D-d}{D-2} \dfrac{\kappa_D^2}{d-2} \, \Bigl[ T_{00} - \tilde{g}^{ij} T_{ij} + \dfrac{d-2}{D-d} \, \e^{-2 \frac{D-2}{d-2} \omega} \bar{g}^{mn} T_{mn} \Bigr] + \dfrac{D-d}{d-2} \, [\ddot{\omega} + (d-1) H \dot{\omega}] \\[0.65ex]
    & + \dfrac{\e^{-2 \frac{D-2}{d-2} \omega}}{d-2} \biggl[ \bar{R} - \dfrac{2 (D-2) - (d-2)}{D-2} \dfrac{\bar{\nabla}^2 \e^{(D-2) \tau}}{\e^{(D-2) \tau}} + (D-1)(D-2) \, \der \tau \, \bar{\cdot} \, \der \tau \biggr] \\
    & - \Bigl[ \ddot{\tau} + (d-1) H \dot{\tau} + (D-2) \dot{\tau}^2 + \dfrac{1}{2} \, (\bar{g}^{mn} \dot{\bar{g}}_{mn}) \,
    \dot{\tau} \Bigr] \\
    & + \Bigl[ \dfrac{d-1}{d-2} \, H + \dfrac{D-d}{d-2} \, \dot{\omega} \Bigr] \Bigl[ \dfrac{1}{2} \, (\bar{g}^{mn} \dot{\bar{g}}_{mn}) + (D-2) \dot{\tau} \Bigr] \\
    & + \dfrac{1}{d-2} \, \biggl[ \dfrac{1}{2} \, (\bar{g}^{mn} \ddot{\bar{g}}_{mn}) - \dfrac{1}{2} \, (\bar{g}^{mn} \dot{\bar{g}}_{np}) (\bar{g}^{pq} \dot{\bar{g}}_{qm}) + (D-2) \ddot{\tau} + \Bigl[ \dfrac{1}{2} \, (\bar{g}^{mn} \dot{\bar{g}}_{mn}) + (D-2) \dot{\tau} \Bigr]^2 \biggr],
\end{split}
\end{equation}
where in particular we also observe the identity
\begin{align*}
    \dfrac{1}{a^{d-1}} \dfrac{1}{\sqrt{\bar{g}_{D-d}} \; \e^{(D-2) \tau}} \, \dfrac{\de}{\de t} \bigl[\sqrt{\bar{g}_{D-d}} \; a^{d-1} \, \e^{(D-2) \tau} \, \dot{\tau}\bigr] = \ddot{\tau} + (d-1) H \dot{\tau} + (D-2) \dot{\tau}^2 + \dfrac{1}{2} \, (\bar{g}^{mn} \dot{\bar{g}}_{mn}) \,
    \dot{\tau}.
\end{align*}

As well as combinations vanishing by virtue of eqs. (\ref{app. eq.: einstein-frame condition}, \ref{app. eq.: differentiated einstein-frame condition}) when suitably integrated, we see that eqs. (\ref{app. eq.: preliminary master F1-eq.}, \ref{app. eq.: preliminary master F2-eq.}) and eq. (\ref{app. eq.: preliminary master Eomega-eq.}) involve additional recurring combinations of terms.
It is convenient to define the functions
\begin{subequations}
\begin{align}
    \varphi & = \dfrac{1}{2} \, \bar{g}^{mn} \dot{\bar{g}}_{mn} + (D-2) \dot{\tau}, \label{app. eq.: varphi} \\
    \theta^m{}_n & = \dfrac{1}{2} \Bigl[ (\bar{g}^{mp} \dot{\bar{g}}_{pn}) - \dfrac{1}{D-d} \, (\bar{g}^{pq} \dot{\bar{g}}_{pq}) \, \delta^m_n \Bigr]. \label{app. eq.: theta}
\end{align}
\end{subequations}
In particular, $\varphi$ is simply the integrand of the weighted integral defining the Einstein-frame in eq. (\ref{app. eq.: einstein-frame condition}).
On the other hand, $\smash{\theta^m{}_n}$ represents the would-be square root of the kinetic term for the volume-preserving deformations.
By construction, it is traceless, i.e. $\smash{\theta^m{}_m = 0}$.
Several useful arrangements follow.
\begin{itemize}
    \item Because
    \begin{align*}
        \dfrac{\de}{\de t} \, (\bar{g}^{mp} \dot{\bar{g}}_{pn}) = \bar{g}^{mp} \ddot{\bar{g}}_{pn} - (\bar{g}^{mp} \dot{\bar{g}}_{pq}) (\bar{g}^{qr} \dot{\bar{g}}_{rn}),
    \end{align*}
    we have
    \begin{align*}
        \dfrac{1}{2} \, (\bar{g}^{mn} \ddot{\bar{g}}_{mn}) - \dfrac{1}{2} \, (\bar{g}^{mn} \dot{\bar{g}}_{np}) (\bar{g}^{pq} \dot{\bar{g}}_{qm}) + (D-2) \ddot{\tau} = \dot{\varphi}.
    \end{align*}
    \item After observing the identities
    \begin{align*}
        \dfrac{D-d}{d-2} \, \dfrac{D-2}{d-2} \, \dot{\omega}^2 - \dfrac{D-2}{d-2} \, \dot{\tau} \, [\dot{\tau} + 2 \dot{\omega}] & = \dfrac{D-d}{d-2} \, \dfrac{D-2}{d-2} \, \Bigl[ \dot{\omega} - \dfrac{d-2}{D-d} \, \dot{\tau} \Bigr]^2 - \dfrac{D-2}{D-d} \dfrac{D-2}{d-2} \, \dot{\tau}^2, \\
        \theta^m{}_n \theta^n{}_m & = \dfrac{1}{4} \Bigl[ (\bar{g}^{mn} \dot{\bar{g}}_{np}) (\bar{g}^{pq} \dot{\bar{g}}_{qm}) - \dfrac{1}{D-d} \, (\bar{g}^{mn} \dot{\bar{g}}_{mn})^2 \Bigr],
    \end{align*}
    we can write
    \begin{align*}
        & - \dfrac{D-d}{d-2} \, \dfrac{D-2}{d-2} \, \dot{\omega}^2 + \dfrac{D-2}{d-2} \, \dot{\tau} \, [\dot{\tau} + 2 \dot{\omega}] - \dfrac{1}{4} \, \dfrac{1}{d-2} \, (\bar{g}^{mn} \dot{\bar{g}}_{np}) (\bar{g}^{pq} \dot{\bar{g}}_{qm}) \\
        = & - \dfrac{D-d}{d-2} \, \dfrac{D-2}{d-2} \, \Bigl[ \dot{\omega} - \dfrac{d-2}{D-d} \, \dot{\tau} \Bigr]^2 - \dfrac{1}{d-2} \, \theta^m{}_n \theta^n{}_m - \dfrac{1}{d-2} \dfrac{1}{D-d} \, \varphi [\varphi - 2 (D-2) \dot{\tau}].
    \end{align*}
\end{itemize}
Therefore, we can conveniently express eqs. (\ref{app. eq.: preliminary master F1-eq.}, \ref{app. eq.: preliminary master F2-eq.}) as
\begin{subequations}
\begin{align}
    0 & =
    \begin{aligned}[t]
        & - \dfrac{2 \kappa_D^2}{d-2} \, T_{00} + (d-1) \Bigl[ H^2 + \dfrac{k}{a^2} \Bigr] + \dfrac{\e^{-2 \frac{D-2}{d-2} \omega}}{d-2} \biggl[ \bar{R} - 2 \, \dfrac{\bar{\nabla}^2 \e^{(D-2) \tau}}{\e^{(D-2) \tau}} + (D-1)(D-2) \, \der \tau \, \bar{\cdot} \, \der \tau \biggr] \\
        & - \dfrac{D-d}{d-2} \, \dfrac{D-2}{d-2} \, \Bigl[ \dot{\omega} - \dfrac{d-2}{D-d} \, \dot{\tau} \Bigr]^2 - \dfrac{1}{d-2} \, \theta^m{}_n \theta^n{}_m \\
        & + \dfrac{D-d-1}{D-d} \dfrac{\varphi^2}{d-2} + \dfrac{2}{d-2} \, \biggl[ (d-1) H - \dfrac{D-2}{d-2} \, \Bigl[ \dot{\omega} - \dfrac{d-2}{D-d} \, \dot{\tau} \Bigr] \biggr] \varphi,
    \end{aligned} \label{app. eq.: master F1-eq.} \\
    0 & =
    \begin{aligned}[t]
        & - \dfrac{\kappa_D^2}{d-2} \, \Bigl[ T_{00} + \dfrac{\tilde{g}^{ij} T_{ij}}{d-1} \Bigr] + \Bigl[ \epsilon H^2 + \dfrac{k}{a^2} \Bigr] - \dfrac{D-d}{d-2} \, \dfrac{D-2}{d-2} \, \Bigl[ \dot{\omega} - \dfrac{d-2}{D-d} \, \dot{\tau} \Bigr]^2 - \dfrac{1}{d-2} \, \theta^m{}_n \theta^n{}_m \\
        & - \dfrac{1}{D-d} \dfrac{\varphi^2}{d-2} - \dfrac{\dot{\varphi}}{d-2} + \dfrac{1}{d-2} \, \biggl[ H - 2 \, \dfrac{D-2}{d-2} \, \Bigl[ \dot{\omega} - \dfrac{d-2}{D-d} \, \dot{\tau} \Bigr] \biggr] \varphi,
    \end{aligned} \label{app. eq.: master F2-eq.}
\end{align}
\end{subequations}
and eq. (\ref{app. eq.: preliminary master Eomega-eq.}) as
\begin{equation} \label{app. eq.: master Eomega-eq.}
\begin{split}
    0 = & - \dfrac{D-d}{D-2} \dfrac{\kappa_D^2}{d-2} \, \Bigl[ T_{00} - \tilde{g}^{ij} T_{ij} + \dfrac{d-2}{D-d} \, \e^{-2 \frac{D-2}{d-2} \omega} \bar{g}^{mn} T_{mn} \Bigr] \\[1.45ex]
    & + \dfrac{D-d}{d-2} \, \biggl[ \Bigl[ \ddot{\omega} - \dfrac{d-2}{D-d} \, \ddot{\tau} \Bigr] + (d-1) H \Bigl[ \dot{\omega} - \dfrac{d-2}{D-d} \, \dot{\tau} \Bigr] \biggr] \\[0.65ex]
    & + \dfrac{\e^{-2 \frac{D-2}{d-2} \omega}}{d-2} \biggl[ \bar{R} - \dfrac{2 (D-2) - (d-2)}{D-2} \dfrac{\bar{\nabla}^2 \e^{(D-2) \tau}}{\e^{(D-2) \tau}} + (D-1)(D-2) \, \der \tau \, \bar{\cdot} \, \der \tau \biggr] \\
    & + \dfrac{1}{d-2} \, \bigl[ \dot{\varphi} + \varphi^2 \bigr] + \biggl[ \dfrac{d-1}{d-2} \, H + \dfrac{D-d}{d-2} \, \Bigl[ \dot{\omega} - \dfrac{d-2}{D-d} \, \dot{\tau} \Bigr] \biggr] \varphi.
\end{split}
\end{equation}

\subsubsection*{Additional cosmological equations}

An independent class of equations is contained in the traceless part of the $(D-d)$-dimensional Einstein equations.
From the trace-reversed $(D-d)$-dimensional components of the Einstein equations, we have
\begin{align*}
    R_{mn} - \dfrac{1}{D-d} \, g_{mn} (g^{pq} R_{pq}) = \kappa_D^2 \, \Bigl[ T_{mn} - \dfrac{1}{D-d} \, g_{mn} (g^{pq} T_{pq}) \Bigr].
\end{align*}
By direct inspection, we are then able to write
\begin{equation} \label{app. eq.: premilinary master traceless internal Einstein eq.}
\begin{split}
    0 = & - \kappa_D^2 \, \e^{-2 \frac{D-2}{d-2} \omega} \Bigl[ T_{mn} - \dfrac{1}{D-d} \, (\bar{g}^{pq} T_{pq}) \, \bar{g}_{mn} \Bigr] \\
    & + \dfrac{1}{2} \, \Bigl[ \ddot{\bar{g}}_{mn} - \dot{\bar{g}}_{mq} (\bar{g}^{qr} \dot{\bar{g}}_{rn}) - \dfrac{1}{D-d} \, [(\bar{g}^{pq} \ddot{\bar{g}}_{pq}) - (\bar{g}^{pq} \dot{\bar{g}}_{qr}) (\bar{g}^{rs} \dot{\bar{g}}_{sp})] \, \bar{g}_{mn} \Bigr] \\
    & + \dfrac{1}{2} \, \Bigl[ \dot{\bar{g}}_{mn} - \dfrac{1}{D-d} \, (\bar{g}^{pq} \dot{\bar{g}}_{pq}) \bar{g}_{mn} \Bigr] \Bigl[ (d-1) H + \dfrac{1}{2} \, (\bar{g}^{pq} \dot{\bar{g}}_{pq}) + (D-2) \dot{\tau} \Bigr] \\
    & + \e^{-2 \frac{D-2}{d-2} \omega} \Bigl[ \bar{R}_{mn} - (D-2) [\bar{\nabla}_m \der_n \tau - \der_m \tau \der_n \tau] - \dfrac{1}{D-d} \, \bigl[ \bar{R} - (D-2) [\bar{\nabla}^2 \tau - \der \tau \, \bar{\cdot} \, \der \tau] \bigr] \, \bar{g}_{mn} \Bigr].
\end{split}
\end{equation}
Noticeably, the equation contains no explicit $\dot{\omega}$-dependent contribution: this is because all the $\dot{\omega}$-dependent terms in $R_{mn}$ are proportional to $\smash{\bar{g}_{mn}}$, and therefore they cancel out when restricting to the traceless part.
Contrarily to the non-traceless equation, the traceless one retains information on the dynamics of the volume-preserving internal deformations.
Importantly, we notice the identity
\begin{align*}
    \dot{\theta}^m{}_n = \dfrac{1}{2} \Bigl[ \bar{g}^{mp} \ddot{\bar{g}}_{pn} - (\bar{g}^{mp} \dot{\bar{g}}_{pq}) (\bar{g}^{qr} \dot{\bar{g}}_{rn}) - \dfrac{1}{D-d} \, [(\bar{g}^{pq} \ddot{\bar{g}}_{pq}) - (\bar{g}^{pq} \dot{\bar{g}}_{qr}) (\bar{g}^{rs} \dot{\bar{g}}_{sp})] \, \delta^m_n \Bigr].
\end{align*}
Moreover, we define the Bakry-Émery Ricci tensor \cite{DeLuca:2021mcj}
\begin{equation} \label{app. eq.: BE tensor}
    \bar{R}^{\mathrm{BE}}_{mn} = \bar{R}_{mn} - (D-2) (\bar{\nabla}_m \der_n \tau - \der_m \tau \der_n \tau).
\end{equation}
Therefore, in terms of the function defined in eqs. (\ref{app. eq.: varphi}, \ref{app. eq.: theta}), we may simply write eq. (\ref{app. eq.: premilinary master traceless internal Einstein eq.}) as
\begin{equation} \label{app. eq.: master traceless internal Einstein eq.}
    \dot{\theta}^m{}_n + [(d-1) H + \varphi] \, \theta^m{}_n = - \e^{-2 \frac{D-2}{d-2} \omega} \biggl[ \bar{g}^{mp} \bar{R}^{\mathrm{BE}}_{pn} - \dfrac{1}{D-d} \, \delta^m_n \bar{R}^{\mathrm{BE}} - \kappa_D^2 \, \Bigl[ \bar{g}^{mp} T_{pn} - \dfrac{1}{D-d} \, \delta^m_n (\bar{g}^{pq} T_{pq}) \Bigr] \biggr].
\end{equation}
In fact, we can exploit this to write a more useful differential condition for $\smash{\theta^m{}_n}$, exploiting the tracelessness condition.
Indeed, a $\theta$-contraction allows us to write
\begin{equation} \label{app. eq.: theta2 eq.}
    \dot{\theta}^m{}_n \theta^n{}_m + [(d-1) H + \varphi] \, \theta^m{}_n \theta^n{}_m + \e^{-2 \frac{D-2}{d-2} \omega} \bar{g}^{mp} [\bar{R}_{pn}^{\mathrm{BE}} - \kappa_D^2 T_{pn}] \theta^n{}_m = 0.
\end{equation}

Based on eq. (\ref{app. eq.: nabla2}), we observe that we the $D$-dimensional Laplace operator takes the specific form
\begin{equation} \label{app. eq.: KK decomposition}
    \nabla^2 f = - \e^{-2 \tau} \, \e^{2 \frac{D-d}{d-2} \omega} \Bigl[ \ddot{f} + [(d-1) H + \varphi] \dot{f} - \e^{-2 \frac{D-2}{d-2} \omega} \bigl[ \bar{\nabla}^2 f + (D-2) \, \der \tau \, \bar{\cdot} \, \der f \bigr] \Bigr],
\end{equation}
where $\der_i f = 0$.
Again, we may define the Bakry-Émery negative Laplacian $\smash{\bar{\Delta}_{\mathrm{BE}} f = - [\bar{\nabla}^2 f + (D-2) \, \der \tau \, \bar{\cdot} \, \der f]}$.

A further relation that we may expand is the $D$-dimensional continuity equation $\smash{\nabla_M T^{MN} = 0}$, which however is not independent of the $D$-dimensional Einstein equations analyzed before.
A constraint comes from the off-diagonal trace-reversed Einstein equation $\smash{R_{\mu n} = \kappa_D^2 T_{\mu n}}$, related to the off-diagonal $D$-dimensional Ricci tensor components in eq. (\ref{app. eq.: Rmun}), where 
FLRW symmetry requires $\smash{T_{in} = 0}$, while $\smash{T_{0n}}$ may be non-zero.
These mixed equations are constraints on the full solution, but they are not needed explicitly for the averaged relations that we analyze.

\subsection{Averaged-out equations} \label{subapp.: averaged-out equations}
From now on, we assume the internal $(D-d)$-dimensional space to fulfill two properties.
\begin{itemize}
    \item The internal $(D-d)$-dimensional space is compact and boundaryless.
    By the Stokes theorem, this ensures that the volume integral of a covariant divergence vanishes, and therefore for any vector field $f_m$ we have
    \begin{equation} \label{app. eq.: stokes}
        \int \de^{D-d} y \, \sqrt{\bar{g}_{D-d}} \; \bar{\nabla} \, \bar{\cdot} \, f = 0.
    \end{equation}
    \item The $\smash{\bar{g}_{mn}}$-metric is Riemannian.
    This implies a fundamental non-negativity condition, as follows.
    Let $s_{mn} = s_{nm}$ be a symmetric tensor.
    Then, we have
    \begin{equation} \label{app. eq.: riemannian gbar}
        s^{mk} \bar{g}_{kl} s^{ln} \bar{g}_{mn} \geq 0.
    \end{equation}
    Now, in full generality, we may write the trace/traceless decomposition
    \begin{align*}
        \dot{\bar{g}}_{mn} = \dfrac{1}{D-d} \, (\bar{g}^{pq} \dot{\bar{g}}_{pq}) \, \bar{g}_{mn} + \Bigl[ \dot{\bar{g}}_{mn} - \dfrac{1}{D-d} \, (\bar{g}^{pq} \dot{\bar{g}}_{pq}) \, \bar{g}_{mn} \Bigr],
    \end{align*}
    whence by direct inspection we obtain
    \begin{align*}
        (\bar{g}^{mn} \dot{\bar{g}}_{np}) (\bar{g}^{pq} \dot{\bar{g}}_{qm}) = \dfrac{1}{D-d} \, (\bar{g}^{mn} \dot{\bar{g}}_{mn}) (\bar{g}^{pq} \dot{\bar{g}}_{pq}) + \bar{g}^{mn} \bar{g}^{pq} \Bigl[ \dot{\bar{g}}_{np} - \dfrac{\bar{g}^{r s} \dot{\bar{g}}_{r s}}{D-d} \, \bar{g}_{np} \Bigr] \Bigl[ \dot{\bar{g}}_{qm} - \dfrac{\bar{g}^{k l} \dot{\bar{g}}_{k l}}{D-d} \, \bar{g}_{qm} \Bigr].
    \end{align*}
    Then, ineq. (\ref{app. eq.: riemannian gbar}) implies the condition
    \begin{equation} \label{app. eq.: deformation-mode ineq.}
        (\bar{g}^{mn} \dot{\bar{g}}_{np}) (\bar{g}^{pq} \dot{\bar{g}}_{qm}) \geq \dfrac{1}{D-d} \, (\bar{g}^{mn} \dot{\bar{g}}_{mn}) (\bar{g}^{pq} \dot{\bar{g}}_{pq}).
    \end{equation}
    This implies the non-negativity of the combination $\smash{\theta^m{}_n \theta^n{}_m \geq 0}$.
\end{itemize}

\subsubsection{Weighted integrals}

It is convenient to define the time-independent positive integral
\begin{equation} \label{app. eq.: K}
    K = \int \de^{D-d} y \, \sqrt{\bar{g}_{D-d}} \; \e^{(D-2) \tau} > 0.
\end{equation}
Moreover, we define the generic weighted integral of a function $f$ as
\begin{equation} \label{app. eq.: vevf}
    \vev{f} K = \int \de^{D-d} y \, \sqrt{\bar{g}_{D-d}} \; \e^{(D-2) \tau} f,
\end{equation}
As will be also crucially emphasized later on, in terms of the function $\varphi$ defined in eq. (\ref{app. eq.: varphi}), we notice the constraint $\smash{\vev{\varphi} = 0}$, as well as the identity $\smash{\de \vev{f} / \de t = \vev{\dot{f}} + \vev{\varphi f}}$.

Now, we define the geometric integrals
\begin{subequations}
\begin{align}
    I_{\mathrm{c}} K & = \dfrac{\e^{-2 \frac{D-2}{d-2} \omega}}{d-2} \int \de^{D-d} y \, \sqrt{\bar{g}_{D-d}} \; \e^{(D-2) \tau} \, \bar{R}, \label{app. eq.: Ic} \\
    I_{\mathrm{w}} K & = (D-1)(D-2) \, \dfrac{\e^{-2 \frac{D-2}{d-2} \omega}}{d-2} \int \de^{D-d} y \, \sqrt{\bar{g}_{D-d}} \; \e^{(D-2) \tau} \, \der \tau \, \bar{\cdot} \, \der \tau \geq 0, \label{app. eq.: Iw}
\end{align}
\end{subequations}
and
\begin{subequations}
\begin{align}
    I_\theta K & = \dfrac{1}{d-2} \, \int \de^{D-d} y \, \sqrt{\bar{g}_{D-d}} \; \e^{(D-2) \tau} \, \theta^m{}_n \theta^n{}_m \geq 0, \label{app. eq.: Itheta} \\
    I_\delta K & = \dfrac{D-2}{D-d} \int \de^{D-d} y \, \sqrt{\bar{g}_{D-d}} \; \e^{(D-2) \tau} \, \Bigl[ \dfrac{\varphi}{d-2} - (\dot{\tau} - \vev{\dot{\tau}}) \Bigr]^2 \geq 0, \label{app. eq.: Idelta}
\end{align}
\end{subequations}
as well as the matter integrals
\begin{subequations}
\begin{align}
    I_{0} K & = \dfrac{\kappa_D^2}{d-2} \int \de^{D-d} y \, \sqrt{\bar{g}_{D-d}} \; \e^{(D-2) \tau} \, T_{00}, \label{app. eq.: I(0)} \\
    I_{d-1} K & = \dfrac{\kappa_D^2}{d-2} \int \de^{D-d} y \, \sqrt{\bar{g}_{D-d}} \; \e^{(D-2) \tau} \, \tilde{g}^{ij} T_{ij}, \label{app. eq.: I(d-1)} \\
    I_{D-d} K & = \dfrac{\kappa_D^2 \, \e^{-2 \frac{D-2}{d-2} \omega}}{d-2} \int \de^{D-d} y \, \sqrt{\bar{g}_{D-d}} \; \e^{(D-2) \tau} \, \bar{g}^{mn} T_{mn}. \label{app. eq.: I(D-d)}
\end{align}
\end{subequations}
For convenience, we also define the combination
\begin{equation} \label{app. eq.: dotomegatau}
    \dot{\Omega} = \dot{\omega} - \dfrac{d-2}{D-d} \, \vev{\dot{\tau}}.
\end{equation}
In the main text, to save space, we have defined $\smash{I_{\mathrm{cw}} = I_{\mathrm{c}} + I_{\mathrm{w}}}$, $\smash{I_{\theta \delta} = I_\theta + I_\delta}$, and $\smash{\dot{\chi} = [(D-d)/(d-2)] \, \dot{\Omega}}$.

\subsubsection{Integrated Friedmann equations}

Based on eq. (\ref{app. eq.: einstein-frame condition}), we have the condition
\begin{equation} \label{app. eq.: vevvarphi=0}
    \vev{\varphi} = 0.
\end{equation}
By a time derivative, we therefore also see the identiy
\begin{equation} \label{app. eq.: vevdotvarphi}
    \vev{\dot{\varphi}} = - \vev{\varphi^2}.
\end{equation}

A few useful identities now follow from the weighted-integral notion in eq. (\ref{app. eq.: vevf}).
\begin{itemize}
    \item By direct substitution, we observe that
    \begin{align*}
        & - \dfrac{D-d}{d-2} \, \dfrac{D-2}{d-2} \, \Bigl\langle \Bigl[ \dot{\omega} - \dfrac{d-2}{D-d} \, \dot{\tau} \Bigr]^2 \Bigr\rangle - \dfrac{1}{d-2} \dfrac{1}{D-d} \, [\vev{\varphi^2} - 2 (D-2) \vev{\varphi \dot{\tau}}] + \dfrac{1}{d-2} \, \vev{\varphi^2} \\
        = {}& - \dfrac{D-d}{d-2} \, \dfrac{D-2}{d-2} \, \dot{\Omega}^2 - I_\delta + \dfrac{d-1}{d-2} \, \dfrac{\vev{\varphi^2}}{d-2}.
    \end{align*}
    \item An additional useful identity is
    \begin{align*}
        \dfrac{D-d}{d-2} \, \Bigl[ \ddot{\omega} - \dfrac{d-2}{D-d} \, \vev{\ddot{\tau}} \Bigr] = \dfrac{D-d}{d-2} \, \ddot{\Omega} + \vev{\varphi \dot{\tau}}.
    \end{align*}
\end{itemize}

In view of the weighted integrals defined in eq. (\ref{app. eq.: vevf}), eqs. (\ref{app. eq.: Ic}, \ref{app. eq.: Iw}), eqs. (\ref{app. eq.: Itheta}, \ref{app. eq.: Idelta}), eqs. (\ref{app. eq.: I(0)}, \ref{app. eq.: I(d-1)}, \ref{app. eq.: I(D-d)}), and eq. (\ref{app. eq.: dotomegatau}), and thanks to the crucial identities in eqs. (\ref{app. eq.: vevvarphi=0}, \ref{app. eq.: vevdotvarphi}), the weighted internal-integrated eqs. (\ref{app. eq.: master F1-eq.}, \ref{app. eq.: master F2-eq.}) can be respectively expressed as
\begin{subequations}
\begin{align}
    (d-1) \Bigl[ H^2 + \dfrac{k}{a^2} \Bigr] & = \dfrac{D-d}{d-2} \, \dfrac{D-2}{d-2} \, \dot{\Omega}^2 + I_\theta + I_\delta - \dfrac{d-1}{d-2} \, \dfrac{\vev{\varphi^2}}{d-2} + 2 I_0 - (I_{\mathrm{c}} + I_{\mathrm{w}}), \label{app. eq.: integrated master F1-eq.} \\
    \epsilon H^2 + \dfrac{k}{a^2} & = \dfrac{D-d}{d-2} \, \dfrac{D-2}{d-2} \, \dot{\Omega}^2 + I_\theta + I_\delta - \dfrac{d-1}{d-2} \, \dfrac{\vev{\varphi^2}}{d-2} + \dfrac{1}{d-1} \, [(d-1) I_0 + I_{d-1}], \label{app. eq.: integrated master F2-eq.}
\end{align}
\end{subequations}
and eq. (\ref{app. eq.: master Eomega-eq.}) as
\begin{equation} \label{app. eq.: integrated master Eomega-eq.}
    \dfrac{D-d}{d-2} \, [\ddot{\Omega} + (d-1) H \dot{\Omega}] = - (I_{\mathrm{c}} + I_{\mathrm{w}}) + \dfrac{d-2}{D-2} \, \Bigl[ \dfrac{D-d}{d-2} \, (I_0 - I_{d-1}) + I_{D-d} \Bigr].
\end{equation}

Thanks to eqs. (\ref{app. eq.: einstein-frame condition}, \ref{app. eq.: differentiated einstein-frame condition}), the field equations have simplified dramatically.
Yet, in their form as in eqs. (\ref{app. eq.: integrated master F1-eq.}, \ref{app. eq.: integrated master F2-eq.}), these prove to be not workable for the proof we have in mind, along the lines of the one where the Einstein-frame condition is imposed locally as in sec. \ref{app.: no-horizon proof for point-wise Einstein frame condition}, since they see an unavoidable competition between positive- and negative-semi-definite terms.
These averaged scalar relations do not exhaust the full Einstein equations: additional information is contained in the traceless internal components and in the mixed components.
Although these generally control additional dynamics and constraints, they turn out to have complicated expressions even after averaging.

\subsection{Higher-dimensional SEC and lower-dimensional physics} \label{subapp.: higher-dimensional SEC and lower-dimensional physics}

Here, we review in detail the $D$-dimensional SEC and its implications.
By definition, for any time-like vector $t^M$, i.e. such that $t^2 < 0$, a theory fulfilling the $D$-dimensional SEC is such that
\begin{equation} \label{app. eq.: SEC}
    \Bigl[ T_{MN} - \dfrac{1}{D-2} \, g_{MN} T \Bigr] t^M t^N \geq 0.
\end{equation}
This has very neat implications, as we detail below.
\begin{itemize}
    \item If we consider the time-like vector $t^M = \delta^M_0$, then we immediately find
    \begin{equation} \label{app. eq.: SEC ineq. 1}
        T_{00} - \dfrac{1}{D-2} \, g_{00} T \geq 0.
    \end{equation}
    \item Let $k^M$ be an arbitrary null vector, and let $l^M$ be another null vector such that $k \cdot l < 0$.
    Then, $t^M(\alpha) = k^M/\alpha + \alpha \, l^M$ is a time-like vector for every non-zero constant $\alpha$.
    In particular, SEC must hold for arbitrarily small values of $\alpha$, by which we can write \cite{Kontou:2020bta}
    \begin{align*}
        \lim_{\alpha \to 0} \bigl[ \alpha^2 R_{MN} t^M(\alpha) t^N(\alpha) \bigr] = R_{MN} k^M k^N + \lim_{\alpha \to 0} \bigl[ \alpha^2 (2 R_{MN} k^M l^N + \alpha^2 R_{MN} l^M l^N) \bigr] = R_{MN} k^M k^N \geq 0.
    \end{align*}
    If we choose the null vector $\smash{k^M = \delta^M_0 + \delta^M_I n^I}$, where $n^I$ is a $(D-1)$-dimensional space-like vector such that $\smash{g_{IJ} n^I n^J = - g_{00}}$ and $\smash{g_{0J} n^J = 0}$, we find
    \begin{equation} \label{app. eq.: SEC ineq. 2}
        T_{00} + 2 T_{0J} n^J + T_{IJ} n^I n^J \geq 0.
    \end{equation}
\end{itemize}

Below, we study in detail the implications of ineqs. (\ref{app. eq.: SEC ineq. 1}, \ref{app. eq.: SEC ineq. 2}) for the metric ansatz in eq. (\ref{app. eq.: metric}).

To start, we study the implications of ineq. (\ref{app. eq.: SEC ineq. 1}).
By expanding the trace $T = g^{\mu \nu} T_{\mu \nu} + g^{mn} T_{mn}$ out, we see that this is equivalent to the condition
\begin{equation} \label{app. eq.: basic SEC}
    T_{00} \geq - \dfrac{1}{D-3} \, \bigl[ \tilde{g}^{ij} T_{ij} + \e^{-2 \frac{D-2}{d-2} \omega} \bar{g}^{mn} T_{mn} \bigr].
\end{equation}
Our goal is to exploit this condition in the Einstein/Friedmann equations.
In particular, we can rearrange ineq. (\ref{app. eq.: basic SEC}) to obtain the inequalities
\begin{align} \label{app. eq.: E0-eq. SEC constraint}
    (d-3) \, T_{00} + \tilde{g}^{ij} T_{ij} & \geq \dfrac{D-d}{D-3} \, \tilde{g}^{ij} T_{ij} - \dfrac{d-3}{D-3} \, \e^{-2 \frac{D-2}{d-2} \omega} \bar{g}^{mn} T_{mn}, \\
    \dfrac{D-d}{D-2} \, \Bigl[ - T_{00} + \tilde{g}^{ij} T_{ij} - \dfrac{d-2}{D-d} \, \e^{-2 \frac{D-2}{d-2} \omega} \bar{g}^{mn} T_{mn} \Bigr] & \leq \dfrac{D-d}{D-3} \, \tilde{g}^{ij} T_{ij} - \dfrac{d-3}{D-3} \, \e^{-2 \frac{D-2}{d-2} \omega} \bar{g}^{mn} T_{mn}, \label{app. eq.: Eomega-eq. SEC constraint}
\end{align}
where we emphasize that ineqs. (\ref{app. eq.: E0-eq. SEC constraint}, \ref{app. eq.: Eomega-eq. SEC constraint}) define a double chain of inequalities.

Now we turn to the implications of ineq. (\ref{app. eq.: SEC ineq. 2}).
Of course, $T_{\mu \nu}$ represents the $d$-dimensional block components of the $D$-dimensional tensor $T_{MN}$, and not the actual $d$-dimensional energy-momentum tensor; yet, $d$-dimensional FLRW homogeneity and isotropy require that any contribution that is not of the form $\smash{T_{i j} = p \, \tilde{g}_{ij}}$, with $\der_i p = 0$, must be vanishing.
Moreover, they also require $\smash{T_{0j} = 0}$.
Let us pick space-like vectors of the form $\smash{n^I = \delta^I_j \tilde{n}^j}$, where the normalization $\smash{g_{ij} n^i n^j = - g_{00}}$ requires $\smash{\tilde{g}_{ij} \tilde{n}^i \tilde{n}^j = 1}$.
It is also implied that $\smash{g_{0i} n^i = 0}$ and $\smash{T_{0i} n^i = 0}$.
Then, we immediately see that $\smash{T_{IJ} n^I n^J = p}$, and at the same time we might simply see that $\smash{\tilde{g}^{ij} T_{ij} = (d-1) p}$.
Therefore, ineq. (\ref{app. eq.: SEC ineq. 2}) implies the inequality
\begin{equation} \label{app. eq.: d-dim. EC}
    T_{00} + \dfrac{1}{d-1} \, \tilde{g}^{ij} T_{ij} \geq 0.
\end{equation}
This has the same algebraic structure as a $d$-dimensional energy condition; however, the $D$-dimensional tensor components $T_{\mu \nu}$ do not generally coincide with the actual $d$-dimensional energy-momentum tensor for the genuine $d$-dimensional theory.
Moreover, the inequality is reminiscent of the $d$-dimensional strong energy condition, but one one would additionally need $\smash{(d-3) T_{00} + \tilde{g}^{ij} T_{ij} \geq 0}$.
This does not contradict refs. \cite{Bernardo:2022ony, Faruk:2024usu}, whose restrictions on the internal geometry and warp factor are stronger than the integral normalization conditions imposed here.

An additional consequence follows from ineq. (\ref{app. eq.: SEC ineq. 2}).
Let us consider the space-like vectors $\smash{n_\pm^I = \pm \delta^I_m \bar{n}^m}$, where the normalization $\smash{g_{m n} n_\pm^m n_\pm^n = - g_{00}}$ requires $\smash{\bar{g}_{mn} \bar{n}^m \bar{n}^n = \e^{-2 [(D-2)/(d-2)] \omega}}$.
By ineq. (\ref{app. eq.: SEC ineq. 2}), we must have $\smash{T_{00} \pm 2 T_{0n} \bar{n}^n + T_{mn} \bar{n}^m \bar{n}^n \geq 0}$; a simple sum then delivers
\begin{align*}
    T_{00} + T_{mn} \bar{n}^m \bar{n}^n \geq 0.
\end{align*}
Locally, we may define the vielbeins $\smash{\bar{e}_a^m}$, such that $\smash{\bar{g}_{mn} \bar{e}_a^m \bar{e}_b^n = \delta_{ab}}$.
Then, we may construct $D-d$ spatial vectors $\smash{\bar{n}_{(a)}^m = \e^{- [(D-2)/(d-2)] \omega} \bar{e}_a^m}$, where for $\smash{\bar{n}_{(a)}^m}$ we should understand $a = 1, \dots, D-d$ as a label.
For each $a$-label, we may write $\smash{T_{00} + T_{mn} \bar{n}_{(a)}^m \bar{n}_{(a)}^n \geq 0}$.
Therefore, summing over all labels, we find
\begin{equation} \label{app. eq.: (D-d)-dim. EC}
    T_{00} + \dfrac{1}{D-d} \, \e^{-2 \frac{D-2}{d-2} \omega} \bar{g}^{mn} T_{mn} \geq 0.
\end{equation}

\subsubsection*{Integrated SEC implications}

If we require the theory to fulfill the $D$-dimensional SEC, then, we get the bounds in eq. (\ref{app. eq.: basic SEC}) and eqs. (\ref{app. eq.: E0-eq. SEC constraint}, \ref{app. eq.: Eomega-eq. SEC constraint}), as well as eq. (\ref{app. eq.: d-dim. EC}) and eq. (\ref{app. eq.: (D-d)-dim. EC}).
In terms of the weighted internal integrals defined in eqs. (\ref{app. eq.: I(0)}, \ref{app. eq.: I(d-1)}, \ref{app. eq.: I(D-d)}), they can be expressed as simple linear inequalities.

First, eq. (\ref{app. eq.: basic SEC}) and eqs. (\ref{app. eq.: E0-eq. SEC constraint}, \ref{app. eq.: Eomega-eq. SEC constraint}), respectively, take the form
\begin{equation} \label{app. eq.: integrated basic SEC}
    (D-3) I_0 + I_{d-1} + I_{D-d} \geq 0,
\end{equation}
and
\begin{align}
    (d-3) \, I_0 + I_{d-1} & \geq \dfrac{1}{D-3} \, [(D-d) \, I_{d-1} - (d-3) \, I_{D-d}], \label{app. eq.: integrated E0-eq. SEC constraint} \\
    - \dfrac{d-2}{D-2} \Bigl[ \dfrac{D-d}{d-2} \, (I_0 - I_{d-1}) + I_{D-d} \Bigr] & \leq \dfrac{1}{D-3} \bigl[ (D-d) \, I_{d-1} - (d-3) \, I_{D-d} \bigr], \label{app. eq.: integrated Eomega-eq. SEC constraint}
\end{align}
which also imply the bound
\begin{equation} \label{app. eq.: integrated d-dim. SEC-like condition}
    - \dfrac{d-2}{D-2} \Bigl[ \dfrac{D-d}{d-2} \, (I_0 - I_{d-1}) + I_{D-d} \Bigr] \leq (d-3) \, I_0 + I_{d-1}.
\end{equation}

Second, the conditions in eq. (\ref{app. eq.: d-dim. EC}) and eq. (\ref{app. eq.: (D-d)-dim. EC}) respectively take the form
\begin{equation} \label{app. eq.: integrated d-dim. EC}
    (d-1) I_0 + I_{d-1} \geq 0,
\end{equation}
and
\begin{equation} \label{app. eq.: integrated (D-d)-dim. EC}
    (D-d) I_0 + I_{D-d} \geq 0.
\end{equation}

\subsection{SSEC}

Before presenting the statement of the SSEC, we set the stage by introducing a few important elements.
\begin{itemize}
    \item In the $d$-dimensional spacetime, the future-directed unit velocity corresponding to observers at rest is $\smash{\tilde{u}^\mu = \delta^\mu_0}$.
    With respect to the $D$-dimensional metric, the corresponding unit vector is
    \begin{equation}
        u^M = \e^{- \tau} \, \e^{\frac{D-d}{d-2} \omega} \, \delta^M_\mu \tilde{u}^\mu.
    \end{equation}
    \item Similarly, the external $(d-1)$-dimensional spatial vielbeins $\smash{\tilde{e}_a{}^i}$ are defined through the identity $\smash{\tilde{e}_a{}^i \tilde{e}_b{}^j \delta^{ab} = \tilde{g}^{ij}}$.
    Then, for the $D$-dimensional metric, we define
    \begin{equation}
        e_a{}^M = \e^{- \tau} \, \e^{\frac{D-d}{d-2} \omega} \, \tilde{e}_a{}^i \delta^M_i,
    \end{equation}
    which fulfill the identities $\smash{g_{MN} u^M e_a{}^N = 0}$ and $\smash{g_{MN} e_a{}^M e_b{}^N = \delta_{ab}}$.
\end{itemize}
Then, we define the projector on the external $(d-1)$-dimensional space, i.e.
\begin{equation} \label{app. eq.: (d-1)-space projector}
    P^{MN} = \delta^{ab} e_a{}^M e_b{}^N = \e^{-2\tau} \, \e^{2 \frac{D-d}{d-2} \omega} \tilde{g}^{ij} \delta_i^M \delta_j^N.
\end{equation}
By noticing that $\smash{P_{MN} = \e^{2\tau} \, \e^{-2 [(D-d)/(d-2)] \omega} \tilde{g}_{ij} \delta^i_M \delta^j_N}$, we can also verify the properties $\smash{P_{MN} (\delta^N_0 + \delta^N_n c^n) = 0}$, $\smash{P_{MQ} P^{QN} = P_M{}^N = \delta_M^i \delta^N_i}$, and $\smash{P_M{}^M = d-1}$.

In the rest of this work, we postulate a strengthened version of the SEC, as follows.
Let us consider any future-directed unit time-like vector $v^M$ such that $\smash{P_{MN} v^N = 0}$.
Then, we say that the theory fulfills the SSEC if the inequality holds
\begin{equation} \label{app. eq.: SSEC}
    \biggl[ T_{MN} - \dfrac{1}{D-2} \, g_{MN} (g^{PQ} T_{PQ}) \biggr] v^M v^N \geq (1-\lambda) \, T_{MN} \Bigl[ v^M v^N + \dfrac{1}{d-1} \, P^{MN} \Bigr].
\end{equation}
We have referred to this condition as a strengthened version of the $D$-dimensional SEC for two fundamental reasons.
\begin{enumerate}[label=(\roman*)]
    \item While the SEC statement holds for arbitrary time-like vectors, the SSEC statement is formulated for time-like vectors orthogonal to the external $(d-1)$-dimensional spatial directions.
    Moreover, $\smash{\lambda = 1}$ would formally resemble the SEC statement.
    \item The SEC statement is a requirement formulated as the non-negativity of the SSEC left-hand side, and that can also be exploited to guarantee the non-negativity of the SSEC right-hand side in FLRW reductions, i.e.
    \begin{align*}
        \biggl[ T_{MN} - \dfrac{1}{D-2} \, g_{MN} (g^{PQ} T_{PQ}) \biggr] v^M v^N & \geq 0, \\
        T_{MN} \Bigl[ v^M v^N + \dfrac{1}{d-1} \, P^{MN} \Bigr] & \geq 0,
    \end{align*}
    but it does not establish any general hierarchy among these.
\end{enumerate}
By the $D$-dimensional Einstein equations, the SSEC statement in eq. (\ref{app. eq.: SSEC}) is equivalent requiring
\begin{equation} \label{app. eq.: ricci-tensor SSEC}
    R_{MN} v^M v^N \geq \dfrac{1-\lambda}{\lambda} \, \dfrac{1}{d-1} \, R_{MN} P^{MN}.
\end{equation}
Here, $\lambda$ is a constant in the interval
\begin{equation} \label{app. eq.: lambda proof interval}
    \dfrac{1}{2} \dfrac{D-d-1+\sqrt{D-d+1}}{D-2} < \lambda \leq \dfrac{1}{2} \Bigl( 1 + \dfrac{\sqrt{d-2}}{\sqrt{D-2}} \Bigr).
\end{equation}
Throughout this work, whenever we mention SSEC, it is understood that we mean it with a $\lambda$-value belonging to the interval in eq. (\ref{app. eq.: lambda proof interval}).

\subsection{No-horizon proof for point-wise Einstein frame condition} \label{app.: no-horizon proof for point-wise Einstein frame condition}

A fully general proof of the no-horizon theorem is technically lengthy, and relegated to app. \ref{app.: no-horizon proof}.
Here, we focus on a special case: for this subsection only, we assume that the Einstein-frame condition does not only take place as an integral condition, as in eq. (\ref{app. eq.: einstein-frame condition}), but that it actually holds point-wise in the internal space, i.e.
\begin{equation} \label{app. eq.: point-wise einstein frame condition}
    \varphi = \dfrac{1}{2} \, \bar{g}^{mn} \dot{\bar{g}}_{mn} + (D-2) \dot{\tau} = 0.
\end{equation}
This is stronger than the integral Einstein-frame condition: even for $\smash{\dot{\tau} = 0}$, singling out the overall volume mode $\omega$ does not require $\smash{\bar{g}^{mn} \dot{\bar{g}}_{mn} = 0}$ point-wise.
Yet, such solutions exist, and eq. (\ref{app. eq.: point-wise einstein frame condition}) allows for an illuminating no-horizon proof.

If the condition in eq. (\ref{app. eq.: point-wise einstein frame condition}) is in place, we can formulate a very simple proof of the no-horizon theorem because in this case we do not only have $\smash{\vev{\varphi} = 0}$, but also $\smash{\vev{\varphi^2} = 0}$.
Else, the presence of the non-positive term $\smash{-\vev{\varphi^2} \leq 0}$ in eqs. (\ref{app. eq.: integrated master F1-eq.}, \ref{app. eq.: integrated master F2-eq.}) forces us to consider the much more sophisticated local methods in app. \ref{app.: no-horizon proof}.
With $\smash{\vev{\varphi^2} = 0}$, the cosmological dynamics are governed by simplified versions of eqs. (\ref{app. eq.: integrated master F1-eq.}, \ref{app. eq.: integrated master F2-eq.}) and eq. (\ref{app. eq.: integrated master Eomega-eq.}), i.e. respectively
\begin{subequations}
\begin{align}
    (d-1) \Bigl[ H^2 + \dfrac{k}{a^2} \Bigr] & = \dfrac{D-d}{d-2} \, \dfrac{D-2}{d-2} \, \dot{\Omega}^2 + I_\theta + I_\delta + 2 I_0 - (I_{\mathrm{c}} + I_{\mathrm{w}}), \label{app. eq.: point-wise integrated master F1-eq.} \\
    \epsilon H^2 + \dfrac{k}{a^2} & = \dfrac{D-d}{d-2} \, \dfrac{D-2}{d-2} \, \dot{\Omega}^2 + I_\theta + I_\delta + \dfrac{1}{d-1} \, [(d-1) I_0 + I_{d-1}], \label{app. eq.: point-wise integrated master F2-eq.}
\end{align}
\end{subequations}
and
\begin{equation} \label{app. eq.: point-wise integrated master Eomega-eq.}
    \dfrac{D-d}{d-2} \, [\ddot{\Omega} + (d-1) H \dot{\Omega}] = - (I_{\mathrm{c}} + I_{\mathrm{w}}) + \dfrac{d-2}{D-2} \, \Bigl[ \dfrac{D-d}{d-2} \, (I_0 - I_{d-1}) + I_{D-d} \Bigr].
\end{equation}
For this simplified proof, we also make additional assumptions, which are however not needed in the general proof: we assume the $D$-dimensional SEC to hold, as in eq. (\ref{app. eq.: SEC}), and we consider non-positive FLRW spatial curvature $k=0,-1/\ell^2$.

To start, we observe one obvious implication of the $D$-dimensional SEC.
By the energy condition in ineq. (\ref{app. eq.: integrated d-dim. EC}), and thanks to the non-negativity of $\smash{\dot{\Omega}^2, I_\theta, I_\delta \geq 0}$, eq. (\ref{app. eq.: point-wise integrated master F2-eq.}) implies that $\smash{\epsilon H^2 + k/a^2 \geq 0}$.
This also means that
\begin{equation} \label{app. eq.: point-wise SEC 0-implication}
    \epsilon \geq 0,
\end{equation}
i.e. the Hubble parameter is a non-increasing function.
This would not be implied if eq. (\ref{app. eq.: point-wise einstein frame condition}) were not holding.

\subsubsection{Relevance of the integrated SSEC}

By subtracting out eqs. (\ref{app. eq.: point-wise integrated master F1-eq.}, \ref{app. eq.: point-wise integrated master F2-eq.}), we find the identity
\begin{align*}
    (1-\epsilon) H^2 + (d-2) \Bigl[ H^2 + \dfrac{k}{a^2} \Bigr] = \dfrac{1}{d-1} [(d-1) I_0 - I_{d-1}] - (I_{\mathrm{c}} + I_{\mathrm{w}}).
\end{align*}
By plugging in eq. (\ref{app. eq.: point-wise integrated master Eomega-eq.}), we find
\begin{align*}
    (1-\epsilon) H^2 + (d-2) \Bigl[ H^2 + \dfrac{k}{a^2} \Bigr] - \dfrac{D-d}{d-2} \, \dfrac{1}{a^{d-1}} \dfrac{\de}{\de t} \, [a^{d-1} \dot{\Omega}] = \dfrac{d-2}{D-2} \, \Bigl[ I_0 + \Bigl( \dfrac{D-2}{d-1} - 1 \Bigr) I_{d-1} - I_{D-d} \Bigr].
\end{align*}
A simple manipulation then allows us to write the identity
\begin{equation} \label{app. eq.: integrated Einstein-equation master combination}
\begin{split}
    & \dfrac{1}{d-2} \dfrac{1}{a} \dfrac{\de}{\de t} \Bigl[ a \Bigl( H - \dfrac{D-d}{d-2} \, \dot{\Omega} \Bigr) \Bigr] + H \Bigl( H - \dfrac{D-d}{d-2} \, \dot{\Omega} \Bigr) + \dfrac{k}{a^2} \\
    = \, & \dfrac{1}{d-1} \, [(d-1) I_0 + I_{d-1}] - \dfrac{1}{D-2} \, [(D-3) I_0 + I_{d-1} + I_{D-d}].
\end{split}
\end{equation}
This is the crucial equation for the proof.
\begin{enumerate*}[label=]
    \item On the left hand-side, we recognize purely geometric terms.
    \item On the right hand-side, we recognize the difference of two terms that are non-negative in view of the $D$-dimensional SEC, i.e. ineq. (\ref{app. eq.: integrated basic SEC}) and ineq. (\ref{app. eq.: integrated d-dim. EC}).
\end{enumerate*}

Unfortunately, the sign of the energy-momentum terms in eq. (\ref{app. eq.: integrated Einstein-equation master combination}) is not fixed by the $D$-dimensional SEC.
This is precisely the place where SSEC enters into play.
In its integrated form, it takes the form
\begin{equation} \label{app. eq.: integrated SSEC}
    \dfrac{1}{D-2} \, [(D-3) I_0 + I_{d-1} + I_{D-d}] \geq \dfrac{1 - \lambda}{d-1} \, [(d-1) I_0 + I_{d-1}].
\end{equation}

From now, we assume ineq. (\ref{app. eq.: integrated SSEC}) to hold.
Then, eq. (\ref{app. eq.: integrated Einstein-equation master combination}) and ineq. (\ref{app. eq.: integrated SSEC}) provide the condition
\begin{align*}
    & \dfrac{1}{d-2} \dfrac{1}{a} \dfrac{\de}{\de t} \Bigl[ a \Bigl( H - \dfrac{D-d}{d-2} \, \dot{\Omega} \Bigr) \Bigr] + H \Bigl( H - \dfrac{D-d}{d-2} \, \dot{\Omega} \Bigr) + \dfrac{k}{a^2} \leq \dfrac{\lambda}{d-1} \, [(d-1) I_0 + I_{d-1}].
\end{align*}
By eq. (\ref{app. eq.: point-wise integrated master F2-eq.}), and thanks to the non-negativity of $\smash{I_\theta, I_\delta \geq 0}$, as long as $\lambda \geq 0$ we can replace the source term and obtain
\begin{align*}
    \dfrac{1}{d-2} \dfrac{1}{a} \dfrac{\de}{\de t} \Bigl[ a \Bigl( H - \dfrac{D-d}{d-2} \, \dot{\Omega} \Bigr) \Bigr] + H \Bigl( H - \dfrac{D-d}{d-2} \, \dot{\Omega} \Bigr) + \dfrac{k}{a^2} \leq \lambda \, \Bigl[ \epsilon H^2 + \dfrac{k}{a^2} - \dfrac{D-d}{d-2} \dfrac{D-2}{d-2} \, \dot{\Omega}^2 \Bigr].
\end{align*}
In view of the identity $\smash{\de (a H) / \de t = (1 - \epsilon) a H^2}$, we can rearrange the inequality and complete a square, obtaining
\begin{align*}
    \dfrac{1}{d-2} \dfrac{1}{a} \dfrac{\de}{\de t} \Bigl[ a \Bigl(H - \dfrac{D-d}{d-2} \, \dot{\Omega} \Bigr) \Bigr] \leq & - \dfrac{\lambda}{a} \dfrac{\de}{\de t} (a H) - (1- \lambda) \dfrac{k}{a^2} \\
    & - \Bigl[ \Bigl( 1 - \lambda - \dfrac{1}{4 \lambda} \dfrac{D-d}{D-2} \Bigr) H^2 + \dfrac{1}{4 \lambda} \dfrac{D-d}{D-2} \Bigl( H - 2 \lambda \dfrac{D-2}{d-2} \, \dot{\Omega} \Bigr)^{\! 2} \Bigr].
\end{align*}
Clearly, the $\smash{\dot{\Omega}}$-dependent squared term is non-negative.
One can also see that the coefficient of the $H^2$-term is non-negative when $\lambda$ is in the interval defined in eq. (\ref{app. eq.: lambda proof interval}).\footnote{In fact, the current step requires the weaker condition
\begin{align*}
    \dfrac{1}{2} \Bigl(1 - \dfrac{\sqrt{d-2}}{\sqrt{D-2}}\Bigr) \leq \lambda \leq \dfrac{1}{2} \Bigl(1 + \dfrac{\sqrt{d-2}}{\sqrt{D-2}}\Bigr).
\end{align*}
The interval in eq. (\ref{app. eq.: lambda proof interval}) is contained in this interval.
The requirement for the restricted interval in eq. (\ref{app. eq.: lambda proof interval}) with respect to the current one comes from the general proof in app. \ref{app.: no-horizon proof}.}
Therefore, as long as $\lambda$ lies within the interval defined in eq. (\ref{app. eq.: lambda proof interval}), we may always write the inequality
\begin{equation} \label{app. eq.: optimized differential proof ineq.}
    \dfrac{1}{d-2} \dfrac{\de}{\de t} \Bigl[ a \Bigl(H - \dfrac{D-d}{d-2} \, \dot{\Omega} \Bigr) \Bigr] \leq - \lambda \, \dfrac{\de}{\de t} (a H) - (1- \lambda) \dfrac{k}{a}.
\end{equation}

\subsubsection{The simplified proof}

Here begins the actual proof by contradiction.
If the solution is such that $\smash{\int_{t_\star}^\infty \de t/a(t) = \infty}$, then there is no future cosmic horizon.
By contradiction, let us therefore assume, instead, that it is
\begin{equation} \label{app. eq.: point-wise would-be horizon condition}
    \int_{t_\star}^\infty \dfrac{\de t}{a(t)} < \infty,
\end{equation}
where $t_\star$ is an initial time.
Under ineq. (\ref{app. eq.: point-wise would-be horizon condition}), the non-negativity of $\smash{\epsilon \geq 0}$ by ineq. (\ref{app. eq.: point-wise SEC 0-implication}) implies $H>0$.
Indeed, if were $H(t_0) \leq 0$ at some time $t_0$, then we would have $H(t) \leq 0$ and hence $a(t) \leq a(t_0)$ for all times $t\geq t_0$, immediately contradicting the horizon hypothesis with $\int_{t_0}^{\infty}\de t/a(t)=\infty$.
We proceed in steps.
\begin{itemize}
    \item A couple of fundamental would-be implications follow immediately.
    \begin{itemize}
        \item By integrating ineq. (\ref{app. eq.: optimized differential proof ineq.}), we find
        \begin{align*}
            \dfrac{1}{d-2} \Bigl(H - \dfrac{D-d}{d-2} \, \dot{\Omega} \Bigr) + \lambda H \leq \dfrac{1}{d-2} \dfrac{\gamma}{a} .
        \end{align*}
        where we have defined the function
        \begin{align*}
            \dfrac{\gamma}{d-2} = a_\star \Bigl[ \dfrac{1}{d-2} \Bigl(H_\star - \dfrac{D-d}{d-2} \, \dot{\Omega}_\star \Bigr) + \lambda H_\star \Bigr] - (1-\lambda) \int_{t_\star}^t \de s \, \dfrac{k}{a(s)}.
        \end{align*}
        In view of ineq. (\ref{app. eq.: point-wise would-be horizon condition}), $\gamma$ is bounded from above by a finite constant.
        Therefore, we may always select a finite positive constant $\Gamma > 0$ such that
        \begin{equation} \label{app. eq.: optimized integrated proof ineq.}
            \dfrac{1}{d-2} \Bigl(H - \dfrac{D-d}{d-2} \, \dot{\Omega} \Bigr) + \lambda H \leq \dfrac{1}{d-2} \, \dfrac{\Gamma}{a}.
        \end{equation}
        
        \item Because the Hubble parameter is non-increasing, by ineq. (\ref{app. eq.: point-wise SEC 0-implication}), for every time $t_\infty \geq t$ we have $\smash{a_\infty = a(t) \, \e^{\int_t^{t_\infty} \de s \, H(s)} \leq a(t) \, \e^{H(t) (t_\infty-t)}}$.
        As a consequence, we may write the chain of (in)equalites
        \begin{align*}
            \int_t^{t_\infty} \dfrac{\de s}{a(s)} \geq \dfrac{1}{a(t)} \int_t^{t_\infty} \de s \, \e^{- H(t) (s-t)} = \dfrac{1 - \e^{-H(t) (t_\infty-t)}}{a(t) H(t)}.
        \end{align*}
        Our interest is in taking $\smash{t_\infty = \infty}$.
        In view of the horizon condition in ineq. (\ref{app. eq.: point-wise would-be horizon condition}), we have immediately
        \begin{equation} \label{app. eq.: aH asymptotic bound}
            \lim_{t \to \infty} a(t)H(t) = \infty.
        \end{equation}
    \end{itemize}

    \item In view of ineq. (\ref{app. eq.: optimized integrated proof ineq.}), and being $\lambda > 0$, we can write
    \begin{align*}
        \dfrac{D-d}{d-2} \dfrac{\dot{\Omega}}{H} \geq 1 + \lambda (d-2) - \dfrac{\Gamma}{aH} > 1 - \dfrac{\Gamma}{aH}.
    \end{align*}
    Then, thanks to eq. (\ref{app. eq.: aH asymptotic bound}), we infer the asymptotic constraint
    \begin{equation} \label{app. eq.: dotomegatau asymptotic proof bound}
        \liminf_{t \to \infty} \, \dfrac{\dot{\Omega}(t)}{H(t)} > \dfrac{d-2}{D-d}.
    \end{equation}
    At the same time, eq. (\ref{app. eq.: point-wise integrated master F2-eq.}), ineq. (\ref{app. eq.: integrated d-dim. EC}), the non-negativity of $\smash{\dot{\Omega}^2, I_\theta, I_\delta \geq 0}$, and the condition $k \leq 0$ imply the inequality
    \begin{align*}
        \epsilon H^2 \geq \dfrac{D-d}{d-2} \dfrac{D-2}{d-2} \, \dot{\Omega}^2.
    \end{align*}
    Therefore, in view of ineq. (\ref{app. eq.: dotomegatau asymptotic proof bound}), there must exist a finite time $t_0$ such that, at all later times $t \geq t_0$, we have
    \begin{equation}
        - \dfrac{\dot{H}}{H^2} \geq \dfrac{D-2}{D-d}.
    \end{equation}
    A time integration gives $\smash{H_0/H \geq 1 + [(D-2)/(D-d)] \, H_0 (t-t_0)}$.
    After choosing the time gauge such that $\smash{t_0 H_0 = (D-d)/(D-2)}$, without loss of generality, an additional time integration gives
    \begin{align*}
        a \leq a_0 \, \Bigl( \dfrac{t}{t_0} \Bigr)^{\frac{D-d}{D-2}}.
    \end{align*}
    Because $\smash{(D-d)/(D-2) < 1}$, this implies
    \begin{align*}
        \int_{t_0}^\infty \dfrac{\de t}{a(t)} \geq \dfrac{1}{a_0} \int_{t_0}^\infty \de t \, \Bigl( \dfrac{t_0}{t} \Bigr)^{\frac{D-d}{D-2}} = \infty,
    \end{align*}
    contradicting the initial assumption in ineq. (\ref{app. eq.: point-wise would-be horizon condition}).
\end{itemize}
This proves the absence of a future cosmic horizon.

\section{No-horizon proof} \label{app.: no-horizon proof}

Here, we prove the no-horizon theorem in full generality, with the integral Einstein-frame condition.
Our assumptions are that the $(D-d)$-dimensional reduction space is Riemannian, compact, and boundaryless, and that the $D$-dimensional theory fulfills SSEC.
Our conclusion is that, for expanding solutions to $d$-dimensional FLRW spacetimes, i.e. with $H>0$, there cannot exist a future cosmic horizon.

By definition of the integral Einstein-frame condition, in eq. (\ref{app. eq.: einstein-frame condition}), the integral $\smash{K = \int \de^{D-d} y \, \sqrt{\bar{g}_{D-d}} \; \e^{(D-2) \tau}}$ is time-independent.
Therefore, for any time $t \geq t_\star$, where $t_\star$ is an initial time, we can write
\begin{align*}
    \int_{t_\star}^t \de s \int \de^{D-d} y \, \sqrt{\bar{g}_{D-d}} \; \e^{(D-2) \tau} \, a^{d-3} = K \int_{t_\star}^t \de s \, a^{d-3}(s).
\end{align*}
Intuitively, in dimension $d>2$, the existence of a horizon requires a scale factor that does not vanish at infinity, which forces this quantity to diverge as $t \to \infty$.
What we will show is that SSEC, instead, would bound it from above by a constant if the remaining conformal time were finite.
By contradiction, therefore, from now on we will assume that there exist a horizon, i.e. that there exist a finite maximal comoving distance $L_\star$ that light can travel, defined as
\begin{equation} \label{app.-proof: would-be horizon condition}
    L_\star = \int_{t_\star}^\infty \dfrac{\de t}{a(t)} < \infty.
\end{equation}
For the proof, our idea consists in defining a family of geodesics, with respect to a suitable metric, that cover all of spacetime.
It is following this family, i.e. defining coordinates along it, that we will able to bring out evidence for an inherent contradiction with the horizon assumption of eq. (\ref{app.-proof: would-be horizon condition}).
The geodesic family will be constructed in ssec. \ref{app.-proof: a useful family of geodesics}.
Building on this, the fundamental inequality implied by SSEC will be worked out in ssec. \ref{app.-proof: fundamental inequality}.
Finally, the proof by contradiction will be completed in ssec. \ref{app.-proof: proof by contradiction}.
As will be apparent, the proof rests on tools that can be understood in terms of a generalized version of the Raychaudhuri equation.
However, given the heaviness of the mathematical machinery involved, we defer this interpretation until sec. \ref{app.-proof: raychaudhuri-equation interpretation of the no-horizon proof}.

\subsection{A useful family of geodesics} \label{app.-proof: a useful family of geodesics}

At fixed time and internal position, i.e. $\smash{\de t = 0}$ and $\smash{\de y^m = 0}$, the line element associated to the metric in eq. (\ref{app. eq.: metric}) is $\smash{d l_{d-1}^2 = a^2 \, \e^{2 \tau} \, \e^{- 2 [(D-d)/(d-2)] \omega} d \mathrm{l}_{d-1}^2}$.
To facilitate the rest of the discussion, it is therefore convenient to define
\begin{equation} \label{app.-proof: alpha}
    \e^\A = \e^{\tau - \frac{D-d}{d-2} \omega} \dfrac{a}{a_\star},
\end{equation}
by which we can write the line element as $\smash{d l_{d-1}^2 = a_\star^2 \, \e^{2 \A} d \mathrm{l}_{d-1}^2}$.
From a physical perspective, we will build our argument around the particle action
\begin{equation} \label{app.-proof: cosmological-time action}
    S = - \mu_0 \int \de t \, \sqrt{1 - \e^{2 \frac{D-2}{d-2} \omega} \bar{g}_{mn} \dot{y}^m \dot{y}^n} \, \dfrac{a_\star}{a} \, \e^{\frac{\A}{\lambda}},
\end{equation}
where $\mu_0>0$ is a positive constant.
With respect to cosmological time, we may interpret this as the action of a particle moving in the reduced $(1,D-d)$-dimensional internal space, with metric $\smash{d \breve{s}_{1,D-d}^2 = -\de t^2 + \e^{2 [(D-2)/(d-2)] \omega} d \bar{s}_{D-d}^2}$, with a spacetime-dependent mass
\begin{equation} \label{app.-proof: particle mass}
    \mu = \mu_0 \, \dfrac{a_\star}{a} \, \e^{\frac{\A}{\lambda}}.
\end{equation}

\subsubsection{Observers with respect to the original metric}

In the following discussion, our interest is in time-like trajectories $x^M = x^M(\rho)$, at fixed external spatial coordinates $\smash{\de x^i = 0}$, that extremize the particle action in eq. (\ref{app.-proof: cosmological-time action}), which we now write as
\begin{equation} \label{app.-proof: particle action}
    S = - \mu_0 \int \de \rho \, \e^{\frac{1-\lambda}{\lambda} \A},
\end{equation}
where $\rho$ is a convenient time parameter.
Again, this action describes a particle with a spacetime-dependent mass.

We organize our reasoning in different steps.
\begin{itemize}
    \item To start, we derive the equation of motion of the particle.
    Given the mass spacetime dependence, we find the modified geodesic equation
    \begin{equation} \label{app.-proof: modified goedesic equation}
        \dfrac{\de^2 x^M}{\de \rho^2} + \Gamma^M_{PQ} \dfrac{\de x^P}{\de \rho} \dfrac{\de x^Q}{\de \rho} + \dfrac{1-\lambda}{\lambda} \Bigl[ g^{MN} + \dfrac{\de x^M}{\de \rho} \dfrac{\de x^N}{\de \rho} \Bigr] \der_N \A = 0.
    \end{equation}
    
    \item For our purposes, it is convenient to work with the unit tangent vector field $v^M = \de x^M/\de \rho$, normalized as $\smash{g_{MN} v^M v^N = -1}$.
    As initial conditions, at each spatial $\smash{(x_\star^i, y_\star^m)}$-point on the slice $t=t_\star$, we choose unit future-pointing vectors.
    In other words, $v^M$ is the family of $D$-dimensional vectors that solve the equations
    \begin{subequations}
    \begin{align}
        v^N \nabla_N v^M & = - \dfrac{1-\lambda}{\lambda}\,(g^{MN} + v^M v^N) \der_N \A, \label{app.-proof: v dynamics} \\
        v^M(t_\star,y_\star) & = \e^{-\tau(t_\star,y_\star)} \, \e^{\frac{D-d}{d-2}\omega(t_\star)} \, \delta^M_0, \label{app.-proof: v initial conditions}
    \end{align}
    \end{subequations}
    where each element of the family is identified by the initial conditions.
    Two fundamental implications follow.
    \begin{itemize}
        \item The unit normalization $\smash{g_{MN} v^M v^N = -1}$ is preserved at all times.
        This is indeed an invariant subspace of the solutions, as follows immediately from eq. (\ref{app.-proof: v dynamics}), being
        \begin{align*}
            \dfrac{1}{2} \, v^N \nabla_N (g_{PQ} v^P v^Q) & = - \dfrac{1-\lambda}{\lambda} \, (1 + g_{PQ} v^P v^Q) v^N \der_N \A.
        \end{align*}
        \item The initial conditions in eq. (\ref{app.-proof: v initial conditions}) have vanishing external components.
        Therefore, these remain zero throughout the evolution at all times.
        This is manifest, being $v^i = 0$ another invariant subspace of the solutions.
        Indeed, because $\smash{\der_i \omega = 0}$, $\smash{\der_i \tau = 0}$ and $\smash{\der_i \bar{g}_{mn} = 0}$, we have $\smash{\Gamma^i_{00} = 0}$, $\smash{\Gamma^i_{0n} = 0}$ and $\smash{\Gamma^i_{mn} = 0}$, with the only non-zero components being $\smash{\Gamma^i_{j0} = \delta^i_j \dot{\A}}$, $\smash{\Gamma^i_{jn} = \delta^i_j \der_n \A}$, and $\smash{\Gamma^i_{jk} = \Upgamma^i_{jk}}$; at the same time, we also have $\smash{\der_i \A = 0}$.
        Therefore eq. (\ref{app.-proof: v dynamics}) provides
        \begin{align*}
            v^N \der_N v^i + 2 (v^0 \dot{\A} + v^m \der_m \A) v^i + \Upgamma^i_{jl} v^j v^l & = - \dfrac{1-\lambda}{\lambda} \, (v^0 \dot{\A} + v^m \der_m \A) v^i.
        \end{align*}
    \end{itemize}
    As $\smash{g_{MN} v^M v^N = -1}$, with $v^0 = \de x^0/\de \rho > 0$, and $\smash{v^i = 0}$, the vector $v^M$ is entirely fixed at all times.
    After writing $\smash{v^m = \dot{y}^m v^0}$,\footnote{Here and below, $\dot{y}^m=\de y^m/\de t$ is the coordinate velocity of a trajectory, while the dot-notation on a spacetime-dependent field still means $\smash{\dot{} = \der_0}$.} the normalization condition fixes $\smash{(v^0)^2 = \e^{-2 \tau} \, \e^{2 [(D-d)/(d-2)] \omega}/[1 - \e^{2 [(D-2)/(d-2)] \omega} \bar{g}_{mn} \dot{y}^m \dot{y}^n]}$.
    In the end, we find the unique solution
    \begin{equation} \label{app.-proof: v}
        v^M = \dfrac{\e^{-\tau} \, \e^{\frac{D-d}{d-2}\omega}}{\sqrt{1 - \e^{2 \frac{D-2}{d-2} \omega} \bar{g}_{pq} \dot{y}^p \dot{y}^q}} \, (\delta^M_0 + \dot{y}^m \delta^M_m).
    \end{equation}
    To match the initial condition in eq. (\ref{app.-proof: v initial conditions}), we must fix $\smash{\dot{y}^m(t_\star, y_\star) = 0}$.
\end{itemize}

For future convenience, we can define the new time variable
\begin{equation} \label{app.-proof: dsigma}
    \de \sigma = \sqrt{1 - \e^{2 \frac{D-2}{d-2} \omega} \bar{g}_{mn} \dot{y}^m \dot{y}^n} \, \dfrac{a_\star}{a} \, \de t,
\end{equation}
where for each trajectory we can define $\smash{\sigma_\star = 0}$ without loss of generality.
An immediate consequence of the horizon hypothesis in eq. (\ref{app.-proof: would-be horizon condition}) is that there is a uniform bound, meaning that it is valid for all initial conditions $y_\star^m$, on the maximal extent of the $\sigma$-coordinate, which we write as
\begin{equation} \label{app.-proof: sigma-horizon inequality}
    0 \leq \sigma(t;y_\star) \leq \int_{t_\star}^t \de s \, \dfrac{a_\star}{a(s)} \leq a_\star L_\star.
\end{equation}
We also note that for every scalar function we can write
\begin{align*}
    v^M \der_M f = \dfrac{\e^{-\tau} \, \e^{\frac{D-d}{d-2}\omega}}{\sqrt{1 - \e^{2 \frac{D-2}{d-2} \omega} \bar{g}_{pq} \dot{y}^p \dot{y}^q}} \, \Bigl[ \dfrac{\der f}{\der x^0} + \dfrac{\der y^m}{\der x^0} \dfrac{\der f}{\der x^m} \Bigr] = \dfrac{\e^{-\tau} \, \e^{\frac{D-d}{d-2}\omega}}{\sqrt{1 - \e^{2 \frac{D-2}{d-2} \omega} \bar{g}_{pq} \dot{y}^p \dot{y}^q}} \, \dfrac{\de f}{\de t}.
\end{align*}
i.e.
\begin{equation} \label{app.-proof: sigma derivative}
    v^M \der_M f = \e^{-\A} \dfrac{\de f}{\de \sigma}.
\end{equation}
We may also write $\smash{v^M \der_M f = \de f/\de \rho}$, i.e. $\smash{\de \rho = \e^{\A} \de \sigma}$.

\subsubsection{Observers with respect to the auxiliary metric}

In the following, we will work in terms of an auxiliary metric $\smash{d \ubreve{s}_{1,D-1}^2}$ that allows us to deal with proper time-like geodesic curves.
Physically, comparing with the action in eq. (\ref{app.-proof: cosmological-time action}), we can describe a constant particle mass $\mu_0$ in the Weyl-transformed such that $\smash{d \ubreve{s}_{1,D-d}^2 = (\mu/\mu_0)^2 d \breve{s}_{1,D-d}^2}$, i.e.
\begin{align*}
    d \ubreve{s}_{1,D-d}^2 = \Bigl( \dfrac{\mu}{\mu_0} \Bigr)^{\!2} \bigl[- \de t^2 + \e^{2 \frac{D-2}{d-2} \omega} d \bar{s}_{D-d}^2 \bigr].
\end{align*}
Equivalently, with respect to the general metric $\smash{d s_{1,D-1}^2}$ in eq. (\ref{app. eq.: metric}), we define the auxiliary metric as
\begin{equation} \label{app.-proof: auxiliary metric}
    d \ubreve{s}_{1,D-1}^2 = \e^{2 \frac{1-\lambda}{\lambda} \A} d s_{1,D-1}^2.
\end{equation}
Indeed, in terms of the time reparameterization $\smash{\de \ubreve{\rho} = (\mu/\mu_0) \, \de t}$ along each trajectory, the particle action in this frame may be expressed as
\begin{align*}
    S = - \int \de t \; \mu \, \sqrt{1 - \e^{2 \frac{D-2}{d-2} \omega} \bar{g}_{mn} \dot{y}^m \dot{y}^n} = - \mu_0 \int \de \ubreve{\rho} \, \sqrt{1 - \ubreve{g}{}_{mn} \dfrac{\de y^m}{\de \ubreve{\rho}} \dfrac{\de y^n}{\de \ubreve{\rho}}}.
\end{align*}
Consistently with the constant-mass interpretation, we can see that the curves extremizing the particle action are time-like geodesics of the auxiliary metric.
Indeed, if we define the vector field $\smash{\ubreve{v}^M = \e^{- [(1-\lambda)/\lambda] \A} v^M}$, then, in view of eqs. (\ref{app.-proof: v dynamics}, \ref{app.-proof: v initial conditions}), we may verify the equation
\begin{align*}
    \ubreve{v}^N \ubreve{\nabla}{}_N \ubreve{v}^M = \e^{-2 \frac{1-\lambda}{\lambda} \A} \Bigl[ v^N \nabla_N v^M + \dfrac{1-\lambda}{\lambda}\,(g^{MN} + v^M v^N) \der_N \A \Bigr] = 0.
\end{align*}
Moreover, the normalization $\smash{g_{MN} v^M v^N = -1}$ is translated into the normalization $\smash{\ubreve{g}{}_{MN} \ubreve{v}^M \ubreve{v}^N = -1}$.
Therefore, the vector $\smash{\ubreve{v}^M}$ represents the unit tangent vector to geodesic curves in the auxiliary $\smash{\ubreve{g}{}_{MN}}$-metric.
We also notice that $\smash{\ubreve{v}^i = 0}$, as a consequence of eq. (\ref{app.-proof: v}).

In summary, the solution to the equations
\begin{subequations}
\begin{align}
    \ubreve{v}^N \ubreve{\nabla}{}_N \ubreve{v}^M & = 0, \label{app.-proof: t-hat dynamics} \\
    \ubreve{v}^M(t_\star; y_\star) & = \e^{- \frac{1}{\lambda} \A(t_\star; y_\star)}\, \delta^M_0, \label{app.-proof: t-hat initial conditions}
\end{align}
\end{subequations}
is
\begin{equation} \label{app.-proof: t-hat}
    \ubreve{v}^M = \dfrac{a}{a_\star} \, \dfrac{\e^{- \frac{\A}{\lambda}}}{\sqrt{1 - \e^{2 \frac{D-2}{d-2} \omega} \bar{g}_{pq} \dot{y}^p \dot{y}^q}} \, (\delta^M_0 + \dot{y}^m \delta^M_m).
\end{equation}
On top of this, we notice that, given the auxiliary-metric proper time $\smash{\sigmahat}$, inferred through the definition $\smash{\ubreve{v}^M = \de x^M/\de \sigmahat}$, is related to the time parameter $\sigma$ defined via eq. (\ref{app.-proof: dsigma}) as
\begin{equation} \label{app.-proof: dsigmahat}
    \de \sigmahat = \e^{\frac{\A}{\lambda}} \de \sigma, 
\end{equation}
as follows immediately from the condition $\smash{\de/\de \sigma = \e^\A v^M \der_M}$, which implies $\smash{\de/\de \sigma = \e^{\A/\lambda} \ubreve{v}^M \der_M}$.
The geometrical advantage of the auxiliary metric is that the trajectories are now ordinary time-like geodesics.
We also highlight the relationship $\smash{\de \sigmahat = \e^{[(1-\lambda)/\lambda] \A} \de \rho}$.

As anticipated, our goal is to define a coordinate system that allows us to describe the family of geodesics in a simple way.
At each time $t$, we have a map $\smash{y_\star^m \to y^m = y^m(t;y_\star)}$.
From now on, we will focus on families of solutions such that the matrix $\smash{j^m{}_n = \der y^m(t;y_\star)/\der y_\star^n}$ is non-singular, i.e.
\begin{equation} \label{app.-proof: trajectory jacobian}
    j = \det j = \det \, \dfrac{\der y^m(t;y_\star)}{\der y_\star^n} \neq 0.
\end{equation}
Because the geodesics are future-directed, $\de\sigmahat/\de t>0$ along each of them.
Therefore, for these families the auxiliary proper time $\sigmahat$ and the initial positions $y_\star^m$ provide good local coordinates.

Hence, our goal is to determine how the auxiliary metric, and in return the original metric, is pulled back onto the family of geodesics as parameterized by the variables $\smash{(\sigmahat,y_\star^m)}$, at each fixed $x^i$-locus.
In other words, we want to perform the pullback
\begin{equation} \label{app.-proof: pullback}
    (\sigmahat, x^i, y_\star^m) \overset{\phi}{\to} (t(\sigmahat, y_\star), x^i, y^m(\sigmahat, y_\star)).
\end{equation}
To proceed, we organize our argument in separate steps.
\begin{itemize}
    \item By construction, we can compute
    \begin{subequations}
    \begin{align}
        (\phi \ubreve{g})_{\sigmahat \sigmahat} & = \ubreve{g}{}_{MN} \dfrac{\der x^M}{\der \sigmahat} \dfrac{\der x^N}{\der \sigmahat} = \ubreve{g}{}_{MN} \ubreve{v}^M \ubreve{v}^N = -1, \\
        (\phi \ubreve{g})_{\sigmahat j} & = \ubreve{g}{}_{MN} \dfrac{\der x^M}{\der \sigmahat} \dfrac{\der x^N}{\der x^j} = \ubreve{g}{}_{Mj} \ubreve{v}^M = 0, \\
        (\phi \ubreve{g})_{i j} & = \ubreve{g}{}_{MN} \dfrac{\der x^M}{\der x^i} \dfrac{\der x^N}{\der x^j} = \ubreve{g}{}_{ij}.
    \end{align}
    \end{subequations}
    Here, we highlight a slight abuse of notation: for each fixed $\smash{y_\star^m}$, earlier we defined $\smash{\ubreve{v}^M = \de x^M / \de \sigmahat}$; now, we are formally considering the geodesics as $\smash{y_\star^m}$, i.e. $\smash{x^M = x^M(\sigmahat,y_\star^m)}$, whence the partial derivative.
    Similarly, we find
    \begin{subequations}
    \begin{align}
        (\phi \ubreve{g})_{y_\star^m y_\star^n} & = \ubreve{g}{}_{MN} \dfrac{\der x^M}{\der y_\star^m} \dfrac{\der x^N}{\der y_\star^n} = \hat{\ubreve{\gamma}}{}_{mn}, \\
        (\phi \ubreve{g})_{y_\star^m j} & = \ubreve{g}{}_{MN} \dfrac{\der x^M}{\der y_\star^m} \dfrac{\der x^N}{\der x^j} = \ubreve{g}{}_{ij} \dfrac{\der x^i}{\der y_\star^m} = 0.
    \end{align}
    \end{subequations}
    Here, $\hat{\ubreve{\gamma}}{}_{mn}$ is just a convenient notation.
    This is a positive-definite matrix, since it is the metric induced by $\ubreve{g}{}_{MN}$ on the spatial directions spanned by variations of the initial positions at fixed $\sigmahat$.
    
    An important property is the following.
    The initial hypersurface is defined via the identifications $\sigmahat=0$ and $\smash{y^m(t_\star;y_\star)=y_\star^m}$, for every trajectory.
    Therefore, we may write
    \begin{align*}
        \dfrac{\der t}{\der y_\star^n} \Bigl\ab_{\sigmahat=0} & = 0, \\
        \dfrac{\der y^m}{\der y_\star^n} \Bigl\ab_{\sigmahat=0} & = \delta^m_n,
    \end{align*}
    and we find immediately
    \begin{equation} \label{app.-proof: initial gamma-hat}
        \hat{\ubreve{\gamma}}{}_{mn}\ab_{\sigmahat = 0} = \ubreve{g}{}_{MN} \dfrac{\der x^M}{\der y_\star^m} \dfrac{\der x^N}{\der y_\star^n} \Bigr\ab_{\sigmahat = 0} = \ubreve{g}{}_{mn} \ab_{\sigmahat = 0}.
    \end{equation}
    
    \item Let us consider an arbitrary one-parameter variation $y_\star^m(\zetahat)$ of the initial point, and denote the corresponding family of geodesics by
    \begin{align*}
        x^M = x^M(\sigmahat,\zetahat).
    \end{align*}
    Along the resulting two-dimensional surface in spacetime, we define the two vectors
    \begin{align*}
        \ubreve{v}^M & = \dfrac{\der x^M}{\der\sigmahat}, \\
        \ubreve{l}^M & = \dfrac{\der x^M}{\der \zetahat}.
    \end{align*}
    By a slight abuse of notation, $\smash{\ubreve{v}^M}$ denotes the tangent vector along each $\smash{\zetahat}$-fixed curve.
    On other hand, $\smash{\ubreve{l}^M}$ measures how the spacetime point $x^M$ changes if we move to a neighboring geodesic while keeping the same value of $\smash{\sigmahat}$.
    By construction, in a torsion-free spacetime we have
    \begin{align*}
        \ubreve{v}^M \ubreve{\nabla}{}_M \ubreve{l}^N - \ubreve{l}^M \ubreve{\nabla}{}_M \ubreve{v}^N = \ubreve{v}^M \der_M \ubreve{l}^N - \ubreve{l}^M \der_M \ubreve{v}^N = \dfrac{\der}{\der \sigmahat}
        \Bigl(\dfrac{\der x^N}{\der \zetahat}\Bigr) - \dfrac{\der}{\der \zetahat} \Bigl(\dfrac{\der x^N}{\der \sigmahat}\Bigr) = 0.
    \end{align*}
    Therefore, we may compute the Lie derivative along each $\smash{\zetahat}$-fixed curve of the product $\smash{\ubreve{g}{}_{MN} \ubreve{v}^M \ubreve{l}^N}$ and verify that is invariant, being
    \begin{align*}
        \ubreve{v}^M \ubreve{\nabla}{}_M (\ubreve{g}{}_{PQ} \ubreve{v}^P \ubreve{l}^Q) = \ubreve{v}^M \ubreve{g}{}_{PQ} \ubreve{v}^P \ubreve{\nabla}{}_M \ubreve{l}^Q = \ubreve{l}^M \ubreve{g}{}_{PQ} \ubreve{v}^P \ubreve{\nabla}{}_M \ubreve{v}^Q = \dfrac{1}{2} \, \ubreve{l}^M \ubreve{\nabla}{}_M (\ubreve{g}{}_{PQ} \ubreve{v}^P \ubreve{v}^Q) = 0.
    \end{align*}
    The initial condition in eq. (\ref{app.-proof: t-hat initial conditions}) is exactly the unit future-directed normal to the hypersurface $t=t_\star$.
    Therefore, the vectors $\smash{\ubreve{v}^M}$ form the congruence of auxiliary-metric time-like geodesics emitted orthogonally from the initial hypersurface.
    At the same time, the vectors $\smash{\ubreve{l}^M}$ are initially tangent to the initial hypersurface $\smash{\sigma_\star = 0}$, i.e. $t=t_\star$, because the $\smash{\zetahat}$-variation is defined at fixed $\smash{\sigmahat}$, i.e. it only changes the initial position.
    As a consequence, the vectors $\smash{\ubreve{l}^M}$ are orthogonal to the vectors $\ubreve{v}^M$ throughout the curve, i.e.
    \begin{align*}
        \ubreve{g}{}_{MN} \ubreve{v}^M \ubreve{l}^N = 0.
    \end{align*}
    Since the variation $y_\star^m(\zetahat)$ is arbitrary, this conclusion holds for every direction obtained by varying the internal starting points $y_\star^m$ at $\smash{\sigmahat_\star = 0}$.
    The natural conclusion of this reasoning is that, for the coordinate change $\smash{(t,y^m) \to (\sigmahat,y_\star^m)}$, we find
    \begin{equation}
        (\phi \ubreve{g})_{\sigmahat m} = \ubreve{g}{}_{MN} \dfrac{\der x^M}{\der\sigmahat} \dfrac{\der x^N}{\der y_\star^m} = 0.
    \end{equation}
\end{itemize}
To summarize, with respect to the auxiliary metric in eq. (\ref{app.-proof: auxiliary metric}), we have the pullback
\begin{equation} \label{app.-proof: pulled-back auxiliary metric}
    (\phi \ubreve{g})_{MN} \de x^M \de x^N = - \de \sigmahat^2 + a_\star^2 \, \e^{\frac{2 \A}{\lambda}} \de \mathrm{l}_{d-1}^2 + \hat{\ubreve{\gamma}}{}_{mn} \de y_\star^m \de y_\star^n.
\end{equation}
Below, we immediately switch back to the general metric.

\subsubsection{Geodesic normal coordinates}

In view of eq. (\ref{app.-proof: pulled-back auxiliary metric}), after a change of coordinates the general metric in eq. (\ref{app. eq.: metric}) can be expressed in the form
\begin{equation} \label{app.-proof: pulled-back metric}
    (\phi g)_{MN} \de x^M \de x^N = - \e^{-2 \frac{1-\lambda}{\lambda} \A} \de \sigmahat^2 + a_\star^2 \, \e^{2 \A} d \mathrm{l}_{d-1}^2 + \hat{\gamma}_{mn} \de y_\star^m \de y_\star^n,
\end{equation}
where we have conveniently defined the rescaled pulled-back internal metric as $\smash{\hat{\gamma}_{mn} = \e^{-2 [(1-\lambda)/\lambda] \A} \hat{\ubreve{\gamma}}{}_{mn}}$.
For completeness, we emphasize the slight abuse of notation where $\smash{m}$-type indices denote both $\smash{y^m}$-coordinates, with respect to the metric in eq. (\ref{app. eq.: metric}), and $\smash{y_\star^m}$-indices, with respect to the metric in eq. (\ref{app.-proof: pulled-back metric}).

\subsubsection*{A simple vector field}
By definition, we have $\smash{v^M = \e^{[(1-\lambda)/\lambda] \A} \ubreve{v}^M}$, and $\smash{v_M = \e^{- [(1-\lambda)/\lambda] \A}  \ubreve{g}{}_{MN} \ubreve{v}^N}$.
The pullback of this vector field is extremely simple.
\begin{itemize}
    \item It is easy to compute the two pullback components
    \begin{align*}
        (\phi v)_{\sigmahat} & = \dfrac{\der x^M}{\der \sigmahat} v_M = \e^{- \frac{1-\lambda}{\lambda} \A} v^M v_M = - \e^{-\frac{1-\lambda}{\lambda} \A}, \\
        (\phi v)_i & = \delta^j_i v_j = 0.
    \end{align*}
    \item Let us consider again $\smash{\zetahat}$-deformations, defining $\smash{l^M = \e^{[(1-\lambda)/\lambda] \A} \ubreve{l}^M}$.
    Therefore, we can write $\smash{g_{MN} v^M l^N = 0}$, i.e.
    \begin{align*}
        v_M \dfrac{\der x^M}{\der \zetahat} = v_M \dfrac{\der x^M}{\der y_\star^m} \dfrac{\der y_\star^m}{\der \zetahat} = (\phi v)_{y_\star^m} \dfrac{\der y_\star^m}{\der \zetahat} = 0.
    \end{align*}
    Since $\smash{y_\star^m(\zetahat)}$ parameterizes arbitrary variations, this condition requires $\smash{(\phi v)_{y_\star^m} = 0}$.
\end{itemize}
In summary, we may write
\begin{equation} \label{app.-proof: phiv}
    (\phi v)^M = \e^{\frac{1-\lambda}{\lambda} \A} \delta^M_{\sigmahat}.
\end{equation}
In particular, for any scalar function $f$, we may write
\begin{equation} \label{app.-proof: sigma-hat derivative}
    v^M \der_M f = \e^{\frac{1-\lambda}{\lambda} \A} \dfrac{\der f}{\der \sigmahat}.
\end{equation}

\subsubsection*{Orthogonal-hyperplane projector}
From now on, for notational ease, we drop the pullback notation.
Let the projector on the subspace orthogonal to $v^M$ be defined as
\begin{equation} \label{app.-proof: h-projector}
    h_{MN} = g_{MN} + v_M v_N,
\end{equation}
as can be seen from the identities $\smash{h_{MN} v^N = 0}$, $\smash{h_{MQ} h^{QN} = h_M{}^N}$, and $h^M{}_M = D-1$.
In our coordinate choice, it is manifest that the projector effectively defines the metric tensor on the total non-time-like $(D-1)$-dimensional space, orthogonal to the $\smash{\sigmahat}$-time direction.
Indeed, if we denote collectively all the spatial directions orthogonal to the $\smash{\sigmahat}$-time direction as $\smash{x^\alpha = x^i, y_\star^m}$, we see that $\smash{h_{\alpha \beta} = g_{\alpha \beta}}$, with $\smash{h_{\sigmahat N} = 0}$.
Of course, we need to be careful when considering $h_{\alpha \beta}$ as the restricted metric.
For instance, if we write the $(D-1)$-spatial metric as
\begin{align*}
    d \check{s}_{D-1}^2 = \check{h}_{\alpha \beta} \de x^\alpha \de x^\beta = a_\star^2 \, \e^{2 \A} d \mathrm{l}_{d-1}^2 + \hat{\gamma}_{mn} \de y_\star^m \de y_\star^n
\end{align*}
from eq. (\ref{app.-proof: pulled-back metric}), where we defined $\smash{\check{h}_{\alpha \beta} = h_{\alpha \beta} = g_{\alpha \beta}}$ to emphasize that the $\smash{\check{h}_{\alpha \beta}}$-metric acts on a $(D-1)$-dimensional space, then on a vector $\smash{f_M}$ we have $\smash{\nabla_\alpha f_\beta = \check{\nabla}_{\alpha} f_\beta - (1/2) \, \e^{2 [(1-\lambda)/\lambda] \A} f_{\sigmahat} \, (\der \check{h}_{\alpha \beta}/\der \sigmahat)}$.
To avoid notational cluttering, below we will keep working in terms of the $g_{MN}$-metric unless differently stated.

Let us also define the spatially-projected $v^M$-gradient, i.e. 
\begin{equation} \label{app.-proof: K-tensor}
    K_{MN} = h_M{}^P h_N{}^Q \nabla_P v_Q = \nabla_M v_N - \dfrac{1-\lambda}{\lambda} \, v_M h_N{}^P \der_P \A,
\end{equation}
where we have exploited eq. (\ref{app.-proof: v dynamics}).
By construction, in particular, we have both $\smash{K_{MN} v^N = 0}$ and $\smash{v^M K_{MN} = 0}$.
This is consistent with the observation that we may just focus on the orthogonal subspace to $\smash{v^M}$, i.e. all spatial directions, being $\smash{K_{\sigmahat N} = 0}$ and $\smash{K_{M \sigmahat} = 0}$.
Then, we may thus just focus on the components $\smash{K_{\alpha \beta} = \nabla_\alpha v_\beta}$.
By eq. (\ref{app.-proof: phiv}), we can actually write $\smash{K_{\alpha \beta} = - \Gamma^{\sigmahat}_{\alpha \beta} v_{\smash{\sigmahat}} = (1/2) \, \e^{[(1-\lambda)/\lambda] \A} \, (\der g_{\alpha \beta} / \der \sigmahat)}$.
This expression also allows to see that $\smash{K_{\alpha \beta} = K_{\beta \alpha}}$, whence the full symmetry property
\begin{equation} \label{app.-proof: K-tensor symmetry}
    K_{MN} = K_{NM}.
\end{equation}
Then, for simplicity we define
\begin{subequations}
\begin{align}
    \h & = v^M \der_M \A = \e^{\frac{1-\lambda}{\lambda} \A} \dfrac{\der \A}{\der \sigmahat}, \label{app.-proof: h} \\
    \B_{mn} & = K_{mn} = \dfrac{1}{2} \, \e^{\frac{1-\lambda}{\lambda} \A} \dfrac{\der}{\der \sigmahat} \hat{\gamma}_{mn}, \label{app.-proof: B}
\end{align}
\end{subequations}
by which we may further write $\smash{K_{ij} = \h \, g_{ij}}$.
Moreover, we define
\begin{equation} \label{app.-proof: vartheta}
    \expansion = \nabla_M v^M = g^{MN} K_{MN}.
\end{equation}
In terms of eqs. (\ref{app.-proof: h}, \ref{app.-proof: B}) and eq. (\ref{app.-proof: vartheta}), by a straightforward calculation, we see that the trace of $\smash{K_{\alpha \beta}}$-tensor and the contraction of the $\smash{K_{\alpha \beta}}$-tensor with itself can be written as
\begin{subequations}
\begin{align}
    K_{MN} g^{MN} & = (d-1) \h + \hat{\gamma}^{mn} \B_{mn}, \label{app.-proof: K-trace} \\
    K_{MN} K^{MN} & = (d-1) \h^2 + 2 \, \B \, \hat{\cdot} \, \B. \label{app.-proof: K2}
\end{align}
\end{subequations}
where we are employing the notation $\smash{\B \, \hat{\cdot} \, \B = \B_{mn} \B_{pq} \hat{\gamma}^{mp} \hat{\gamma}^{nq}/2}$.

Before concluding, we emphasize one obvious aspect related to the Raychaudhuri-equation interpretation that exists for this discussion, as will be detailed in ssec. \ref{app.-proof: raychaudhuri-equation interpretation of the no-horizon proof}.
By direct inspection, we can see that $\expansion$ has the interpretation of a generalized expansion scalar, being
\begin{align*}
    \expansion = g^{MN} K_{MN} = \dfrac{1}{2} \, \e^{\frac{1-\lambda}{\lambda} \A} \check{h}^{\alpha \beta} \dfrac{\der}{\der \sigmahat} \check{h}_{\alpha \beta} = \e^{\frac{1-\lambda}{\lambda} \A} \dfrac{\der}{\der \sigmahat} \ln \sqrt{\vphantom{h^h}\smash{\check{h}_{D-1}}}.
\end{align*}
In other words, $\expansion$ controls the relative variation of the volume of the $(D-1)$-dimensional space orthogonal to the $\smash{\sigmahat}$-direction along the family of auxiliary-metric geodesics.

\subsubsection*{Global and local transformations}

Up to this point, we have constructed a mapping between the global $\smash{(t,x^i,y^m)}$-coordinates, as in the metric of eq. (\ref{app. eq.: metric}), and the local $\smash{(\sigmahat,x^i,y_\star^m)}$-coordinates describing suitable families of geodesic curves, as in the metric of eq. (\ref{app.-proof: pulled-back metric}).
Locally, we may write the total $D$-dimensional volume elements equivalently in either way.
In detail, we observe the following the following structures.
\begin{itemize}
    \item In terms of the global $\smash{(t,x^i,y^m)}$-coordinates, we may write
    \begin{align*}
        \de \mathrm{vol}_D = \e^{D \tau} \, \e^{-2 \frac{D-d}{d-2} \omega} a^{d-1} \, \sqrt{\mathrm{g}_{d-1}} \; \sqrt{\bar{g}_{D-d}} \; \de t \, \de^{d-1} x \, \de^{D-d} y.
    \end{align*}
    For our purposes, it is convenient to single out the measure by which we are able to consistently define the Einstein frame, via eq. (\ref{app. eq.: einstein-frame condition}), i.e.
    \begin{equation} \label{app.-proof: dmu}
        \de \mu = \e^{(D-2) \tau} \sqrt{\bar{g}_{D-d}} \; \de^{D-d} y,
    \end{equation}
    and thus to write
    \begin{equation} \label{app.-proof: global-coordinate volume element}
        \de \mathrm{vol}_D = \e^{2 \tau} \, \e^{-2 \frac{D-d}{d-2} \omega} a^{d-1} \, \sqrt{\mathrm{g}_{d-1}} \; \de t \, \de^{d-1} x \, \de \mu = \e^{2 \A} a^{d-1} \Bigl( \dfrac{a_\star}{a} \Bigr)^{2} \, \sqrt{\mathrm{g}_{d-1}} \; \de t \, \de^{d-1} x \, \de \mu.
    \end{equation}
    \item In terms of the local $\smash{(\sigmahat,x^i,y_\star^m)}$-coordinates, we may write
    \begin{align*}
        \de \mathrm{vol}_D = \e^{[(d-1) - \frac{1-\lambda}{\lambda}] \A} a_\star^{d-1} \sqrt{\mathrm{g}_{d-1}} \; \sqrt{\hat{\gamma}_{D-d}} \; \de \sigmahat \, \de^{d-1} x \, \de^{D-d} y_\star.
    \end{align*}
    In this case, it is useful to exploit the time parameter $\sigma$, defined in eq. (\ref{app.-proof: dsigma}), through the identification of eq. (\ref{app.-proof: dsigmahat}), and write
    \begin{equation} \label{app.-proof: local-coordinate volume element}
        \de \mathrm{vol}_D = \e^{d \A} a_\star^{d-1} \sqrt{\mathrm{g}_{d-1}} \; \sqrt{\hat{\gamma}_{D-d}} \; \de \sigma \, \de^{d-1} x \, \de^{D-d} y_\star.
    \end{equation}
\end{itemize}

Therefore, via a straightforward identification of eqs. (\ref{app.-proof: global-coordinate volume element}, \ref{app.-proof: local-coordinate volume element}), we find
\begin{equation} \label{app.-proof: volume-element transformation}
    \Bigl( \dfrac{a}{a_\star} \Bigr)^{d-3} \, \de t \, \de \mu = \e^{(d-2) \A} \sqrt{\hat{\gamma}_{D-d}} \; \de \sigma \, \de^{D-d} y_\star.
\end{equation}
Before proceeding, we also run a consistency check.
On the initial time slice $\smash{\sigma = 0}$ we have $\smash{\hat{\gamma}_{mn} \ab_{\sigmahat = 0} = [\e^{2 \tau} \, \e^{2 \omega} \bar{g}_{mn}] \ab_{\sigmahat = 0}}$ by suitably adapting the identification eq. (\ref{app.-proof: initial gamma-hat}) from the auxiliary-metric expression.
This precisely agrees with equating the volume elements on the initial slice $\smash{\sigma = 0}$, correctly providing $\smash{\de \mu_\star = [\e^{(D-2) \tau} \sqrt{\bar{g}_{D-d}}]\ab_{\sigma=0} \, \de^{D-d} y_\star}$.
Equivalently, we may write $\smash{[\e^{(D-2) \tau} \sqrt{\bar{g}_{D-d}}]\ab_{\sigma=0} = [\e^{(d-2) \A} \sqrt{\hat{\gamma}_{D-d}}]\ab_{\sigma=0}}$.

As an additional useful relationship, we can relate the determinants of the internal-space metric.
Again, the starting point is the identification in eq. (\ref{app.-proof: volume-element transformation}).
By exploiting the $\sigma$-time parameterization by eq. (\ref{app.-proof: dsigma}), and the identity $\smash{\de^{D-d} y = \ab j \ab \, \de^{D-d} y_\star}$, where $\smash{j^m{}_n = \der y^m(t;y_\star)/\der y_\star^n}$, we find
\begin{equation} \label{app.-proof: internal-metric transformation}
    \sqrt{\hat{\gamma}_{D-d}} = \dfrac{\e^{(D-d) (\omega + \tau)} \ab j \ab}{\sqrt{1 - \e^{2 \frac{D-2}{d-2} \omega} \bar{g}_{mn} \dot{y}^m \dot{y}^n}} \,  \sqrt{\bar{g}_{D-d}}.
\end{equation}
This is quite a useful identity because it allows us to relate the present proof to the discussion of the function $\smash{\varphi = \der_t \ln [\e^{(D-2) \tau} \sqrt{\bar{g}_{D-d}}]}$, defined in eq. (\ref{app. eq.: varphi}).
By the Jacobi formula, as long as the regularity condition in eq. (\ref{app.-proof: trajectory jacobian}) holds, we have
\begin{align*}
    \der_t \ln \ab j \ab = \mathrm{tr} (j^{-1} \der_t j) = \dfrac{\der y_\star^m}{\der y^n} \dfrac{\der \dot{y}^n}{\der y_\star^m} = \der_m \dot{y}^m.
\end{align*}
Then, we find
\begin{equation} \label{app.-proof: log-phi-hat derivative}
\begin{split}
    \der_t \ln \sqrt{\hat{\gamma}_{D-d}} = \varphi & + (D-d) \Bigl[ \dot{\omega} - \dfrac{d-2}{D-d} \, \dot{\tau} \Bigr] + \Bigl[ \dfrac{1}{2} \, (\bar{g}^{pq} \der_m \bar{g}_{pq}) + (D-d) \der_m \tau \Bigr] \dot{y}^m + \der_m \dot{y}^m \\
    & + \dfrac{\e^{2 \frac{D-2}{d-2} \omega}}{1 - \e^{2 \frac{D-2}{d-2} \omega} \bar{g}_{rs} \dot{y}^r \dot{y}^s} \Bigl[ \dfrac{D-2}{d-2} \, \dot{\omega} \, \bar{g}_{pq} \dot{y}^p + \dfrac{1}{2} \, \dot{\bar{g}}_{pq} \dot{y}^p + \bar{g}_{pq} \ddot{y}^p + \dfrac{1}{2} \, \dot{y}^m \der_m \bar{g}_{pq} \dot{y}^p \Bigr] \dot{y}^q.
\end{split}
\end{equation}
We emphasize that we are respecting to the following conventions.
\begin{enumerate*}[label=(\roman*)]
    \item For global-coordinate fields, such as the terms appearing in the $g_{MN}$-metric in eq. (\ref{app. eq.: metric}), partial $t$-derivatives are understood at fixed $y^m$-coordinates.
    In other words, for a field $\smash{f=f(t,y)}$, we have $\smash{\dot f=\der_t f\ab_y}$.
    When this same field is evaluated along the geodesic trajectories, we interpret it as $\smash{f(t,y(t,y_\star))}$, and its $t$-derivative (at fixed $\smash{y_\star^m}$) is instead $\smash{\de f/\de t = \dot{f} + \dot{y}^m \der_m f}$.
    \item Partial $t$-derivatives are meant as acting at fixed $\smash{y_\star^m}$-coordinates for terms appearing in the pulled-back $\smash{(\phi g)_{MN}}$-metric in eq. (\ref{app.-proof: pulled-back metric}), which are understood as functions $\smash{f = f(t, y_\star)}$.
\end{enumerate*}
For instance, we have $\smash{\dot{\tau} = \der_t \tau\ab_y}$, and $\smash{\der_t \hat{\gamma}_{D-d} = \der_t [\hat{\gamma}_{D-d}\ab_{y_\star}]}$.

For notational convenience, we may define the Jacobian factor
\begin{equation} \label{app.-proof: J}
    J = \dfrac{\e^{(d-2) \A} \sqrt{\hat{\gamma}_{D-d}}}{\bigl[ \e^{(d-2) \A} \sqrt{\hat{\gamma}_{D-d}} \bigr] \ab_{\sigma=0}}
\end{equation}
and write the volume-element transformation in eq. (\ref{app.-proof: volume-element transformation}) as
\begin{equation} \label{app.-proof: jacobian volume-element transformation}
    \Bigl( \dfrac{a}{a_\star} \Bigr)^{d-3} \, \de t \, \de \mu = J \, \de \sigma \, \de \mu_\star.
\end{equation}
We emphasize that $J$ is a scalar with respect to diffeomorphisms $\smash{y_\star^m \to y_\star'{}^m}$: indeed, it is a ratio of scalar densities in $\smash{y_\star^m}$-space, and hence a scalar.
Thanks to eq. (\ref{app.-proof: sigma-hat derivative}), we also compute the derivative
\begin{equation} \label{app.-proof: log-J derivative}
\begin{split}
    \dfrac{\dot{J}}{J} = \varphi & + (d-2) H + \Bigl[ \dfrac{1}{2} \, (\bar{g}^{pq} \der_m \bar{g}_{pq}) + (D-2) \der_m \tau \Bigr] \dot{y}^m + \der_m \dot{y}^m \\
    & + \dfrac{\e^{2 \frac{D-2}{d-2} \omega}}{1 - \e^{2 \frac{D-2}{d-2} \omega} \bar{g}_{rs} \dot{y}^r \dot{y}^s} \Bigl[ \dfrac{D-2}{d-2} \, \dot{\omega} \, \bar{g}_{pq} \dot{y}^p + \dfrac{1}{2} \, \dot{\bar{g}}_{pq} \dot{y}^p + \bar{g}_{pq} \ddot{y}^p + \dfrac{1}{2} \, \dot{y}^m \der_m \bar{g}_{pq} \dot{y}^p \Bigr] \dot{y}^q,
\end{split}
\end{equation}
where we have exploited the fact that the surface at $\sigma = 0$ is the initial spatial slice defined by constant $\smash{t=t_\star}$.

\subsection{Fundamental inequality} \label{app.-proof: fundamental inequality}

Here, we exploit the geometric structure defined above to derive the fundamental inequality to carry out the proof.

\subsubsection{Fundamental identity}

With these tools at hand, we are ready to write two non-trivial equations.
Our goal is that of finding the dynamical equations describing the evolution of the pulled-back Ricci tensor.
In particular, we want to determine the purely-temporal and purely-external components.
\begin{itemize}
    \item Our basic strategy to extract the purely-temporal pulled-back Ricci components is to exploit the directional vector $\smash{v^M}$, noticing that $\smash{R_{\sigmahat \sigmahat} = \e^{- 2 [(1-\lambda)/\lambda] \A} R_{MN} v^M v^N}$.
    
    To start, we notice that we can write $\smash{v^M \nabla_M (\nabla_N v^N) = v^M \nabla_N \nabla_M v^N - R_{MN} v^M v^N}$, as follows from the basic definition $\smash{[\nabla_M, \nabla_N] v^P = R^P{}_{QMN} v^Q}$.
    Therefore, we have
    \begin{align*}
        v^M \nabla_M (\nabla_N v^N) = \nabla_N (v^M \nabla_M v^N) - (\nabla_N v^M) (\nabla_M v^N) - R_{MN} v^M v^N.
    \end{align*}
    In view of the defining equation for $v^M$, i.e. eq. (\ref{app.-proof: v dynamics}), we may thus write
    \begin{align*}
        R_{MN} v^M v^N = & - v^M \nabla_M (\nabla_N v^N) - (\nabla_N v^M) (\nabla_M v^N) \\
        & - \dfrac{1 - \lambda}{\lambda} \, [\nabla^2 \A + (\nabla_M v^M) (v^N \der_N \A) + v^M \nabla_M (v^N \der_N \A)],
    \end{align*}
    We simplify this expression in steps.
    \begin{itemize}
        \item According to the definitions in eqs. (\ref{app.-proof: h}, \ref{app.-proof: B}, \ref{app.-proof: vartheta}), and thanks to the identity in eq. (\ref{app.-proof: K2}), we may write immediately
        \begin{align*}
            - v^M \nabla_M (\nabla_N v^N) - (\nabla_N v^M) (\nabla_M v^N) = - v^M \der_M \expansion - (d-1) \h^2 - 2 \, \B \, \hat{\cdot} \, \B.
        \end{align*}
        \item In view of eqs. (\ref{app.-proof: v dynamics}, \ref{app.-proof: h-projector}), we may write
        \begin{align*}
            \nabla^2 \A + v^M \nabla_M (v^N \der_N \A) = h^{MN} \nabla_M \der_N \A - \dfrac{1 - \lambda}{\lambda} \, h^{MN} \der_M \A \der_N \A.
        \end{align*}
        Just for this step, let us recall the purely spatial $(D-1)$-dimensional metric $\smash{d \check{s}_{D-1}^2 = \check{h}_{\alpha \beta} \de x^\alpha \de x^\beta}$.
        Then, we may compute the chain of identities
        \begin{align*}
            h^{MN} \nabla_M \der_N \A & = h^{\alpha \beta} \nabla_\alpha \der_\beta \A = \check{h}^{\alpha \beta} \check{\nabla}_\alpha \der_\beta \A - \check{h}^{\alpha \beta} \Gamma^{\sigmahat}_{\alpha \beta} \dfrac{\der \A}{\der \sigmahat} \\
            & = \dfrac{1}{\displaystyle \sqrt{\vphantom{h^h}\smash{\check{h}_{D-1}}}} \, \der_\alpha [\sqrt{\vphantom{h^h}\smash{\check{h}_{D-1}}} \; \check{h}^{\alpha \beta} \der_\beta \A] - h^{\alpha \beta} K_{\alpha \beta} \h \\
            & = \hat{\nabla}^2 \A + (d-1) \, (\der \A \, \hat{\cdot} \, \der \A) - \expansion \h.
        \end{align*}
    \end{itemize}
    To conclude, we get
    \begin{equation} \label{app.-proof: temporal pulled-back RMN}
    \begin{split}
        R_{MN} v^M v^N = - v^M \der_M \expansion - (d-1) \h^2 - 2 \, \B \, \hat{\cdot} \, \B - \dfrac{1 - \lambda}{\lambda} \, \biggl[ \hat{\nabla}^2 \A + \Bigl[(d-1) - \dfrac{1-\lambda}{\lambda} \Bigr] \, (\der \A \, \hat{\cdot} \, \der \A) \biggr].
    \end{split}
    \end{equation}
    
    \item In order to extract the external pulled-back Ricci components, we look directly at the corresponding Einstein equations.
    \begin{itemize}
        \item By direct inspection, in view of eq. (\ref{app. eq.: KK decomposition}), we can compute the $D$-dimensional Laplacian as
        \begin{align*}
            \nabla^2 \A = \e^{-2 \tau} \, \e^{2 \frac{D-d}{d-2} \omega} \biggl[ \epsilon H^2 - (d - 1) H^2 + \dfrac{D-d}{d-2} \Bigl[ \Bigl( \ddot{\omega} - \dfrac{d-2}{D-d} \, \ddot{\tau} \Bigr) + (d-1) H \Bigl( \dot{\omega} - \dfrac{d-2}{D-d} \, \dot{\tau} \Bigr) & \Bigr] \\
            + \e^{-2 \frac{D-2}{d-2} \omega} \bigl[ \bar{\nabla}^2 \tau + (D-2) \, \der \tau \, \bar{\cdot} \, \der \tau \bigr] - \varphi \Bigl[ H - \dfrac{D-d}{d-2} \Bigl( \dot{\omega} - \dfrac{d-2}{D-d} \, \dot{\tau} \Bigr) & \Bigr] \biggr].
        \end{align*}
        \item Again by direct inspection, thanks to eqs. (\ref{app. eq.: E0-eq.}, \ref{app. eq.: E_d-eq.}), we can expand the condition $(E_0 + E_d)/(d-1) = 0$ to infer the equation
        \begin{align*}
            \dfrac{\tilde{g}^{ij} R_{ij}}{d-1} - (d-2) \, \dfrac{k}{a^2} = - \epsilon H^2 + (d - 1) H^2 - \dfrac{D-d}{d-2} \Bigl[ \Bigl( \ddot{\omega} - \dfrac{d-2}{D-d} \, \ddot{\tau} \Bigr) + (d-1) H \Bigl( \dot{\omega} - \dfrac{d-2}{D-d} \, \dot{\tau} \Bigr) & \Bigr] \\
            - \e^{-2 \frac{D-2}{d-2} \omega} \bigl[ \bar{\nabla}^2 \tau + (D-2) \, \der \tau \, \bar{\cdot} \, \der \tau \bigr] + \varphi \Bigl[ H - \dfrac{D-d}{d-2} \Bigl( \dot{\omega} - \dfrac{d-2}{D-d} \, \dot{\tau} \Bigr) & \Bigr].
        \end{align*}
    \end{itemize}
    By comparison, we thus obtain
    \begin{align*}
        \dfrac{\e^{2 \tau} \, \e^{- 2 \frac{D-d}{d-2} \omega}}{d-2} \, \nabla^2 \A = \dfrac{k}{a^2} - \dfrac{1}{d-2} \dfrac{\tilde{g}^{ij} R_{ij}}{d-1}.
    \end{align*}
    As in eq. (\ref{(d-1)-space projector}), we define the projector $\smash{P_{MN}}$ on the $(d-1)$-dimensional external space.
    Then, in view of the definition of $\A$ in eq. (\ref{app.-proof: alpha}), we may write
    \begin{align*}
        \nabla^2 \A = (d-2) \, \dfrac{k}{a_\star^2} \, \e^{-2 \A} - \dfrac{1}{d-1} \, R_{MN} P^{MN}.
    \end{align*}
    
    We now translate the result into the language of the orthogonal-hyperplane projector.
    This is in fact the same step that is done above, and we can write immediately
    \begin{align*}
        \nabla^2 A & = \hat{\nabla}^2 \A + \Bigl[(d-1) - \dfrac{1-\lambda}{\lambda} \Bigr] \, (\der \A \, \hat{\cdot} \, \der \A) - \expansion \h - v^M \der_M \h.
    \end{align*}
    Therefore, we get
    \begin{equation} \label{app.-proof: external pulled-back RMN}
        \dfrac{1}{d-1} \, R_{MN} P^{MN} = (d-2) \, \dfrac{k}{a_\star^2} \, \e^{-2 \A} - \hat{\nabla}^2 \A - \Bigl[(d-1) - \dfrac{1-\lambda}{\lambda} \Bigr] \, (\der \A \, \hat{\cdot} \, \der \A) + \expansion \h + v^M \der_M \h.
    \end{equation}
\end{itemize}

Finally, we are able to compute the fundamental equation for the proof, involving precisely the combination whose sign is fixed as long as SSEC is fulfilled.
By direct comparison of eqs. (\ref{app.-proof: temporal pulled-back RMN}, \ref{app.-proof: external pulled-back RMN}), we are immediately able to write
\begin{equation} \label{app.-proof: fundamental id.}
\begin{split}
    R_{MN} v^M v^N - \dfrac{1-\lambda}{\lambda} \dfrac{1}{d-1} \, R_{MN} P^{MN} = & - \dfrac{1-\lambda}{\lambda} \, (d-2) \, \dfrac{k}{a_\star^2} \, \e^{-2 \A} \\
    & - \biggl[ \dfrac{1-\lambda}{\lambda} \, \expansion + (d-1) \h \biggr] \h - v^M \der_M \biggl[ \expansion + \dfrac{1-\lambda}{\lambda} \, \h \biggr] - 2 \, \B \, \hat{\cdot} \, \B.
\end{split}
\end{equation}

\subsubsection{Fundamental inequality}

Based on eq. (\ref{app.-proof: fundamental id.}), we are going to be able to provide a sharp positivity condition that will lay the core to manifest the contradiction of the horizon assumption in eq. (\ref{app.-proof: would-be horizon condition}).

To start, we will simplify the universal geometric terms.
Our main goal is to simplify the $\smash{v^M \der_M}$-operator in terms of a $\smash{\sigma}$-derivative, according to eqs. (\ref{app.-proof: dsigmahat}, \ref{app.-proof: sigma-hat derivative}), as $\smash{v^M \der_M f = \e^{-\A} \, \der f/\der \sigma}$, and to express all remaining quantities in terms of $\sigma$-derivatives.
For simplicity, we will use the notation $\smash{' = \der/\der \sigma}$.
\begin{itemize}    
    \item To start, we rearrange all terms depending on $\h$ and $\expansion$, defined in eqs. (\ref{app.-proof: h}, \ref{app.-proof: vartheta}).
    \begin{itemize}
        \item By direct inspection, we notice that the Jacobian factor in eq. (\ref{app.-proof: J}) is such that
        \begin{align*}
            \dfrac{J'}{J} = (d-2) \A' + \e^{\A} \hat{\gamma}^{mn} \B_{mn}.
        \end{align*}
    
        \item In view of eqs. (\ref{app.-proof: h}, \ref{app.-proof: K-trace}), we have $\smash{\h = \e^{-\A} \A'}$ and $\smash{\hat{\gamma}^{mn} \B_{mn} = \expansion - (d-1) \h}$.
        Hence, we can also write the identities
        \begin{subequations}
        \begin{align}
            \expansion & = \e^{-\A} [\ln J + \A]', \\ 
            \hat{\gamma}^{mn} \B_{mn} & = \e^{-\A} [\ln J - (d-2) \A]'.
        \end{align}
        \end{subequations}
    \end{itemize}
    In summary, we have
    \begin{align*}
        \e^{2 \A} \biggl[ \dfrac{1-\lambda}{\lambda} \, \expansion + (d-1) \h \biggr] \h + \e^{2 \A} v^M \der_M \biggl[ \expansion + \dfrac{1-\lambda}{\lambda} \, \h \biggr] = \dfrac{1 - 2 \lambda}{\lambda} \, \dfrac{J'}{J} \A' + (d-2) (\A')^2 + \Bigl[ \ln J + \dfrac{\A}{\lambda} \Bigr]''.
    \end{align*}

    \item By definition, in eq. (\ref{app.-proof: B}), the $\B_{mn}$-tensor is symmetric.
    By construction, the tensor $\smash{\hat{\gamma}_{mn}}$ defines a positive-definite $(D-d)$-dimensional metric.
    Therefore, by the standard trace decomposition of a tensor, we may infer the inequality
    \begin{equation} \label{app.-proof: B2 ineq.}
        2 \, \B \, \hat{\cdot} \, \B \geq \dfrac{1}{D-d} \, (\hat{\gamma}^{mn} \B_{mn})^2,
    \end{equation}
    whence
    \begin{align*}
        2 \, \B \, \hat{\cdot} \, \B \geq \dfrac{\e^{-2 \A}}{D-d} \, \Bigl[ \dfrac{J'}{J} - (d-2) \A' \Bigr]^2.
    \end{align*}
\end{itemize}
The conclusion to the reasoning above is that, on the basis of purely geometric considerations, we may exploit eq. (\ref{app.-proof: fundamental id.}) to write the inequality
\begin{equation} \label{app.-proof: geometric fundamental ineq.}
\begin{split}
    \e^{2 \A} \biggl[ R_{MN} v^M v^N - \dfrac{1-\lambda}{\lambda} \dfrac{1}{d-1} \, R_{MN} P^{MN} \biggr] \leq & - \dfrac{1-\lambda}{\lambda} \, (d-2) \, \dfrac{k}{a_\star^2} - \Bigl[ \ln J + \dfrac{\A}{\lambda} \Bigr]'' \\
    & - 2 \Bigl( \dfrac{1}{2 \lambda} - \dfrac{D-2}{D-d} \Bigr) \dfrac{J'}{J} \A' - (d-2) \dfrac{D-2}{D-d} (\A')^2 - \dfrac{1}{D-d} \Bigl( \dfrac{J'}{J} \Bigr)^{\!2}.
\end{split}
\end{equation}
In order to get rid of the mixed $\smash{J' \A'}$-term, we may complete the square.
For future convenience, let us define
\begin{subequations}
\begin{align}
    \psi & = \ln J + \dfrac{\A}{\lambda}, \label{app.-proof: psi} \\
    \chi & = \ln J - \Bigl( D - 2 - \dfrac{D-d}{2 \lambda} \Bigr) \A. \label{app.-proof: chi}
\end{align}
\end{subequations}
In terms of these geometric variables, we may then express ineq. (\ref{app.-proof: geometric fundamental ineq.}) as
\begin{align*}
    \dfrac{\e^{2 \A}}{d-2} \biggl[ R_{MN} v^M v^N - \dfrac{1-\lambda}{\lambda} \dfrac{1}{d-1} \, R_{MN} P^{MN} \biggr]
    \leq & - \dfrac{1-\lambda}{\lambda} \, \dfrac{k}{a_\star^2} - \dfrac{\psi''}{d-2} - \dfrac{1}{d-2} \dfrac{(\chi')^2}{D-d} \\
    & - \dfrac{1}{\lambda^2} \dfrac{D-2}{d-2} \, \Bigl[ \lambda (1 - \lambda) -\dfrac{1}{4} \dfrac{D-d}{D-2} \Bigr] (\A')^2
\end{align*}

At this time, we invoke SSEC.
If the $D$-dimensional theory fulfills SSEC, i.e. ineq. (\ref{ricci-tensor SSEC}), with any value of $\lambda$ in the interval
\begin{equation} \label{app.-proof: temporary lambda interval}
    \dfrac{1}{2} \Bigl( 1 - \dfrac{\sqrt{d-2}}{\sqrt{D-2}} \Bigr) \leq \lambda \leq \dfrac{1}{2} \Bigl( 1 + \dfrac{\sqrt{d-2}}{\sqrt{D-2}} \Bigr),
\end{equation}
which we assume for the time being, then we have the fundamental inequality
\begin{equation} \label{app.-proof: geometric+SSEC fundamental ineq.}
    \dfrac{\psi''}{d-2} + \dfrac{1}{d-2} \dfrac{(\chi')^2}{D-d} + \dfrac{1-\lambda}{\lambda} \, \dfrac{k}{a_\star^2} \leq 0.
\end{equation}
Although in principle we may get rid of the squared term, it will turn out to be fundamental to retain it to carry out the proof.

\subsection{Proof by contradiction} \label{app.-proof: proof by contradiction}

We are now ready to begin the proof by contradiction.
So far, no concrete use has been made of the would-be horizon in ineq. (\ref{app.-proof: would-be horizon condition}).
We have just observed that a consequence of this is that the $\sigma$-variable is bounded from above, as in ineq. (\ref{app.-proof: sigma-horizon inequality}), but no practical use of any of this has been made.
Below, we will exploit this in order to bound the $J$-integral, and then make use of its boundedness to demonstrate the presence of a contradiction.

\subsubsection{Uniform bounds}

The $\smash{\sigma}$-derivative $\smash{\psi'}$ is related to internal-space measure, according to eq. (\ref{app.-proof: log-J derivative}).    
Based on our fundamental inequality in ineq. (\ref{app.-proof: geometric+SSEC fundamental ineq.}), by the non-negativity of $\smash{(\chi')^2 \geq 0}$, we can immediately find the bound
\begin{align*}
    \int_{0}^\sigma \de \rho \, \Bigl[ \dfrac{\psi''}{d-2} + \dfrac{1-\lambda}{\lambda} \, \dfrac{k}{a_\star^2} \Bigr] = \dfrac{\psi' - \psi'_\star}{d-2} + \dfrac{1-\lambda}{\lambda} \, \dfrac{k \sigma}{a_\star^2} \leq 0.
\end{align*}
By eqs. (\ref{app.-proof: dsigma}, \ref{app.-proof: log-J derivative}), and thanks to the initial condition $\smash{\dot{y}^m(t_\star; y_\star) = 0}$ all over the slice $t=t_\star$, we can write
\begin{align*}
    \psi'_\star = \varphi_\star + (d-2) H_\star + \dfrac{1}{\lambda} \Bigl[ H_\star - \dfrac{D-d}{d-2} \Bigl( \dot{\omega}_\star - \dfrac{d-2}{D-d} \, \dot{\tau}_\star \Bigr) \biggr].
\end{align*}
Explicitly, this denotes a $\smash{y_\star^m}$-dependent initial condition $\smash{\psi_\star' = \psi'(\sigma; y_\star) \ab_{\sigma_\star = 0}}$.
Because the internal space is compact, this function admits a finite maximum $\smash{\max_{y_\star} \psi'(\sigma; y_\star) \ab_{\sigma_\star = 0} \in \mathbb{R}}$.
If we define
\begin{equation}
    m = m_0 + \max \bigl\lbrace 0, \max_{y_\star} \psi'(\sigma; y_\star) \ab_{\sigma_\star = 0} \bigr\rbrace > 0,
\end{equation}
where $\smash{m_0 > 0}$ is any positive finite value, then we may write
\begin{equation} \label{app.-proof: psi-derivative uniform bound}
    \dfrac{\psi'(\sigma, y_\star)}{d-2} + \dfrac{1-\lambda}{\lambda} \, \dfrac{k \sigma}{a_\star^2} <\dfrac{m}{d-2}.
\end{equation}
By definition of $m$, this is a uniform bound, valid for all of the $\smash{y_\star^m}$-determined geodesic line.
Thanks to ineq. (\ref{app.-proof: psi-derivative uniform bound}), we can identify the positive function
\begin{equation} \label{app.-proof: F}
    F(\sigma, y_\star) = \dfrac{m}{d-2} - \dfrac{\psi'(\sigma, y_\star)}{d-2} - \dfrac{1-\lambda}{\lambda} \, \dfrac{k \sigma}{a_\star^2} > 0.
\end{equation}
This is just a shorthand for the next inequality.
We emphasize that, on the initial slice $\sigma = \sigma_\star = 0$, for all possible initial $\smash{y_\star^m}$-points, this function is finite, due to compactness, and positively bounded from below, by construction, as
\begin{equation} \label{app.-proof: F uniform bound}
    F(\sigma,y_\star)\ab_{\sigma = 0} \geq \dfrac{m_0}{d-2} > 0.
\end{equation}

Our ultimate goal in this calculation is to bound $J$, and we currently have at our disposal a combination of derivatives of $\psi$ and $\chi$, which also include the factor $\A$.
Therefore, we define the diagonalization such that
\begin{equation} \label{app.-proof: pq definition}
    p \psi + q \chi = \ln J.
\end{equation}
By direct inspection, we find the solutions
\begin{subequations}
\begin{align}
    p & = \dfrac{1}{\lambda - \dfrac{1}{2} \dfrac{D-d}{D-2} + \dfrac{1}{D-2}} \Bigl[ \lambda - \dfrac{1}{2} \dfrac{D-d}{D-2} \Bigr], \\
    q & = \dfrac{1}{\lambda - \dfrac{1}{2} \dfrac{D-d}{D-2} + \dfrac{1}{D-2}} \dfrac{1}{D-2},
\end{align}
\end{subequations}
where we emphasize that $p+q=1$.
Then, thanks to ineq. (\ref{app.-proof: psi-derivative uniform bound}), we may write
\begin{align*}
    \dfrac{J'}{J} = q \chi' + p \psi' = q \chi' + m p - \dfrac{1-\lambda}{\lambda} \, (d-2) \, \dfrac{k}{a_\star^2} \, p \sigma - (d-2) p F,
\end{align*}
where we have singled out the positive term $F$, defined in eq. (\ref{app.-proof: F}).
This identity can be written more smoothly in the form
\begin{align*}
    \dfrac{\de}{\de \sigma} \ln \Bigl[ \e^{- m p \sigma + \frac{1-\lambda}{\lambda} \, (d-2) \, \frac{k}{a_\star^2} \frac{p \sigma^2}{2}} J \Bigr] = q \chi' - (d-2) p F.
\end{align*}
Unfortunately, we do not know a simple way to express $\smash{\chi'}$.
However, we know how $\smash{(\chi')^2}$ is bounded, thanks to ineq. (\ref{app.-proof: geometric fundamental ineq.}), and we further notice that we can write this bound as $\smash{(\chi')^2/[(d-2)(D-d)] \leq F'}$.
Therefore, we proceed with a square completion, writing the chain of identities
\begin{align*}
    \dfrac{\de}{\de \sigma} \ln \Bigl[ \e^{- m p \sigma + \frac{1-\lambda}{\lambda} \, (d-2) \, \frac{k}{a_\star^2} \frac{p \sigma^2}{2}} J \Bigr] = & - (d-2) p F + q \chi' + \dfrac{1}{d-2} \dfrac{(\chi')^2}{D-d} \dfrac{1}{F} - \dfrac{1}{d-2} \dfrac{(\chi')^2}{D-d} \dfrac{1}{F} \\
    = & - (d-2) p \Bigl[ 1 - (D-d) \dfrac{q^2}{4p} \Bigr] F \\
    & - \dfrac{1}{d-2} \dfrac{1}{D-d} \, \dfrac{1}{F} \Bigl[ \chi'- \dfrac{1}{2} \, (d-2) (D-d) q F \Bigr]^2 + \dfrac{1}{d-2} \dfrac{(\chi')^2}{D-d} \dfrac{1}{F}.
\end{align*}
We are currently assuming that $\lambda$ is in the interval of eq. (\ref{app.-proof: temporary lambda interval}).
This does not ensure a definite sign in the term $\smash{p - (D-d) q^2/4}$.
However, if
\begin{equation} \label{app.-proof: b>0}
    \dfrac{b}{d-2} = p - (D-d) \dfrac{q^2}{4} > 0,
\end{equation}
then we can greatly simplify this expression due to the positivity of $\smash{F > 0}$, via eq. (\ref{app.-proof: F}).
A sufficient restriction ensuring both this condition and eq. (\ref{app.-proof: temporary lambda interval}) is the $\smash{\lambda}$-interval defined in eq. (\ref{app. eq.: lambda proof interval}),\footnote{In fact, there also exists the additional interval
\begin{align*}
    \dfrac{1}{2} \Bigl( 1 - \dfrac{\sqrt{d-2}}{\sqrt{D-2}} \Bigr) \leq \lambda < \dfrac{1}{2} \dfrac{D-d-1-\sqrt{D-d+1}}{D-2}.
\end{align*}
As this is non-empty only for $D \geq 22$, we ignore it from now on.} i.e.
\begin{equation} \label{app.-proof: lambda proof interval}
    \dfrac{1}{2} \dfrac{D-d-1+\sqrt{D-d+1}}{D-2} < \lambda \leq \dfrac{1}{2} \Bigl( 1 + \dfrac{\sqrt{d-2}}{\sqrt{D-2}} \Bigr).
\end{equation}
By restricting to this interval from now on, in view of the $\smash{(\chi')^2}$-inequality anticipated above, we may finally write
\begin{align*}
    \dfrac{\de}{\de \sigma} \ln \Bigl[ \e^{- m p \sigma + \frac{1-\lambda}{\lambda} \, (d-2) \, \frac{k}{a_\star^2} \frac{p \sigma^2}{2}} J \Bigr] \leq \dfrac{F'}{F} - b F.
\end{align*}
For our purposes, the best way to write this inequality is
\begin{equation} \label{app.-proof: J-F ineq.}
    \dfrac{\de}{\de \sigma} \biggl[ \e^{- m p \sigma + \frac{1-\lambda}{\lambda} \, (d-2) \, \frac{k}{a_\star^2} \frac{p \sigma^2}{2}} \dfrac{J}{F} \biggr] \leq - b \, \e^{- m p \sigma + \frac{1-\lambda}{\lambda} \, (d-2) \, \frac{k}{a_\star^2} \frac{p \sigma^2}{2}} J.
\end{equation}

Now, for the first time, we invoke the assumption of the horizon existence, in eq. (\ref{app.-proof: would-be horizon condition}).
By its immediate implication that the $\sigma$-coordinate is bounded from above as $\smash{\sigma \leq a_\star L_\star}$, by ineq. (\ref{app.-proof: sigma-horizon inequality}), and thanks to the non-negativity of $b>0$ and positivity of $J \geq 0$, ineq. (\ref{app.-proof: J-F ineq.}) allows us to write
\begin{align*}
    \dfrac{\de}{\de \sigma} \biggl[ \e^{- m p \sigma + \frac{1-\lambda}{\lambda} \, (d-2) \, \frac{k}{a_\star^2} \frac{p \sigma^2}{2}} \dfrac{J}{F} \biggr] \leq - b \, \e^{- m p a_\star L_\star + \frac{1}{2} \frac{1-\lambda}{\lambda} \, (d-2) \, p k_0 L_\star^2} J,
\end{align*}
where for brevity we have defined $\smash{k_0 = \min \lbrace 0, k \rbrace}$.
An integration gives
\begin{align*}
    \e^{- m p \sigma + \frac{1-\lambda}{\lambda} \, (d-2) \, \frac{k}{a_\star^2} \frac{p \sigma^2}{2}} \dfrac{J}{F} - \dfrac{1}{F_\star} \leq - b \, \e^{- m p a_\star L_\star + \frac{1}{2} \frac{1-\lambda}{\lambda} \, (d-2) \, p k_0 L_\star^2} \int_{0}^{\sigma} \de \sigma' \, J(\sigma').
\end{align*}
Thanks to the positivity of $\smash{J, F > 0}$, and thanks to the uniform bound $\smash{F_\star \geq m_0/(d-2)}$, we can rearrange this as
\begin{align*}
    \int_{0}^{\sigma} \de \sigma' \, J(\sigma') \leq \dfrac{1}{F_\star b} \, \e^{m p a_\star L_\star - \frac{1}{2} \frac{1-\lambda}{\lambda} \, (d-2) \, p k_0 L_\star^2}.
\end{align*}
By the compactness of the internal space, we know that $\smash{F_\star}$ is finite and in particular uniformly bounded as $\smash{F_\star \geq m_0/(d-2)}$, via eq. (\ref{app.-proof: F uniform bound}).
In view of the finiteness of $\smash{L_\star < \infty}$, coming from the horizon assumption in ineq. (\ref{app.-proof: would-be horizon condition}), we are therefore certain of the finiteness of the constant
\begin{equation} \label{app.-proof: Gamma}
    \uniformbound = \dfrac{d-2}{m_0 b} \, \e^{m p a_\star L_\star - \frac{1}{2} \frac{1-\lambda}{\lambda} \, (d-2) \, p k_0 L_\star^2}.
\end{equation}
Our key inequality can be simply expressed, explicitly, as
\begin{equation} \label{app.-proof: J-integral key inequality}
    \int_{0}^{\sigma} \de \sigma' \, J(\sigma', y_\star) \leq \uniformbound.
\end{equation}
We stress that this is a universal inequality: it holds, with the same rate $\uniformbound$, irrespective of the initial $\smash{y_\star^m}$-point of the geodesic, and of the upper integration bound.

\subsubsection{Integrations over geodesic families and over global spacetime}

We now show that the trajectories considered above, followed until their possible first zero of the internal-coordinate Jacobian $\smash{j = \det \, [\der y^m(t;y_\star)/\der y_\star^n]}$, cover the spacetime region of interest up to a set of zero volume.
For each $\smash{y_\star^m}$, the first time such that $j=0$ interrupts our argument from eq. (\ref{app.-proof: trajectory jacobian}) onwards.
This will allow us to promote our bounds on trajectory integrals to bounds on global integrals.
Throughout this subsubsection, we will explicitly recall definitions and theorems from ref. \cite{Wald} to ensure maximal mathematical accuracy in our presentation.

Let us consider the same particle action that selects the observers, in eq. (\ref{app.-proof: cosmological-time action}).
For simplicity, let us fix the external coordinates throughout, with $\smash{\de x^i = 0}$, from now on, and let us work explicitly in the auxiliary Weyl-rescaled spacetime where the particle mass is actually constant as in eq. (\ref{app.-proof: auxiliary metric}), i.e. 
\begin{align*}
    d \ubreve{s}_{1,D-d}^2 = \Bigl( \dfrac{\mu}{\mu_0} \Bigr)^{\!2} \bigl[ - \de t^2 + \e^{2 \frac{D-2}{d-2} \omega} d \bar{s}_{D-d}^2 \bigr].
\end{align*}
We emphasize the underlying assumptions that are going to be exploited below.
\begin{enumerate*}[label=\alph*)]
    \item The internal $(D-d)$-dimensional metric is positive-definite, and the internal $(D-d)$-dimensional space is compact and boundaryless.
    \item All the spacetime functions involved in the auxiliary metric are smooth on both time and all internal-space directions.
    In particular, for the time dependence, we may restrict our attention to smoothness on the semi-infinite time domain $\smash{t \in\; ]t_-,\infty[}$, with $t_\star > t_- > - \infty$.
\end{enumerate*}    

To start, we show that the auxiliary spacetime is globally hyperbolic.
\begin{itemize}
    \item To begin, we consider arbitrary future-directed causal curves $\smash{x^M(\rho)}$ in the auxiliary spacetime $\smash{d \ubreve{s}_{1,D-d}^2}$, for some parameter $\rho$: these are curves whose (nonzero) tangent vectors $\smash{\ubreve{t}^M=\de x^M/\de \rho}$ are time-like or null, and lie inside or on the future light cone \cite[sec.~8.1]{Wald}, i.e. $\smash{\ubreve{g}{}_{MN} \ubreve{t}^M \ubreve{t}^N \leq 0}$, and $\smash{\ubreve{t}^0 > 0}$.
    As time $t$ increases along every future-directed causal curve, it can be used as its parameter, with $\de t/\de \rho > 0$.    
    We then highlight two basic observations.
    \begin{itemize}
        \item By definition of causality, along every such curve we have $\smash{1 - \e^{2 [(D-2)/(d-2)] \omega} \bar{g}_{mn} \dot{y}^m \dot{y}^n \geq 0}$.
        \item The $\smash{\e^{2 [(D-2)/(d-2)] \omega} \bar{g}_{mn}}$-metric is positive-definite and smooth, and the internal space is compact.
        Now, for any time-independent Riemannian metric on the internal space, let us consider any unit-length vectors in this metric.
        On any compact time interval within the $\smash{]t_-,\infty[}$-domain, their $\smash{\e^{2 [(D-2)/(d-2)] \omega} \bar{g}_{mn}}$-metric length is a real smooth function over a compact domain, in both the time and internal-space directions.
        Therefore, this length function admits a uniform positive lower bound \cite[th.~A.6]{Wald}, which we may express as the length computed by a specific time-independent Riemannian metric.
        In other words, we may identify a static metric $\smash{\gamma_{mn}}$ such that, throughout the interval, we have $\smash{\e^{2 [(D-2)/(d-2)] \omega} \bar{g}_{mn} v^m v^n \geq \gamma_{mn} \, v^m v^n}$ for any spatial vector $v^m$.
        (The $\gamma_{mn}$-metric may depend on the chosen time interval, and it is generally a function of the internal position, but it is independent of time within that interval.)
    \end{itemize}
    By definition, the distance between two positions cannot exceed the length of any spatial path joining them.
    Let us focus on the $\smash{\gamma_{mn}}$-distance $\smash{d_\gamma (x,y)}$.
    Based on the discussion above, for any two times $t_1, t_2$ within any chosen compact time interval, with $t_1 \leq t_2$, we may write the chain of inequalities
    \begin{align*}
        d_\gamma [y(t_1), y(t_2)] \leq \int_{t_1}^{t_2} \de t \, \sqrt{\gamma_{mn} \, \dot{y}^m \dot{y}^n} \leq t_2-t_1.
    \end{align*}
    With this tool, we are able to show that all causal inextensible curves, i.e. roughly speaking curves that cannot be extended further beyond their domain \cite[sec.~8.1]{Wald}, are curves that span the whole time domain, as follows.
    \begin{itemize}
        \item If the curve is defined over the whole $]t_-,\infty[$-domain, then it is inextendible.
        By continuity, for any time $t_0$ we also have the limit $\smash{y^m(t \to t_0) \to y^m(t_0)}$.
        \item If the curve time range has a finite upper point $t_0>t_-$, then we choose the compact time interval to contain $t_0$.
        For any two times $t$ and $s$ along the curve, with $t<s<t_0$, the distance traveled between them is bounded as $\smash{d_\gamma[y(t), y(s)] \leq s-t < t_0 - t}$, which implies the limit $\smash{d_\gamma[y(t \to t_0), y(s \to t_0)] \to 0}$.
        By compactness of the internal space, we know that this limiting position belongs to the internal space, which we denote as $\smash{y_0^m = \lim_{t \to t_0} y^m}$.
        Then, we can continue the curve in a time-like way by attaching the trajectory $\smash{y^m = y_0^m}$ at all times $t \geq t_0$.
        Therefore, a causal curve whose time range has a finite upper point $t_0>t_-$ is future-extendible, and the analogous statement holds for a curve with a finite lower point in the domain under consideration, which is past-extendible.
    \end{itemize}
    In summary, we can say that all the causal inextendible curves span the full $\smash{]t_-,\infty[}$-domain.
    Because $t_0>t_-$ and the internal space has no boundary, any point $\smash{(t_0,y_0^m)}$ on the curve always lies in the interior of the auxiliary spacetime domain.

    \item A hypersurface is achronal if, for any two points on it, there is no time-like curve joining them \cite[sec. 8.1]{Wald}.
    For an achronal hypersurface, the domains of future and past dependence are the sets of spacetime points such that all past- and future-inextendible causal curves passing through these points, respectively, intersect the given hypersurface \cite[sec. 8.3]{Wald}.
    A closed achronal hypersurface such that the union of the domains of future and past dependence is the whole spacetime itself is a Cauchy surface, and a spacetime that admits a Cauchy surface is globally hyperbolic \cite[sec. 8.3]{Wald}.
    A hypersurface is acausal if, for any two points on it, there is no causal curve joining them \cite{Treude:2012np}: this is a stronger requirement that achronality because it forbids not only time-like curves, but also null curves.

    Now, let us consider constant-$t$ hypersurfaces in the auxiliary metric, such as $t=t_\star$.
    By definition, any such surface is closed and achronal.
    Then, because all the causal inextendible curves span the full time domain, any inextendible curve with $t \lesseqgtr t_\star$ must always pass through $t=t_\star$.
    This means that constant-$t$ hypersurfaces are Cauchy hypersurfaces.
    Therefore, the auxiliary spacetime is globally hyperbolic.
    In particular, the initial-time Cauchy surface $\Sigma_\star=\lbrace t=t_\star\rbrace$ is a compact, space-like and smooth hypersurface because it is simply the compact internal space, endowed with a Riemannian smooth metric.
    This Cauchy surface is also not only (tautologically) achronal, but also acausal, since $t$ is strictly increasing along any causal curve.
\end{itemize}

With the information at our disposal, we can proceed further.
A point on a geodesic belonging to the congruence of geodesics orthogonal to a space-like hypersurface is conjugate to the space-like hypersurface if there exists an orthogonal deviation vector defined for the congruence which is non-zero on the space-like hypersurface but zero at the point \cite[sec. 9.3]{Wald}.
This might also be defined as a focal point \cite{Witten:2019qhl}.

Now, because $\smash{\Sigma_\star}$ is a Cauchy hypersurface, every spacetime point $\smash{\mathrm{P}=(t,y^m)}$ with $t>t_\star$ can be joined from the hypersurface through a future-directed causal curve.
As the auxiliary space-time is globally hyperbolic, for the Cauchy hypersurface $\Sigma_\star$ and any point joined from it by a future-directed causal curve, there exists a future-directed causal curve that maximizes the proper time elapsed along the causal curve \cite[sec. 8.3, th. 9.4.5]{Wald}.
In particular, on the maximizing curve there cannot be a focal point between the hypersurface and the point itself, and the maximizing curve is a geodesic orthogonal to the hypersurface \cite[secs. 8.3, th. 9.4.3]{Wald}.
Importantly, this is where our assumption of boundarylessness enters into play: for the orthogonality result, it is always understood that hypersurfaces are boundaryless \cite[sec. 9.3, app. B.2]{Wald}.
Indeed, this is similar to the intuitive explanation adapted from Riemannian signature \cite[ssec. 4.2]{Witten:2019qhl}.
We emphasize that the absence of earlier focal points applies only to the maximizing geodesic orthogonal to the hypersurface and joining the point, and not to every geodesic orthogonal to the hypersurface and joining the point.

As discussed above, the auxiliary-metric geodesics emanating orthogonally from the initial-time surface $t=t_\star$ are the trajectories with unit time-like velocity $\smash{\ubreve{v}^M}$, as defined in eq. (\ref{app.-proof: t-hat}).
Indeed, these are the geodesic curves with initial conditions $\smash{(y^m(t;y_\star),\dot{y}^m(t;y_\star))\ab_{t=t_\star} = (y_\star^m,0)}$, i.e. the trajectories whose tangent vectors are the solutions to eqs. (\ref{app.-proof: t-hat dynamics}, \ref{app.-proof: t-hat initial conditions}).
(The condition $\dot{y}^m(t_\star)=0$ ensures orthogonality to the initial slice, because the auxiliary metric has no mixed $\de t\,\de y^m$ terms.)
Since a geodesic is uniquely determined by its initial position and velocity \cite[sec.~3.3]{Wald}, we know that the maximizing geodesic belongs to the family of trajectories with velocities $\smash{\ubreve{v}^M}$.
Therefore, we can conclude that each event in the future of the initial-time hypersurface $t=t_\star$ is reached by at least one auxiliary-metric geodesic curve without encountering a strictly earlier focal point.

At any finite time, these geodesics remain well-defined
and depend smoothly on their starting positions.\footnote{This statement concerns continuation as geodesics, whereas the preceding causal-curve construction allowed arbitrary causal continuations. Using $t$ as parameter, the geodesic equations are ordinary equations of motion with smooth coefficients. On a compact time interval, compactness of the internal space and the causal speed bound prevent positions and spatial velocities from escaping their allowed finite ranges. The equations therefore remain regular and continue the solutions through every finite interior time, with smooth dependence on the initial data \cite[sec.~3.3]{Wald}.}
As argued above, for any spacetime point $\smash{(t,y^m)}$ with $t>t_\star$, there is a maximizing geodesic orthogonal to the initial Cauchy hypersurface with no focal points strictly before the point itself; however, as also noted above, there can in principle be geodesics orthogonal to the Cauchy hypersurface with focal points, and focal points may exist for maximizing geodesics at arrival.
Let us review the transformation introduced in eq.~(\ref{app.-proof: pullback}), which assigns to each starting position and elapsed auxiliary proper time the spacetime event reached by the corresponding normal geodesic, i.e.
\begin{align*}
    (\sigmahat, y_\star^m) \overset{\phi}{\to} (t(\sigmahat, y_\star), y^m(\sigmahat, y_\star))
\end{align*}
in the auxiliary spacetime.
Since $\smash{\de t/\de\sigmahat > 0}$, we may factorize this map as $\smash{(\sigmahat, y_\star^m) \to (t, y_\star^m) \to (t, y^m(t; y_\star))}$, with the Jacobian determinant
\begin{align*}
    \det \dfrac{\der (t,y^m)}{\der (\sigmahat,y_\star^n)} = \dfrac{\der t}{\der \sigmahat} \det \dfrac{\der y^m(t;y_\star)}{\der y_\star^n} = \dfrac{\der t}{\der \sigmahat} \, j.
\end{align*}
A focal point is precisely an event where this transformation is singular, i.e. some nonzero infinitesimal variation of the starting position gives zero separation of the geodesics at that event.
Since $\der t/\der\sigmahat>0$, the determinant identity shows that this happens precisely when $j=0$.
In particular, we highlight that a zero of $j$ does not interrupt an individual geodesic: rather, it obstructs the use of starting positions as local coordinates.
By Sard's theorem, the images of the points where a smooth map has a vanishing Jacobian form a measure-zero set in the target space \cite[th.~6.10]{Lee:2013}.
Therefore, for fixed $t$, the Jacobian of the transformation $y_\star \mapsto y(t;y_\star)$ is such that the positions $\smash{y^m(t;y_\star)}$ with $\smash{j(t;y_\star)=0}$ occupy vanishing internal-space volume.

Based on the considerations spelt out above, for any finite time $t>t_\star$, we can follow all trajectories from $\smash{s=t_\star}$ to whichever occurs first between $\smash{s=t}$ and the first zero of $j$, denoted as $\smash{t_{j=0}(y_\star)}$.
This time will be denoted as
\begin{equation} \label{app.-proof: t-end}
    t_\e(t,y_\star) = \min \lbrace t, t_{j=0}(y_\star) \rbrace,
\end{equation}
with $\smash{t_{j=0}(y_\star)=+\infty}$ if no zero occurs.
Now, let us consider any point $\smash{(s,y^m)}$ with $\smash{t_\star < s \leq t}$.
Our goal is to integrate non-negative functions $f = f(t,y)$.
For any $\smash{(s,y^m)}$-point, there is a maximizing geodesic orthogonal to the initial hypersurface that reaches it without an earlier focal point, and hence without an earlier zero of $j$.
In other words, at each fixed time, every internal-space point, except possibly for the zero-volume set of arrival points corresponding to focal points, is reached by a maximizing geodesic before its first focal point.
Non-maximizing geodesics reaching that point before their own first focal point are also included, and therefore count the same point again, while those reaching it after their first focal point have already been truncated, but the point remains covered by a maximizing geodesic.
Therefore, apart from the possible zero-volume set, every point is counted at least once, and any repetitions can only increase the integral of a non-negative function.
In practice, we may write the integral bound
\begin{align*} 
    \int_{t_\star}^t \de s \int \de^{D-d} y \, \sqrt{\bar{g}_{D-d}(s,y)} \; f(s,y) \leq \int \de^{D-d} y_\star \int_{t_\star}^{t_\e(t,y_\star)} \de s \, \sqrt{\bar{g}_{D-d}(s, y(s; y_\star))} \; \ab j(s; y_\star) \ab \, f(s,y(s;y_\star)).
\end{align*}
In the case of interest to us, we employ the measure $\smash{\de \mu = \e^{(D-2) \tau} \sqrt{\bar{g}_{D-d}} \; \de^{D-d} y}$ defined in eq. (\ref{app.-proof: dmu}), i.e. equivalently we consider functions $\phi$ such that $\smash{f = \e^{(D-2) \tau} \phi}$.
After suppressing the arguments for brevity, we may write the chain of implications
\begin{align*}
    \int_{t_\star}^t \de s \int \de \mu \, \phi = \int_{t_\star}^t \de s \int \de^{D-d} y \, \sqrt{\bar{g}_{D-d}} \; \e^{(D-2) \tau} \phi \leq \int \de^{D-d} y_\star \int_{t_\star}^{t_\e(t,y_\star)} \de s \, \sqrt{\bar{g}_{D-d}} \; \ab j \ab \, \e^{(D-2) \tau} \phi.
\end{align*}
In view of the transformation rule $\smash{\de \mu = \e^{(D-2) \tau} \sqrt{\bar{g}_{D-d}} \; \ab j \ab \, \de^{D-d} y_\star}$, and thanks to the volume-element identification $\smash{(a/a_\star)^{d-3} \, \de t \, \de \mu = J \, \de \sigma \, \de \mu_\star}$ in eq. (\ref{app.-proof: jacobian volume-element transformation}), we may write
\begin{align*}
    \int \de^{D-d} y_\star \int_{t_\star}^{t_\e(t,y_\star)} \de s \, \sqrt{\bar{g}_{D-d}} \; \ab j \ab \, \e^{(D-2) \tau} \phi = \int \de \mu_\star \int_{0}^{\sigma_\e(t,y_\star)} \de \sigma \, \Bigl( \dfrac{a_\star}{a} \Bigr)^{d-3} \, J \phi,
\end{align*}
where we have also defined $\smash{\sigma_\e (t, y_\star) = \sigma(t_\e(t,y_\star), y_\star)}$.
Therefore, we obtain the universal integral bound
\begin{equation} \label{app.-proof: local-to-global integral coverage}
    \int_{t_\star}^t \de s \int \de \mu \, \phi \leq \int \de \mu_\star \int_{0}^{\sigma_\e(t,y_\star)} \de \sigma \, \Bigl( \dfrac{a_\star}{a} \Bigr)^{d-3} \, J \phi. 
\end{equation}
This is the definitive inequality allowing us to transform global-spacetime integral bounds, on the left hand-side, into geodesic-family integral bounds, on the right hand-side.

\subsubsection{Contradiction}

Based on the discussion above, we may bound integrals computed in global coordinates from above, by their pulled-back counterparts, with an identification in the absence of overlapping geodesic trajectories.
After recalling the definition of the time-independent integral $K = \int \de \mu$ in eq. (\ref{app. eq.: K}), and taking into account the hierarchy rule between global and local integrals in eq. (\ref{app.-proof: local-to-global integral coverage}), we may write the chain of (in)equalities
\begin{align*}
    K \int_{t_\star}^t \de s \, a(s)^{d-3} = \int_{t_\star}^t \de t' \int \de \mu \, a^{d-3}(t') \leq a_\star^{d-3} \int \de \mu_\star \int_0^{\sigma_\e(t,y_\star)} \de \sigma' J(\sigma', y_\star).
\end{align*}
By the key condition in eq. (\ref{app.-proof: J-integral key inequality}), the integral of $J$ is uniformly bounded as $\smash{\int_{0}^{\sigma_\e (t, y_\star)} \de \sigma' \, J(\sigma', y_\star) \leq \uniformbound}$, where $\uniformbound$ is the $\smash{y_\star^m}$-independent positive finite constant defined in eq. (\ref{app.-proof: Gamma}).
In view of the time independence of $K$, we may as well write $\smash{K = \int \de \mu_\star}$.
These considerations allow us to write the chain of implications
\begin{align*}
    \int \de \mu_\star \int_0^{\sigma_\e(t,y_\star)} \de \sigma' J(\sigma', y_\star) \leq \uniformbound \int \de \mu_\star = \uniformbound K,
\end{align*}
whence we deduce the definitive inequality
\begin{equation} \label{app.-proof: global integral bound}
    \int_{t_\star}^t \de s \, a^{d-3}(s) \leq a_\star^{d-3} \uniformbound.
\end{equation}
Because $\smash{d \geq 3}$, we may write the trivial condition
\begin{align*}
    1 \leq \Bigl( \dfrac{a}{a_\star} \Bigr)^{d-3} + \dfrac{a_\star}{a}.
\end{align*}
By ineq. (\ref{app.-proof: global integral bound}), a time integration immediately gives us the inequality $\smash{t-t_\star \leq \uniformbound + \int_{t_\star}^t \de s \, a_\star/a}$.
As a consequence of the horizon hypothesis in eq. (\ref{app.-proof: would-be horizon condition}), the integral of the inverse scale factor is finite, and in particular $\smash{\int_{t_\star}^t \de s/a \leq L_\star}$.
Therefore, the inequality must hold
\begin{equation}
    t - t_\star \leq \uniformbound + a_\star L_\star.
\end{equation}
By finiteness of $\smash{L_\star < \infty}$, this implies that the constant term $\uniformbound$ must be arbitrarily large.
This contradicts the finiteness of $\uniformbound$ demonstrated for eq. (\ref{app.-proof: Gamma}).
Therefore, the horizon assumption in eq. (\ref{app.-proof: would-be horizon condition}) is inherently inconsistent.
It is necessary to conclude, instead, that there is no horizon, i.e.
\begin{equation} \label{app.-proof: no horizon}
    \int_{t_\star}^{\infty} \dfrac{\de t}{a(t)} = \infty.
\end{equation}
This concludes our proof.
The reduction of a theory fulfilling SSEC over a compact space cannot accommodate a lower-dimensional future cosmic horizon.

\subsection{Raychaudhuri-equation interpretation of the no-horizon proof} \label{app.-proof: raychaudhuri-equation interpretation of the no-horizon proof}

The tool through which one is usually able to bound the behavior of families of geodesics in view of a given energy condition is the Raychaudhuri equation, as reviewed e.g. in refs. \cite{Wald, Carroll}.
Although we have not presented our proof in these terms so far, the machinery we have built allows for a straightforward interpretation of the no-horizon statement as a consequence of SSEC along these lines.

\subsubsection{Auxiliary-metric Raychaudhuri equation}

To mirror the basic derivation of the Raychaudhuri equation, we seek a family of geodesics.
For the auxiliary metric in eq. (\ref{app.-proof: auxiliary metric}), we have the family of geodesics defined through their tangent vectors $\smash{\ubreve{v}^M}$ eqs. (\ref{app.-proof: t-hat dynamics}, \ref{app.-proof: t-hat initial conditions}).
We will work in terms of the coordinate frame defined in eq. (\ref{app.-proof: pullback}), i.e. $\smash{(\sigmahat, x^i, y_\star^m) \overset{\phi}{\to} (t(\sigmahat, y_\star), x^i, y^m(\sigmahat, y_\star))}$, with the metric as in eq. (\ref{app.-proof: pulled-back metric}).
By direct inspection, we can immediately compute the auxiliary-metric orthogonal projector and transport tensor, i.e. respectively
\begin{subequations}
\begin{align}
    \ubreve{h}{}_{MN} & = \ubreve{g}{}_{MN} + \ubreve{v}{}_M \ubreve{v}{}_N = \e^{2 \frac{1-\lambda}{\lambda} \A} h_{MN}, \label{app.-proof: auxiliary-frame h-projector} \\
    \ubreve{B}{}_{MN} & = \ubreve{\nabla}{}_N \ubreve{v}{}_M = \e^{\frac{1-\lambda}{\lambda} \A} \Bigl[ K_{NM} + \dfrac{1 - \lambda}{\lambda} \, h_{NM} \, v^P \der_P \A \Bigr], \label{app.-proof: auxiliar-frame uB-tensor}
\end{align}
\end{subequations}
where we have taken advantage of the definitions of the vector family $\smash{v^M = \e^{[(1-\lambda)/\lambda] \A}} \ubreve{v}^M$, and of the generalized orthogonal-subspace projector $h_{MN}$ and the corresponding spatially-projected velocity-gradient tensor $\smash{K_{MN}}$ in eqs. (\ref{app.-proof: h-projector}, \ref{app.-proof: K-tensor}).
In view of the definitions in eqs. (\ref{app.-proof: h}, \ref{app.-proof: B}, \ref{app.-proof: vartheta}), of the identity in eq. (\ref{app.-proof: K-trace}), and of the symmetry property in eq. (\ref{app.-proof: K-tensor symmetry}), we may immediately determine the auxiliary-metric expansion $\smash{\ubreve{\theta} = \ubreve{h}^{MN} \ubreve{B}{}_{MN} = \ubreve{\nabla}_M \ubreve{v}^M}$, the auxiliary-metric shear $\smash{\ubreve{\sigma}{}_{MN} = \ubreve{B}{}_{(MN)} - \ubreve{\theta} \, \ubreve{h}{}_{MN}/(D-1)}$, and the auxiliary-metric twist $\smash{\ubreve{\omega}{}_{MN} = \ubreve{B}{}_{[MN]}}$, respectively, as
\begin{subequations}
\begin{align}
    \ubreve{\theta} & = \e^{- \frac{1 - \lambda}{\lambda} \A} \Bigl[ \expansion + \dfrac{1-\lambda}{\lambda} \, (D-1) \h \Bigr], \label{app.-proof: auxiliary-metric expansion} \\
    \ubreve{\sigma}{}_{MN} & = \e^{\frac{1 - \lambda}{\lambda} \A} \Bigl[ K_{MN} - \dfrac{1}{D-1} \, \expansion \, h_{MN} \Bigr], \label{app.-proof: auxiliary-metric shear} \\[0.85ex]
    \ubreve{\omega}{}_{MN} & = 0, \label{app.-proof: auxiliary-metric twist}
\end{align}
\end{subequations}
where in particular we may further write
\begin{align*}
    \ubreve{\sigma}{}_{ij} & = \e^{\frac{1 - \lambda}{\lambda} \A} \Bigl[ \h - \dfrac{\expansion}{D-1} \Bigr] \, g_{ij}, \\
    \ubreve{\sigma}{}_{mn} & = \e^{\frac{1 - \lambda}{\lambda} \A} \Bigl[ \B_{mn} - \dfrac{1}{D-1} \, \expansion \, \hat{\gamma}_{mn} \Bigr],
\end{align*}
with $\smash{\ubreve{\sigma}{}_{\sigmahat N} = 0}$ and $\smash{\ubreve{\sigma}{}_{in} = 0}$.
Then, being the vectors $\smash{\ubreve{v}^M}$ tangent to geodesics by eq. (\ref{app.-proof: t-hat dynamics}), we may compute the variation of $\smash{\ubreve{B}_{MN}}$ along the curves as
\begin{align*}
    \ubreve{v}^P \ubreve{\nabla}_P \ubreve{B}_{MN} & = \ubreve{v}^P \ubreve{\nabla}_P \ubreve{\nabla}_N \ubreve{v}_M = \ubreve{v}^P \ubreve{\nabla}_N \ubreve{\nabla}_P \ubreve{v}_M + \ubreve{g}{}_{MS} \ubreve{R}^S{}_{QPN} \ubreve{v}^P \ubreve{v}^Q = - (\ubreve{\nabla}_N \ubreve{v}^P) (\ubreve{\nabla}_P \ubreve{v}_M) + \ubreve{g}{}_{MS} \ubreve{R}^S{}_{QPN} \ubreve{v}^P \ubreve{v}^Q.
\end{align*}
By a $\smash{\ubreve{g}{}_{MN}}$-trace, we obtain
\begin{align*}
    \ubreve{v}^N \ubreve{\nabla}_N \ubreve{\nabla}_M \ubreve{v}^M = - (\ubreve{\nabla}_N \ubreve{v}^M) (\ubreve{\nabla}_M \ubreve{v}^N) -\ubreve{R}{}_{MN} \ubreve{v}^M \ubreve{v}^N.
\end{align*}
This is simply the general auxiliary-metric time-like Raychaudhuri equation, which we may write as
\begin{equation} \label{app.-proof: auxiliary-metric time-like raychaudhuri eq.}
    \dfrac{\der \ubreve{\theta}}{\der \sigmahat} = - \dfrac{\ubreve{\theta}^2}{D-1} - 2 \, \ubreve{\sigma} \, \ubreve{\cdot} \, \ubreve{\sigma} + 2 \, \ubreve{\omega} \, \ubreve{\cdot} \, \ubreve{\omega} - \ubreve{R}{}_{MN} \ubreve{v}^M \ubreve{v}^N,
\end{equation}
where we are employing the notation whereby $\smash{f \, \ubreve{\cdot} \, f = f_{MN} f_{PQ} \, \ubreve{g}^{MP} \ubreve{g}^{NQ}/2}$ for any rank-2 tensor $f_{MN}$.
Our goal is to express eq. (\ref{app.-proof: auxiliary-metric time-like raychaudhuri eq.}) in terms of the functions computed for the actual metric, i.e. eqs. (\ref{app.-proof: h}, \ref{app.-proof: B}, \ref{app.-proof: vartheta}).
Thanks to previous calculations, this is actually a simple task.
\begin{itemize}
    \item First, let us focus on the terms depending on $\smash{\ubreve{\theta}}$, $\smash{\ubreve{\sigma}{}_{MN}}$ and $\ubreve{\omega}{}_{MN}$, defined in eqs. (\ref{app.-proof: auxiliary-metric expansion}, \ref{app.-proof: auxiliary-metric shear}).
    In view of eq. (\ref{app.-proof: sigma-hat derivative}), we can write
    \begin{align*}
        \dfrac{\der \ubreve{\theta}}{\der \sigmahat} & = \e^{- \frac{1-\lambda}{\lambda} \A} v^M \der_M \Bigl[ \e^{- \frac{1-\lambda}{\lambda} \A} \Bigl( \expansion + \dfrac{1-\lambda}{\lambda} \, (D-1) \h \Bigr) \Bigr] \\
        & = \e^{- 2 \frac{1-\lambda}{\lambda} \A} \Bigl[ - \dfrac{1-\lambda}{\lambda} \, \expansion \h - (D-1) \Bigl( \dfrac{1-\lambda}{\lambda} \Bigr)^{\!2} \h^2 + v^M \der_M \expansion + (D-1) \, \dfrac{1-\lambda}{\lambda} \, v^M \der_M \h \Bigr],
    \end{align*}
    and, at the same time, thanks to eqs. (\ref{app.-proof: K-trace}, \ref{app.-proof: K2}) we infer that
    \begin{align*}
        \dfrac{\ubreve{\theta}^2}{D-1} + 2 \, \ubreve{\sigma} \, \ubreve{\cdot} \, \ubreve{\sigma} = \e^{- 2 \frac{1-\lambda}{\lambda} \A} \Bigl[ (D-1) \Bigl( \dfrac{1-\lambda}{\lambda} \Bigr)^{\!2} \h^2 + 2 \, \dfrac{1-\lambda}{\lambda} \, \expansion \h + (d-1) \h^2 + 2 \, \B \, \hat{\cdot} \, \B \Bigr].
    \end{align*}
    A combination of these identities gives immediately
    \begin{equation} \label{app.-proof: auxiliary-metric time-like raychaudhuri eq., term 1}
        \e^{2 \frac{1-\lambda}{\lambda} \A} \Bigl[ \dfrac{\der \ubreve{\theta}}{\der \sigmahat} + \dfrac{\ubreve{\theta}^2}{D-1} + 2 \, \ubreve{\sigma} \, \ubreve{\cdot} \, \ubreve{\sigma} \Bigr] = v^M \der_M \Bigl[ \expansion + \dfrac{1-\lambda}{\lambda} \, (D-1) \h \Bigr] + \dfrac{1-\lambda}{\lambda} \, \expansion \h + (d-1) \h^2 + 2 \, \B \, \hat{\cdot} \, \B.
    \end{equation}
    \item Second, let us expand the Ricci-tensor term.
    In view of the Weyl rescaling giving $\smash{\ubreve{g}{}_{MN} = \e^{2 [(1-\lambda)/\lambda] \A} g_{MN}}$ and $\smash{\ubreve{v}^M = \e^{- [(1-\lambda)/\lambda] \A} v^M}$, we can write
    \begin{align*}
        \e^{2 \frac{1-\lambda}{\lambda} \A} \ubreve{R}{}_{MN} \ubreve{v}^M \ubreve{v}^N = R_{MN} v^M v^N & - (g_{MN} v^M v^N) \, \dfrac{1-\lambda}{\lambda} \, \nabla^2 \A \\
        &  - (D-2) \, \dfrac{1-\lambda}{\lambda} \, v^M v^N \biggl[ \nabla_M \der_N \A - \dfrac{1-\lambda}{\lambda} \, \Bigl[ \der_M \A \der_N \A - g_{MN} \, (\der \A \cdot \der \A) \Bigr] \biggr].
    \end{align*}
    To proceed, it is convenient to arrange our reasoning into two steps.
    \begin{itemize}
        \item By direct inspection, thanks to eq. (\ref{app.-proof: v dynamics}), we see the chain of identities
        \begin{align*}
            v^M v^N \nabla_M \der_N \A = {}& v^M [\nabla_M (v^N \der_N \A) - (\nabla_M v^N) \der_N \A] \\
            = {}& v^M \nabla_M (v^N \der_N \A) + \dfrac{1-\lambda}{\lambda} \, [(\der \A \cdot \der \A) + (v^M \der_M \A) (v^N \der_N \A)].
        \end{align*}
        Therefore, thanks to the normalization $\smash{g_{MN} v^M v^N = -1}$, we obtain
        \begin{equation} \label{app.-proof: auxiliary-metric ricci-tensor expansion}
            \e^{2 \frac{1-\lambda}{\lambda} \A} \ubreve{R}{}_{MN} \ubreve{v}^M \ubreve{v}^N = R_{MN} v^M v^N + \dfrac{1-\lambda}{\lambda} \, \nabla^2 \A - (D-2) \, \dfrac{1-\lambda}{\lambda} \, v^M \nabla_M (v^N \der_N \A).
        \end{equation}
        \item It is now possible to perform a crucial substitution.
        For our metric ansatz, the external Ricci trace determines the Laplacian of the generalized warp factor $\A$ through a purely geometric identity.
        As already established in deriving eq. (\ref{app.-proof: external pulled-back RMN}), we have
        \begin{equation} \label{app.-proof: warp-factor field eq.}
            \nabla^2 \A = - \dfrac{1}{d-1} \, g^{ij} R_{ij} + (d-2) \, \dfrac{k}{a_\star^2} \, \e^{-2 \A}.
        \end{equation}
        In terms of the $(d-1)$-dimensional external-space projector $P^{MN}$, defined in eq. (\ref{(d-1)-space projector}), we can identify $\smash{g^{ij} R_{ij} = R_{MN} P^{MN}}$.
    \end{itemize}
    A combination of eqs. (\ref{app.-proof: auxiliary-metric ricci-tensor expansion}, \ref{app.-proof: warp-factor field eq.}) eventually gives
    \begin{equation} \label{app.-proof: auxiliary-metric time-like raychaudhuri eq., term 2}
    \begin{split}
        \e^{2 \frac{1-\lambda}{\lambda} \A} \ubreve{R}{}_{MN} \ubreve{v}^M \ubreve{v}^N = & + R_{MN} v^M v^N - \dfrac{1-\lambda}{\lambda} \dfrac{1}{d-1} \, R_{MN} P^{MN} \\
        & + \dfrac{1-\lambda}{\lambda} \, \biggl[ (d-2) \, \dfrac{k}{a_\star^2} \, \e^{-2 \A} - (D-2) \, v^M \der_M \h \biggr].
    \end{split}
    \end{equation}
    This is a key equation because it shows the geometric origin of the SSEC combination: the $\lambda$-dependent combination in ineq. (\ref{app. eq.: SSEC}) emerges naturally as a consequence of the Weyl rescaling from the auxiliary metric to the physical metric, thanks to eq. (\ref{app.-proof: auxiliary-metric ricci-tensor expansion}), after expanding the traced $(d-1)$-dimensional external Ricci-tensor components, thanks to eq. (\ref{app.-proof: warp-factor field eq.}).
\end{itemize}
Finally, thanks to eqs. (\ref{app.-proof: auxiliary-metric time-like raychaudhuri eq., term 1}, \ref{app.-proof: auxiliary-metric time-like raychaudhuri eq., term 2}), we are able to express the auxiliary-metric time-like Raychaudhuri equation, i.e. eq. (\ref{app.-proof: auxiliary-metric time-like raychaudhuri eq.}), as
\begin{equation} \label{app.-proof: physical-variables auxiliary-metric time-like raychaudhuri eq.}
\begin{split}
    v^M \der_M \Bigl[ \expansion + \dfrac{1-\lambda}{\lambda} \, \h \Bigr] = & - \dfrac{1-\lambda}{\lambda} \, (d-2) \, \dfrac{k}{a_\star^2} \, \e^{-2 \A} - \biggl[ R_{MN} v^M v^N - \dfrac{1-\lambda}{\lambda} \dfrac{1}{d-1} \, R_{MN} P^{MN} \biggr] \\
    & - \dfrac{1-\lambda}{\lambda} \, \expansion \h - (d-1) \h^2 - 2 \, \B \, \hat{\cdot} \, \B.
\end{split}
\end{equation}
As it must be, we have recovered a known result, i.e. the fundamental identity in eq. (\ref{app.-proof: fundamental id.}).
From here onwards, one may proceed with the proof as provided in ssecs. \ref{app.-proof: fundamental inequality} and \ref{app.-proof: proof by contradiction}.

\section{SEC and SSEC for higher-dimensional SUGRA-like sources} \label{app.: SEC and SSEC for higher-dimensional SUGRA-like sources}

Here, we discuss the relationship between the $D$-dimensional SEC and SSEC for the standard bosonic SUGRA general-rank tensor fields and localized membranes.
We consider field strengths that individually inherit the external FLRW spatial symmetries, with either no external $(d-1)$-dimensional spatial component or all of them at the same time, and membranes filling the external $d$-dimensional spacetime directions.

For completeness, let us recall the SSEC statement.
Let $P^{MN}$ be the projector onto the external spatial directions, as defined in eq. (\ref{app. eq.: (d-1)-space projector}), i.e. $\smash{P^{MN} = \e^{-2\tau} \, \e^{2 [(D-d)/(d-2)] \omega} \tilde{g}^{ij} \delta_i^M \delta_j^N}$.
Let $v^M$ be any unit time-like vector orthogonal to the $(d-1)$-dimensional external space, i.e. such that $\smash{P_{MN} v^N = 0}$.
In particular, $v^M$ may have arbitrary internal components.
Then, in principle for a generic value of $\lambda$, we define the local SSEC combination
\begin{equation} \label{app. eq.: local SSEC combination}
    S[v] = \biggl[ T_{MN} - \dfrac{1}{D-2} \, g_{MN} (g^{PQ} T_{PQ}) \biggr] v^M v^N - (1-\lambda) \, T_{MN} \Bigl[ v^M v^N + \dfrac{1}{d-1} \, P^{MN} \Bigr].
\end{equation}
For simplicity, exploiting the normalization $\smash{g_{MN} v^M v^N = -1}$, we may write this as
\begin{align*}
    S[v] = \lambda \, T_{MN} v^M v^N - \dfrac{1-\lambda}{d-1} \, T_{MN} P^{MN} + \dfrac{1}{D-2} \, (g^{PQ} T_{PQ}),
\end{align*}
or equivalently, thanks to the $D$-dimensional Einstein equations, as
\begin{equation}
    \dfrac{1}{\lambda} \, \kappa_D^2 S[v] = R_{MN} v^M v^N - \dfrac{1-\lambda}{\lambda} \, \dfrac{1}{d-1} \, R_{MN} P^{MN}.
\end{equation}
This is the formulation exploited directly in the proof of the horizon theorem, in ineq. (\ref{app.-proof: geometric+SSEC fundamental ineq.}).
The statement of the general-observer local SSEC is that, for every choice of a vector $v^M$, one must have
\begin{equation} \label{app. eq.: local SSEC statement}
    S[v] \geq 0
\end{equation}
for a fixed constant value of $\lambda$ sitting in the interval defined in eq. (\ref{app. eq.: lambda proof interval}), i.e. $\smash{\lambda_- < \lambda \leq \lambda_+}$, where
\begin{subequations}
\begin{align}
    \lambda_- & = \dfrac{1}{2} \dfrac{D-d-1+\sqrt{D-d+1}}{D-2}, \label{app. eq.: minimal lambda} \\
    \lambda_+ & = \dfrac{1}{2} \Bigl( 1 + \dfrac{\sqrt{d-2}}{\sqrt{D-2}} \Bigr). \label{app. eq.: maximal lambda}
\end{align}
\end{subequations}
Clearly, as eq. (\ref{app. eq.: local SSEC combination}) is linear in the energy-momentum tensor, any source-by-source fulfillment at a common value of $\lambda$ implies the fulfillment by their sum.
Below, we identify common values of $\lambda$ in this interval for which the SEC-fulfilling sources under consideration also fulfill SSEC, with the restrictions on the number of dimensions specified below.

\subsection{Field strengths}

Let a $q$-form $A_q$ be a dynamical field with field-strength tensor $\smash{F_{q+1} = \de A_q}$ and gauge coupling $g_q$, with $q$ being any integer value in the interval $0 \leq q \leq D-1$.
Additionally, we also accept $F_0$-terms formally corresponding to $q=-1$.
In SUGRA models, the term $\smash{g_q}$ may depend on scalar fields, and the field-strength tensor $F_{q+1}$ may receive Chern-Simons contributions, such that the actual tensor to consider is $\smash{G_{q+1} = \de A_q + \sum_{p} c_{p,q} A_p \wedge F_{q+1-p}}$, for appropriate coefficients $c_{p,q}$; however, the argument below is algebraic in $F_{q+1}$, and therefore it applies irrespectively of any of these details.
The energy-momentum tensor associated to the field-strength tensor $\smash{F_{q+1}}$ is of the form
\begin{equation} \label{app. eq.: form-field energy-momentum tensor}
    T_{MN} = \dfrac{1}{g_q^2} \, \biggl[ F_{M,q} \cdot F_{N,q} - \dfrac{1}{2} \, g_{MN} (F_{q+1} \cdot F_{q+1}) \biggr],
\end{equation}
where we employ the notation such that $\smash{\alpha_p \cdot \beta_p = \alpha_{M_1 \dots M_p} \beta_{N_1 \dots N_p} g^{M_1 N_1} \dots g^{M_p N_p} / p!}$, for any $p$-forms $\alpha_p$ and $\beta_p$.
Here and below, we follow the conventions of ref. \cite{Nakahara}.

To discuss SSEC for field strengths, it is convenient to review the decomposition of the energy-momentum tensor in its electric and magnetic components.
For a given unit time-like vector $u^M$, i.e. such that $\smash{g_{MN} u^M u^N = -1}$, we may define the $1$-form $u_1$.
Then, we may decompose the field-strength tensor as
\begin{equation}
    F_{q+1} = - u_1 \wedge E_q + B_{q+1},
\end{equation}
where the electric and magnetic forms are respectively defined as
\begin{align*}
    E_q & = \I_u F_{q+1}, \\
    B_{q+1} & = u_1 \wedge E_q + F_{q+1}.
\end{align*}
By definition of interior and exterior products, we can readily verify several useful identities.
\begin{itemize}
    \item Because $\smash{\I_u^2 = 0}$, we have that
    \begin{align*}
        \I_u E_q = 0.
    \end{align*}
    \item Because of the distributive property $\smash{\I_u (\alpha_p \wedge \beta_q) = (\I_u \alpha_p) \wedge \beta_q + (-1)^p \alpha_p \wedge (\I_u \beta_q)}$ for any $p$-form $\alpha_p$ and $q$-form $\beta_q$, and thanks to the identity $\smash{\I_u u_1 = -1}$, we have that
    \begin{align*}
        \I_u B_{q+1} = 0.
    \end{align*}
\end{itemize}
In our conventions, we have that $\smash{F_{q+1} \wedge *_{1,D-1} F_{q+1} = (F_{q+1} \cdot F_{q+1}) \, \mathrm{vol}_{1,D-1}}$.
Then, in view of the obvious property by which we can write $\smash{(u_1 \wedge \beta_q) \wedge *_{1,D-1} \gamma_{q+1} = \beta_q \wedge *_{1,D-1} (\I_u \gamma_{q+1})}$, for any $q$-form $\beta_q$ and $(q+1)$-form $\gamma_{q+1}$, and exploiting again the interior-product distributive property, the time-like unit normalization $\I_u u_1 = -1$, and the orthogonality property $\smash{\I_u E_q = 0}$, we may write
\begin{align*}
    F_{q+1} \cdot F_{q+1} = - E_q \cdot E_q + B_{q+1} \cdot B_{q+1}.
\end{align*}
Indeed, we have $\smash{(u_1 \wedge E_q) \wedge *_{1,D-1} B_{q+1} = E_q \wedge *_{1,D-1} (\I_u B_{q+1}) = 0}$, and therefore
\begin{align*}
    F_{q+1} \wedge *_{1,D-1} F_{q+1} & = B_{q+1} \wedge *_{1,D-1} B_{1,q+1} + (u_1 \wedge E_{q}) \wedge *_{1,D-1} (u_1 \wedge E_{q}) \\
    & = B_{q+1} \wedge *_{1,D-1} B_{1,q+1} + E_{q} \wedge *_{1,D-1} [\I_u (u_1 \wedge E_{q})] \\
    & = B_{q+1} \wedge *_{1,D-1} B_{1,q+1} - E_{q} \wedge *_{1,D-1} E_q.
\end{align*}
Because $\smash{u_1}$ is a time-like vector, both $E_q$ and $B_{q+1}$ have purely spatial components.
Therefore, their squares are non-negative, i.e. $\smash{E_q \cdot E_q \geq 0}$ and $\smash{B_{q+1} \cdot B_{q+1} \geq 0}$.

Before discussing SSEC, let us review the discussion of the $D$-dimensional SEC for field strengths.
In view of the identity $\smash{g^{MN} F_{M, q} \cdot F_{N, q} = (q+1) \, F_{q+1} \cdot F_{q+1}}$, we can write a simple expression for the energy-momentum tensor in eq. (\ref{app. eq.: form-field energy-momentum tensor}).
Therefore, the SEC requirement, reviewed in eq. (\ref{app. eq.: SEC}), demands the condition
\begin{equation} \label{app. eq.: form-field SEC condition}
    \biggl[ T_{MN} - \dfrac{1}{D-2} \, g_{MN} (g^{PQ} T_{PQ}) \biggr] u^M u^N
    = \dfrac{1}{g_q^2} \biggl[ \dfrac{D-2-q}{D-2} \, E_q \cdot E_q + \dfrac{q}{D-2} \, B_{q+1} \cdot B_{q+1} \biggr] \geq 0.
\end{equation}
By the non-negativity of the squares $\smash{E_q \cdot E_q \geq 0}$ and $\smash{B_{q+1} \cdot B_{q+1} \geq 0}$, we see that a sufficient condition for field strengths to fulfill the $D$-dimensional SEC is if the integer values of $q$ fall in the interval $0 \leq q \leq D-2$.
Indeed, the only values that are excluded are $q=-1$, which formally corresponds to a SEC-violating positive $D$-dimensional scalar potential, and its Hodge-dual top-form counterpart, with $q=D-1$.

In order to discuss SSEC, we need to analyze in detail the structure of the field-strength tensor.
Under our symmetry assumptions, we may decompose the field-strength tensor as
\begin{equation} \label{app. eq.: FLRW-compatible field-strength decomposition}
    F_{q+1} = \alpha_{q+1} + \mathrm{vol}_{d-1} \wedge \beta_{q-d+2}.
\end{equation}
Here, $\smash{\alpha_{q+1}}$ and $\smash{\beta_{q-d+2}}$ are $(q+1)$- and $(q-d+2)$-forms orthogonal to the $(d-1)$-dimensional external spatial directions, and $\smash{\mathrm{vol}_{d-1}}$ is the $(d-1)$-dimensional spatial volume element.
To discuss SSEC, we need to contract $T_{MN}$ with $v^M$ and with the $P^{MN}$-projector.
\begin{itemize}
    \item For $T_{MN}$, double $v^M$-contractions provide
    \begin{align*}
        T_{MN} v^M v^N = \dfrac{1}{g_q^2} \biggl[ (v^M F_{M, q}) \cdot (v^N F_{N, q}) + \dfrac{1}{2} \, F_{q+1} \cdot F_{q+1} \biggr].
    \end{align*}
    Here, we see that there cannot be any cross $\alpha_{q+1}$-/$\beta_{q-d+2}$-term.
    Indeed, in the products $\smash{(v^M F_{M, q}) \cdot (v^N F_{N, q})}$ and $F_{q+1} \cdot F_{q+1}$ there is no possible contraction between the external $(d-1)$-dimensional space volume form, combined with $\smash{\beta_{q-d+2}}$, and the orthogonal $\smash{\alpha_{q+1}}$-form.
    While the $\smash{\alpha_{q+1}}$-terms are formally easy to deal with, the $\smash{\beta_{q-d+2}}$-terms require more care.
    \begin{itemize}
        \item For a $D$-dimensional spacetime factorized into an $m$- and $n$-dimensional component, with $m+n=D$, the Hodge operator decomposes as
        \begin{align*}
            *_{1,D-1} (\eta_p \wedge \xi_q) = (-1)^{q(m-p)} \, (*_m \eta_p) \wedge (*_n \xi_q).
        \end{align*}
        for any $p$-form $\eta_p$ defined on $m$-dimensional component and $q$-form $\smash{\xi_q}$ defined on the $n$-dimensional component.
        Hence, we may write
        \begin{align*}
            *_{1,D-1} (\mathrm{vol}_{d-1} \wedge \beta_{q-d+2}) = (-1)^{d-1} \, *_{1,D-d} \beta_{q-d+2},
        \end{align*}
        where the emergence of the sign is due to the ordering $\smash{\mathrm{vol}_{1,D-1} = \sqrt{-g_{00}} \; \de t \wedge \mathrm{vol}_{d-1} \wedge \mathrm{vol}_{D-d}}$ that we started with, and then
        \begin{align*}
            (\mathrm{vol}_{d-1} \wedge \beta_{q-d+2}) \wedge *_{1,D-1} (\mathrm{vol}_{d-1} \wedge \beta_{q-d+2}) & = (-1)^{d-1} \, \mathrm{vol}_{d-1} \wedge \beta_{q-d+2} \wedge *_{1,D-d} \beta_{q-d+2} \\
            & = (\beta_{q-d+2} \cdot \beta_{q-d+2}) \, \mathrm{vol}_{1,D-1}.
        \end{align*}
        \item By the interior-product distributive property, we have $\smash{\I_v (\mathrm{vol}_{d-1} \wedge \beta_{q-d+2}) = (-1)^{d-1} \, \mathrm{vol}_{d-1} \wedge \I_v \beta_{q-d+2}}$.
        Based on the identity above, we can also write
        \begin{align*}
            [\I_v (\mathrm{vol}_{d-1} \wedge \beta_{q-d+2})] \wedge *_{1,D-1} [\I_v (\mathrm{vol}_{d-1} \wedge \beta_{q-d+2})] = (\I_v \beta_{q-d+2} \cdot \I_v \beta_{q-d+2}) \, \mathrm{vol}_{1,D-1}
        \end{align*}
    \end{itemize}
    In view of the considerations above, we can write
    \begin{equation} \label{app. eq.: form-field SSEC id. 1}
    \begin{split}
        T_{MN} v^M v^N = \dfrac{1}{g_q^2} \biggl[ (v^M \alpha_{M, q}) \cdot (v^N \alpha_{N, q}) + (v^M \beta_{M, q-d+1}) \cdot (v^N \beta_{N, q-d+1}) & \\
        + \dfrac{1}{2} \, (\alpha_{q+1} \cdot \alpha_{q+1}) + \dfrac{1}{2} \, (\beta_{q-d+2} \cdot \beta_{q-d+2}) & \biggr].
    \end{split}
    \end{equation}

    \item To compute the $P^{MN}$-projection of the $T_{MN}$-tensor, it is convenient to exploit again form notation.
    \begin{itemize}
        \item By definition, we can write in components $\smash{(\de x^M \wedge A_p)_{M_1 \dots M_{p+1}} = (p+1) \, \delta^M_{[M_1} A_{M_2 \dots M_{p+1}]}}$ for any $p$-form $A_p$.
        At the same time, again by definition, we have in components $\smash{(\I_M F_{q+1})_{M_2 \dots M_{q+1}} = F_{M M_2 \dots M_{q+1}}}$, where $\smash{\I_M = \I_{\der_M}}$.
        Therefore, we see that
        \begin{align*}
            F_{q+1} \cdot (\de x^M \wedge \I_N F_{q+1}) = \dfrac{q+1}{(q+1)!} \, F^{M_1 \dots M_{q+1}} \delta^M_{M_1} F_{N M_2 \dots M_{q+1}} = \dfrac{1}{q!} \, F^{M M_2 \dots M_{q+1}} F_{N M_2 \dots M_{q+1}},
        \end{align*}
        i.e.
        \begin{align*}
             F_{q+1} \cdot (\de x^M \wedge \I_N F_{q+1}) = g^{MP} F_{P, q} \cdot F_{N, q}.
        \end{align*}
        \item By definition, $\alpha_{q+1}$ and $\smash{\beta_{q-d+2}}$ are orthogonal to the external $(d-1)$-dimensional space, i.e. $\smash{\I_i \alpha_{q+1} = 0}$ and $\smash{\I_i \beta_{q-d+2} = 0}$.
        We may thus write $\smash{\I_i F_{q+1} = (\I_i \mathrm{vol}_{d-1}) \wedge \beta_{q-d+2}}$.
        Then, by direct inspection we see the identity
        $\smash{\delta_i^j \, \de x^i \wedge \I_j \mathrm{vol}_{d-1} = (d-1) \mathrm{vol}_{d-1}}$, which allows us to write
        \begin{align*}
            P^M{}_N (\de x^N \wedge \I_M F_{q+1}) = \delta^M_i \delta_N^j \, \delta^i_j \, [\de x^N \wedge (\I_M \mathrm{vol}_{d-1}) \wedge \beta_{q-d+2}] = (d-1) \, \mathrm{vol}_{d-1} \wedge \beta_{q-d+2}.
        \end{align*}
    \end{itemize}
    Thanks to the two identities above, we may eventually write
    \begin{align*}
        (F_{M, q} \cdot F_{N, q}) \, P^{MN} & = [g_{MP} \, F_{q+1} \cdot (\de x^P \wedge \I_N F_{q+1})] \, P^{MN} \\
         & = (d-1) F_{q+1} \cdot (\mathrm{vol}_{d-1} \wedge \beta_{q-d+2}) \\
         & = (d-1) \, (\beta_{q-d+2} \cdot \beta_{q-d+2}).
    \end{align*}
    Therefore, we get
    \begin{equation} \label{app. eq.: form-field SSEC id. 2}
        \dfrac{1}{d-1} \, T_{MN} P^{MN} = \dfrac{1}{g_q^2} \, \dfrac{1}{2} \, [- (\alpha_{q+1} \cdot \alpha_{q+1}) + (\beta_{q-d+2} \cdot \beta_{q-d+2})].
    \end{equation}

    \item By the same reasoning above, we can trivially expand the $T_{MN}$-trace as
    \begin{equation} \label{app. eq.: form-field SSEC id. 3}
        \dfrac{1}{D-2} \, (g^{PQ} T_{PQ}) = \dfrac{1}{g_q^2} \, \dfrac{1}{2} \dfrac{2q - (D-2)}{D-2} \, [(\alpha_{q+1} \cdot \alpha_{q+1}) + (\beta_{q-d+2} \cdot \beta_{q-d+2})].
    \end{equation}
\end{itemize}
Finally, by exploiting eqs. (\ref{app. eq.: form-field SSEC id. 1}, \ref{app. eq.: form-field SSEC id. 2}, \ref{app. eq.: form-field SSEC id. 3}), we are able to expand the full SSEC combination in eq. (\ref{app. eq.: local SSEC combination}).
As there are no cross terms, it is convenient to write down the contributions from $\alpha_{q+1}$ and $\beta_{q-d+2}$ separately, as $\smash{S[v] = S[v; \alpha] + S[v; \beta]}$, with
\begin{subequations}
\begin{align}
    g_q^2 S[v; \alpha] & = \lambda \, [(v^M \alpha_{M, q}) \cdot (v^N \alpha_{N, q})] + \dfrac{q}{D-2} \, (\alpha_{q+1} \cdot \alpha_{q+1}), \label{app. eq.: alpha form-field SSEC combination} \\
    g_q^2 S[v; \beta] & = \lambda \, [(v^M \beta_{M, q-d+1}) \cdot (v^N \beta_{N, q-d+1})] + \Bigl[ \lambda - \dfrac{D-2-q}{D-2} \Bigr] \, (\beta_{q-d+2} \cdot \beta_{q-d+2}). \label{app. eq.: beta form-field SSEC combination}
\end{align}
\end{subequations}
Below, thanks to this separation, we discuss the requirements for SSEC to hold separately for field-strength tensors that are purely of the form $\smash{F_{q+1} = \alpha_{q+1}}$ and $\smash{F_{q+1} = \mathrm{vol}_{d-1} \wedge \beta_{q-d+2}}$.
\begin{itemize}
    \item Let us consider field strengths of the form $\smash{F_{q+1} = \alpha_{q+1}}$, with $\smash{0 \leq q \leq D-d}$.
    These denote the $q$-forms of this type that fulfill the $D$-dimensional SEC, and we will comment on $q=-1$ below.
    By choosing $v^M$ as the vector with respect to which we define $\smash{E_q = \I_v F_{q+1}}$ and $\smash{B_{q+1} = v_1 \wedge E_q + F_{q+1}}$, from eq. (\ref{app. eq.: alpha form-field SSEC combination}) we immediately get
    \begin{align*}
        g_q^2 S[v; \alpha] = \Bigl( \lambda - \dfrac{q}{D-2} \Bigr) \, E_q \cdot E_q + \dfrac{q}{D-2} \, B_{q+1} \cdot B_{q+1}.
    \end{align*}
    By the non-negativity of the squares $\smash{E_q \cdot E_q \geq 0}$ and $\smash{B_{q+1} \cdot B_{q+1} \geq 0}$, a sufficient condition for SSEC to hold is the bound $\smash{\lambda \geq q/(D-2)}$.
    Because the integer $q$ can take values up to $\smash{q=D-d}$, the minimal value of $\lambda$ that is able to ensure SSEC for all field-strength tensors of this kind is $\smash{\lambda \geq (D-d)/(D-2)}$.
    
    \item Let us consider field strengths of the form $\smash{F_{q+1} = \mathrm{vol}_{d-1} \wedge \beta_{q-d+2}}$, with $\smash{d-2 \leq q \leq D-2}$.
    These denote the $q$-forms of this class that fulfill the $D$-dimensional SEC, and we will comment on $q=D-1$ below.
    Again, by defining $\smash{E_q = \I_v F_{q+1}}$ and $\smash{B_{q+1} = v_1 \wedge E_q + F_{q+1}}$, from eq. (\ref{app. eq.: beta form-field SSEC combination}) we immediately get
    \begin{align*}
        g_q^2 S[v; \beta] = \dfrac{D-2-q}{D-2} \, E_q \cdot E_q + \Bigl[ \lambda - \dfrac{D-2-q}{D-2} \Bigr] \, B_{q+1} \cdot B_{q+1}.
    \end{align*}
    By the non-negativity of the squares $\smash{E_q \cdot E_q \geq 0}$ and $\smash{B_{q+1} \cdot B_{q+1} \geq 0}$, a sufficient condition for SSEC to hold is $\smash{\lambda \geq (D-2-q)/(D-2)}$.
    Because the integer $q$ can take values as small as $\smash{q=d-2}$, the minimal value of $\lambda$ that is able to ensure SSEC for all field-strength tensors of this kind is again $\smash{\lambda \geq (D-d)/(D-2)}$.
    This is expected because the $\smash{\beta_{q-d+2}}$-type field-strength tensors are the Hodge duals of the $\smash{\alpha_{q+1}}$-type ones.

    By the same expression, a nonzero top-form with $q=D-1$ violates SSEC.
    However, top-forms do not fulfill the $D$-dimensional SEC either.
\end{itemize}
Our analysis singles out a universal requirement for any field strength $\smash{F_{q+1}}$ that fulfills the $D$-dimensional SEC, i.e. such that $0 \leq q \leq D-2$, a sufficient requirement to fulfill SSEC is for $\lambda$ to be bounded from below as
\begin{equation} \label{app. eq.: form-field lambda-bound}
    \lambda \geq \dfrac{D-d}{D-2} = \lambda_0.
\end{equation}
This value falls inside the window that forbids cosmic horizons, defined in eq. (\ref{app. eq.: lambda proof interval}), or equivalently eqs. (\ref{app. eq.: minimal lambda}, \ref{app. eq.: maximal lambda}), as long as $D \leq 4d-6$, and we will comment on this point in detail later.
On the other hand, for the $q$-forms that violate the $D$-dimensional SEC, i.e. with $q=-1,D-1$, this bound does not ensure the fulfillment of SSEC.

Now, we point out an important caveat.
In principle, one might consider an isotropic energy-momentum tensor that is the sum of non-isotropic terms, in the external $(d-1)$-dimensions, which fulfills the $D$-dimensional SEC, but that may violate SSEC.
As an example, in flat $d$-dimensional FLRW spacetime ($k=0$), we might have a configuration with $(d-2)$-form fluxes with components
\begin{align*}
    (F_{d-2})_a = \I_{a} \mathrm{vol}_{d-1} \, f,
\end{align*}
where we define $\smash{\I_a = \I_{e_a}}$, with the vielbeins such that $\smash{e_a{}^i e_b{}^j g_{ij} = \delta_{ab}}$, and $a = 1, \dots, d-1$ labels different terms; see e.g. ref. \cite{Maleknejad:2012fw}.
By direct inspection, we find the total isotropic components
\begin{align*}
    g_{d-3}^2 T_{MN} = \Bigl[ - \dfrac{1}{2} \, (d-1) g_{MN} + (d-2) \, g_{ij} \delta^i_M \delta^j_N \Bigr] f^2,
\end{align*}
despite the individual $a$-labelled components being non-isotropic.
Then, based on ineq. (\ref{app. eq.: SEC}), for any unit time-like vector $u^M$ we find that the condition
\begin{align*}
    g_{d-3}^2 \Bigl[ T_{MN} - \dfrac{1}{D-2} \, g_{MN} T \Bigr] u^M u^N = \Bigl[ \dfrac{d-3}{D-2} \, (d-1) + (d-2) \, g_{ij} u^i u^j \Bigr] f^2 \geq 0
\end{align*}
is always fulfilled, which ensures the fulfillment of the $D$-dimensional SEC.
On the other hand, for unit time-like vectors $v^M$ such that $P_{MN} v^N = 0$, based on eq. (\ref{app. eq.: local SSEC combination}), we find that the condition
\begin{align*}
    g_{d-3}^2 S[v] = \dfrac{d-3}{D-2} \, (d-1) f^2 - (1-\lambda) (d-2) f^2 \geq 0
\end{align*}
requires $\smash{\lambda \geq 1 - [(d-1) (d-3)]/[(d-2) (D-2)]}$, which is not necessarily inside the interval $\smash{\lambda_- < \lambda \leq \lambda_+}$.
If $\lambda > \lambda_+$, then there is no admissible interval within which the no-horizon proof can be carried out.
Throughout this work, we restrict the individual field strengths to the form in eq. (\ref{app. eq.: FLRW-compatible field-strength decomposition}), thereby excluding configurations whose isotropy arises only after summing non-isotropic energy-momentum contributions.

\subsection{Localized sources}

Here, we consider a $p$-brane with tension $\smash{\tau_p}$.
This is a membrane that spans all $d$-dimensional directions and a $\smash{(p-d+1)}$-dimensional internal subspace $\smash{\Sigma_{p-d+1}}$, with $d-1 \leq p \leq D-1$.
For our purposes, the $p$-brane can be thought of as the locus such that $\smash{x^M(\xi) = \xi^\alpha \delta_\alpha^M}$, where $\xi^\alpha$ are the coordinates on the $(1, p)$-dimensional worldvolume in the static gauge.
On the orthogonal $(D-p-1)$-dimensional space $\smash{\Sigma^\perp_{D-p-1}}$, we define the positive Dirac distribution $\smash{\delta_\Sigma}$ such that $\smash{\int_{\Sigma^\perp_{D-p-1}} \de^{D-p-1} y \sqrt{g \ab_{\Sigma^\perp}} \; \delta_\Sigma f = f\ab_{\Sigma}}$.
Then, in orthogonal, the $p$-brane energy-momentum tensor is
\begin{equation} \label{app. eq.: brane energy-momentum tensor}
    T_{MN} = - \tau_p \, \gamma_{\alpha \beta} \delta^\alpha_M \delta^\beta_N \, \delta_\Sigma,
\end{equation}
where $\smash{\gamma_{\alpha \beta} = \delta_\alpha^M \delta_\beta^N g_{MN}}$ represents the induced $(1,p)$-dimensional metric on the brane worldvolume.
The $p$-brane may have a field-dependent tension, and it may have electric and magnetic Chern-Simons couplings.
However, the argument we will make sees the tension just as a proportionality factor whose only significant feature is the sign, and the Chern-Simons couplings are topological and hence transparent in the energy-momentum tensor.
Therefore, none of these details can affect our discussion.

Again, we start by reviewing the conditions for a $p$-brane to fulfill the $D$-dimensional SEC.
For the time being focus is on positive-tension $p$-branes, i.e. with $\smash{\tau_p > 0}$.
Given any unit time-like vector $u^M$, we can write
\begin{align*}
    \biggl[ T_{MN} - \dfrac{1}{D-2} \, g_{MN} (g^{PQ} T_{PQ}) \biggr] u^M u^N & = - \Bigl[ (\gamma_{\alpha \beta} \delta^\alpha_M \delta^\beta_N - g_{MN}) u^M u^N - 1 + \dfrac{p+1}{D-2} \Bigr] \, \tau_p \, \delta_\Sigma.
\end{align*}
Since the vector $u^M$ can be directed arbitrarily as long as it stays time-like, the $\smash{\gamma_{\alpha \beta}}$-contraction depends on the vector details: for instance, if $u^M = \delta^M_0/\sqrt{-g_{00}}$, we have $\smash{(\gamma_{\alpha \beta} \delta^\alpha_M \delta^\beta_N - g_{MN}) u^M u^N = 0}$, but if $\smash{u^M}$ has non-zero spatial components, then this might not be the case.
Yet, by positive-definiteness of the spatial metric components, we can bound the contraction as $\smash{(\gamma_{\alpha \beta} \delta^\alpha_M \delta^\beta_N - g_{MN}) u^M u^N \leq 0}$.
Therefore, the $D$-dimensional SEC, in eq. (\ref{app. eq.: SEC}), demands
\begin{equation} \label{app. eq.: brane SEC condition}
    \biggl[ T_{MN} - \dfrac{1}{D-2} \, g_{MN} (g^{PQ} T_{PQ}) \biggr] u^M u^N \geq \dfrac{D-3-p}{D-2} \, \tau_p \, \delta_\Sigma \geq 0.
\end{equation}
A sufficient condition for the $D$-dimensional SEC to be fulfilled is if $p \leq D-3$.

Let us turn our attention to SSEC.
In this case, we pick a unit time-like vector that is also such that $P_{MN} v^N = 0$.
Term by term, we can compute
\begin{align*}
    T_{MN} v^M v^N & = - (\gamma_{\alpha \beta} \delta^\alpha_M \delta^\beta_N \, v^M v^N) \, \tau_p \, \delta_\Sigma \geq - (g_{MN} v^M v^N) \, \tau_p \, \delta_\Sigma = \tau_p \, \delta_\Sigma, \\
    \dfrac{1}{d-1} \, T_{MN} P^{MN} & = - \dfrac{1}{d-1} \, (\gamma_{\alpha \beta} \delta^\alpha_M \delta^\beta_N \, P^{MN}) \, \tau_p \, \delta_\Sigma = - \tau_p \, \delta_\Sigma, \\
    \dfrac{1}{D-2} \, (g^{PQ} T_{PQ}) & = - \dfrac{p+1}{D-2} \, \tau_p \, \delta_\Sigma,
\end{align*}
and hence
\begin{equation} \label{app. eq.: brane SSEC combination}
    S[v] = \Bigl[ \dfrac{D-3-p}{D-2} - \lambda \, (\gamma_{\alpha \beta} \delta^\alpha_M \delta^\beta_N - g_{MN}) \, v^M v^N \Bigr] \tau_p \delta_\Sigma \geq \dfrac{D-3-p}{D-2} \, \tau_p \, \delta_\Sigma.
\end{equation}
Therefore, every positive-tension SEC-fulfilling brane, i.e. with $p \leq D-3$, also fulfills SSEC automatically.

A negative-tension brane of non-zero codimension, i.e. with $p \leq D-2$, violates both the $D$-dimensional SEC and SSEC, since we can always pick a time-like vector with a sufficiently negative value $\smash{(\gamma_{\alpha \beta} \delta^\alpha_M \delta^\beta_N - g_{MN}) u^M u^N < 0}$.
A negative-tension brane of zero codimension, i.e. with $\smash{p=D-1}$, instead, fulfills both conditions: this is because, formally, it corresponds to a negative-definite scalar potential.
To conclude, let us mention two important remarks.
First of all, branes not filling the external spatial directions and additional worldvolume contributions beyond pure tension require a separate analysis.
Second, here we have not taken into account explicitly the delta-function singularity and the properly backreacted geometry \cite{Blaback:2010sj}: the algebraic SEC/SSEC classification is unchanged if $\delta_\Sigma$ is replaced by a smooth non-negative transverse density preserving the pure-tension tensor structure.

\subsection{Sharpening the SSEC reach}

The general no-horizon proof applies when SSEC holds for a fixed $\lambda$ in the interval $\smash{\lambda_- < \lambda \leq \lambda_+}$ defined in eq. (\ref{app. eq.: lambda proof interval}), or equivalently eqs. (\ref{app. eq.: minimal lambda}, \ref{app. eq.: maximal lambda}).

For field strengths, our analysis has revealed the following claim: a sufficient condition for a field strength fulfilling the $D$-dimensional SEC to fulfill SSEC is that $\lambda$ is bounded as in ineq. (\ref{app. eq.: form-field lambda-bound}), i.e. $\smash{\lambda \geq \lambda_0 = (D-d)/(D-2)}$.
We need to discuss whether this value falls in the the window ensuring no horizons.
\begin{itemize}
    \item the lower bound of the window in eq. (\ref{app. eq.: lambda proof interval}) is always below the form-field critical value, being $\lambda_0 > \lambda_-$ for all values such that $\smash{D > d}$;
    \item the upper bound of the window in eq. (\ref{app. eq.: lambda proof interval}), instead, is above the form-field critical value, i.e. $\smash{\lambda_0 \leq \lambda_+}$, as long as
    \begin{equation} \label{app. eq.: no-horizon form-field bound}
        D \leq 4d-6.
    \end{equation}
\end{itemize}
For localized sources, instead, we have shown that, if they fulfill the $D$-dimensional SEC, they also fulfill SSEC.

In principle, the naive requirement for any field strength to ensure the no-horizon theorem, i.e. ineq. (\ref{app. eq.: no-horizon form-field bound}), can be violated for values of $D$ that are large enough compared to $d$.
However, as we show below, if we consider the tree-level 2-derivative higher-dimensional effective field theories that emerge in the weak-coupling limit of critical string theories, i.e. the 10d type-I, type-II and heterotic SUGRAs, and corresponding SEC-fulfilling membranes, but also any other non-supersymmetric 10d theory of field strengths and membranes, as well as 11d SUGRA, then we find that all field strengths do actually respect the SSEC with a value of $\lambda$ compatible with the no-horizon theorem, as long as $d \geq 4$; 11d SUGRA is in fact incompatible with cosmic horizons even for $d=3$.
More generally, we can argue the following.
\begin{itemize}
    \item For theories of field strengths and membranes with $D=10$, ineq. (\ref{app. eq.: no-horizon form-field bound}) is fulfilled as long as $\smash{d \geq 4}$.
    Therefore, the 10d type-I, type-II and heterotic SUGRAs, as well as any other classical 10d theory of field strengths and membranes do not allow the presence of a future cosmic horizon after their compactification to $d$-dimensional FLRW spacetime, as long as $d \geq 4$.
    On the other hand, a priori, compactifications of these 10d theories down to dimension $d=3$ are not covered by our no-horizon theorem.
    \item For theories of field strengths and membranes in dimension $D=11$, ineq. (\ref{app. eq.: no-horizon form-field bound}) is a priori fulfilled only for $d \geq 5$.
    However, if we consider specifically 11d SUGRA, our statement is much sharper.
    Indeed, the only independent field strength comes with $q=3$.
    Then, if we go back to the analysis following eqs. (\ref{app. eq.: alpha form-field SSEC combination}, \ref{app. eq.: beta form-field SSEC combination}), the sufficient conditions for fulfilling SSEC are $\smash{\lambda \geq 3/9 = 1/3}$, and $\smash{\lambda \geq 6/9 = 2/3}$, i.e. $\smash{\lambda \geq 2/3 = \lambda_{11}}$, independently of $d$.
    \begin{itemize}
        \item With $d=3$, we have $\smash{\lambda_- = 5/9 <2 /3}$, $\smash{\lambda_0 = 8/9}$ and $\smash{\lambda_+ = 2/3}$.
        Being $\smash{\lambda_{11} = \lambda_+ = 2/3}$, the no-horizon proof still applies because all sources fulfill SSEC with a value in the interval that delivers the proof.
        \item With $d=4$, we have $\lambda_- = (3+\sqrt{2})/9 < 2/3$, $\smash{\lambda_0 = 7/9}$ and $\smash{\lambda_+ = (1+\sqrt{2}/3)/2 > 2/3}$.
        Hence, being $\smash{\lambda_{11} < \lambda_+}$, the no-horizon proof still applies, again.
    \end{itemize}
\end{itemize}
To summarize, the universal value $\smash{\lambda_0=(D-d)/(D-2)}$ ensures SSEC for the SEC-fulfilling field strengths considered above, but it does not necessarily lie in the interval required by the no-horizon theorem.
Nonetheless, for 10d SUGRAs, it does as long as $d \geq 4$.
Instead, for 11d SUGRA, the restricted field content allows a common value of $\lambda$ in that interval for every $\smash{d \geq 3}$.

\section{No horizons and de Sitter and transplanckian censorship conjectures} \label{app.: no horizons and de Sitter and transplanckian censorship conjectures}

In this section, we discuss the relationship of our no-horizon theorem with the (refined and strong) dS conjecture \cite{Obied:2018sgi, Ooguri:2018wrx, Rudelius:2021azq} and with the transplanckian censorship conjecture (TCC) \cite{Bedroya:2019snp}.
To do so, we consider a $d$-dimensional cosmological theory including gravity and scalar fields $\smash{\phi^a}$, associated to the arbitrary positive-definite field-space metric $g_{ab}=g_{ab}(\phi)$.

The $d$-dimensional Friedmann equations read
\begin{subequations}
\begin{align}
    (d-1) \Bigl[ H^2 + \dfrac{k}{a^2} \Bigr] & = \dfrac{2 \kappa_d^2}{d-2} \, \Bigl[ \dfrac{1}{2} \, \dot{\phi}_a \dot{\phi}^a + V \Bigr], \label{app. dS-conjecture analysis: F1-eq.} \\
    \epsilon H^2 + \dfrac{k}{a^2} & = \dfrac{\kappa_d^2}{d-2} \, \dot{\phi}_a \dot{\phi}^a, \label{app. dS-conjecture analysis: F2-eq.}
\end{align}
\end{subequations}
where $\kappa_d$ is the $d$-dimensional gravitational coupling, with the cosmological Klein-Gordon equations
\begin{equation} \label{app. dS-conjecture analysis: KG}
    \ddot{\phi}^a + \Gamma^a_{bc} \dot{\phi}^b \dot{\phi}^c + (d-1) H \dot{\phi}^a + \dfrac{\der V}{\der \phi_a} = 0,
\end{equation}
where we emphasize the useful identity $\smash{(1/2) \, \de (\dot{\phi}_a \dot{\phi}^a) / \de t = (\ddot{\phi}^a + \Gamma^a_{bc} \dot{\phi}^b \dot{\phi}^c) g_{ad} \dot{\phi}^d}$.

To start, we consider the contraction $\smash{\dot{\phi}^a \der V /\der \phi^a}$.
By eq. (\ref{app. dS-conjecture analysis: KG}), and thanks to the Cauchy-Schwarz inequality, we immediately find
\begin{equation} \label{app. dS-conjecture analysis: speed-gradient bound}
    \biggl[ \dfrac{1}{a^{d-1}} \, \dfrac{\de}{\de t} \Bigl[ a^{d-1} \sqrt{\vphantom{\bigl\ab}\smash{\dot{\phi}_a \dot{\phi}^a}} \Bigr] \biggr]^2 \leq \Bigl({\dfrac{\der V}{\der\phi^a} \dfrac{\der V}{\der\phi_a}}\Bigr).
\end{equation}
Now, to explore the relationship with the conjectures above, let us assume by contradiction that there exists a constant $\gamma>0$ such that
\begin{equation} \label{app. dS-conjecture analysis: subcritical slope}
    \dfrac{1}{\kappa_d^2 V^2} \Bigl( \dfrac{\der V}{\der\phi^a} \dfrac{\der V}{\der\phi_a} \Bigr) < \gamma^2,
\end{equation}
throughout the field-space region under consideration, with $V>0$.
In particular, defining
\begin{equation} \label{strong-dS: critical slope}
    \gamma_d = \dfrac{2}{\sqrt{d-2}},
\end{equation}
any value $\smash{\gamma < \gamma_d}$ would contradict the strong dS conjecture and TCC for the field-space asymptotics \cite{Bedroya:2019snp, Rudelius:2021azq}.
By ineq. (\ref{app. dS-conjecture analysis: speed-gradient bound}), this implies the upper bound
\begin{equation} \label{app. dS-conjecture analysis: kinetic-energy bound}
    \dfrac{1}{2} \, \dfrac{\de}{\de t} \, (\dot{\phi}_a \dot{\phi}^a) \leq \kappa_d \gamma V \sqrt{\vphantom{\bigl\ab}\smash{\dot{\phi}_a \dot{\phi}^a}} - (d-1) H (\dot{\phi}_a \dot{\phi}^a).
\end{equation}
We can translate the bound in ineq. (\ref{app. dS-conjecture analysis: kinetic-energy bound}) into a differential inequality for the scale factor, as follows.
\begin{itemize}
    \item By differentiating eq. (\ref{app. dS-conjecture analysis: F2-eq.}), we immediately see the identity
    \begin{align*}
        \dfrac{1}{2} \dot{\epsilon} H = \epsilon^2 H^2 + \dfrac{k}{a^2} + \dfrac{1}{2} \dfrac{\kappa_d^2}{d-2} \, \dfrac{1}{H} \dfrac{\de}{\de t} \, (\dot{\phi}_a \dot{\phi}^a).
    \end{align*}
    Then, in view of ineq. (\ref{app. dS-conjecture analysis: kinetic-energy bound}), we get
    \begin{align*}
        \dfrac{1}{2} \, \dot{\epsilon} H \leq \epsilon^2 H^2 + \dfrac{k}{a^2} + \dfrac{\gamma}{\gamma_d} \dfrac{2 \kappa_d^2 V}{d-2} \dfrac{1}{H} \sqrt{\dfrac{\kappa_d^2 \, (\dot{\phi}_a \dot{\phi}^a)}{d-2}} - (d-1) \dfrac{\kappa_d^2 \, (\dot{\phi}_a \dot{\phi}^a)}{d-2}.
    \end{align*}
    
    \item By a combination of eqs. (\ref{app. dS-conjecture analysis: F1-eq.}, \ref{app. dS-conjecture analysis: F2-eq.}), we immediately find the identity
    \begin{align*}
        \dfrac{2 \kappa_d^2 V}{d-2} = (d-1-\epsilon) H^2 + (d-2) \, \dfrac{k}{a^2}.
    \end{align*}
    At the same time, eq. (\ref{app. dS-conjecture analysis: F2-eq.}) also gives
    \begin{align*}
        \dfrac{\kappa_d^2}{d-2} \, (\dot{\phi}_a \dot{\phi}^a) = \epsilon H^2 + \dfrac{k}{a^2}.
    \end{align*}
\end{itemize}
Thanks to the expressions above, we obtain
\begin{align*}
    \dfrac{1}{2} \, \dot{\epsilon} H \leq - \Bigl[ (d-1-\epsilon) \epsilon H^2 + (d-2) \, \dfrac{k}{a^2} \Bigr] + \dfrac{\gamma}{\gamma_d} \dfrac{1}{H} \Bigl[ (d-1-\epsilon) H^2 + (d-2) \, \dfrac{k}{a^2} \Bigr] \sqrt{\epsilon H^2 + \dfrac{k}{a^2}},
\end{align*}
i.e. equivalently
\begin{equation} \label{app. dS-conjecture analysis: epsilon bound}
    \dfrac{1}{2} \, \dot{\epsilon} H \leq (d-1-\epsilon) H \biggl[ \dfrac{\gamma}{\gamma_d} \sqrt{\epsilon H^2 + \dfrac{k}{a^2}} - \epsilon H \biggr] + (d-2) \, \dfrac{k}{a^2} \biggl[ \dfrac{\gamma}{\gamma_d} \dfrac{1}{H} \sqrt{\epsilon H^2 + \dfrac{k}{a^2}} - 1 \biggr].
\end{equation}

From now on, we focus on cases where $k \leq 0$.
It is convenient to introduce the parameter
\begin{equation} \label{app. dS-conjecture analysis: epsilon-phi}
    \epsilon^\phi = \dfrac{\epsilon H^2 + \dfrac{k}{a^2}}{H^2+\dfrac{k}{a^2}} = (d-1) \, \dfrac{\dot{\phi}_a \dot{\phi}^a}{\dot{\phi}_b \dot{\phi}^b + 2V},
\end{equation}
where the additional identification follows immediately from eqs. (\ref{app. dS-conjecture analysis: F1-eq.}, \ref{app. dS-conjecture analysis: F2-eq.}).
We observe that, as a consequence of the positivity of $V>0$, the $\epsilon^\phi$-parameter lies in the range $\smash{0 \leq \epsilon^\phi < d-1}$.
By direct inspection, we can thus rearrange ineq. (\ref{app. dS-conjecture analysis: epsilon bound}) as
\begin{align*}
    \dfrac{1}{1 + \dfrac{k}{a^2 H^2}} \dfrac{\dot{\epsilon}}{2 H} \leq \dfrac{\gamma}{\gamma_d} (d - 1 - \epsilon^\phi) \sqrt{\epsilon^\phi} \sqrt{1 + \dfrac{k}{a^2 H^2}} - (d-2) \dfrac{k}{a^2 H^2} (1 - \epsilon^\phi) - (d - 1 - \epsilon^\phi) \biggl[ 1 - (1 - \epsilon^\phi) \Bigl[ 1 + \dfrac{k}{a^2 H^2} \Bigr] \biggr].
\end{align*}
Then, after observing the identity
\begin{align*}
    \dfrac{\dot{\epsilon}^\phi}{2H} = \dfrac{1}{1 + \dfrac{k}{a^2 H^2}} \dfrac{\dot{\epsilon}}{2H} - \dfrac{k}{a^2 H^2} \dfrac{(1-\epsilon)^2} {\Bigl[1 + \dfrac{k}{a^2 H^2}\Bigr]^2},
\end{align*}
we obtain
\begin{align*}
    \dfrac{\dot{\epsilon}^\phi}{2H} \leq (d-1-\epsilon^\phi) \sqrt{\epsilon^\phi} \biggl[ \dfrac{\gamma}{\gamma_d} \sqrt{1 + \dfrac{k}{a^2 H^2}} - \sqrt{\epsilon^\phi} \biggr].
\end{align*}
By the non-positivity of $k \leq 0$, we thus conclude with the simple inequality
\begin{equation} \label{app. dS-conjecture analysis: epsilon-phi bound}
    \dfrac{\dot{\epsilon}^\phi}{2H} \leq (d-1-\epsilon^\phi) \sqrt{\epsilon^\phi} \biggl[ \dfrac{\gamma}{\gamma_d} - \sqrt{\epsilon^\phi} \biggr].
\end{equation}

Now, let us define a positive constant $\smash{\delta}$ such that
\begin{equation} \label{app. dS-conjecture analysis: epsilonstar}
    \Bigl( \dfrac{\gamma}{\gamma_d} \Bigr)^{\!2} < \delta < 1.
\end{equation}
For an expanding solution with a non-negative potential $V \geq 0$, eqs. (\ref{app. dS-conjecture analysis: F1-eq.}, \ref{app. dS-conjecture analysis: F2-eq.}) imply $\smash{0 \leq \epsilon \leq d-1}$, which in turn gives $\smash{\int_{t_0}^{\infty} H \, \de t = \infty}$ for any initial time $t_0$ \cite{Shiu:2023nph}.
If $\smash{\epsilon_0^\phi \geq \delta}$, then ineq. (\ref{app. dS-conjecture analysis: epsilon-phi bound}) makes $\epsilon^\phi$ decrease, with the fixed negative bound $\smash{\dot{\epsilon}^\phi/(2H) \leq \sqrt{\delta} \, (d-1-\epsilon_0^\phi) (\gamma/\gamma_d - \sqrt{\delta}) < 0}$ for as long as $\smash{\epsilon^\phi \geq \delta}$.
Thus, at some finite time $t_\star \geq t_0$, we have
\begin{equation} \label{app. dS-conjecture analysis: epsilon-phi crossing}
    \epsilon^\phi_\star < \delta.
\end{equation}
Then, $\smash{\epsilon^\phi}$ may start to grow.
At the hypothetical first time $\smash{\tilde{t}}$ when $\epsilon^\phi$ could reach
$\delta$ from below, a smooth evolution would require the sign $\smash{\dot{\epsilon}^\phi \geq 0}$.
Instead, in view of the condition in ineq. (\ref{app. dS-conjecture analysis: epsilonstar}), we see that inserting the limiting behavior of the evolution bound in ineq. (\ref{app. dS-conjecture analysis: epsilon-phi bound}) would force the sign
\begin{align*}
    \lim_{t \to \tilde{t}} \, \dfrac{\dot{\epsilon}^\phi}{2H} \leq (d-1-\delta) \sqrt{\delta} \Bigl[ \dfrac{\gamma}{\gamma_d} \sqrt{1 + \dfrac{k}{\tilde{a}^2 \tilde{H}^2}} - \sqrt{\delta} \Bigr] \leq (d-1-\delta) \sqrt{\delta} \Bigl[ \dfrac{\gamma}{\gamma_d} - \sqrt{\delta} \Bigr] < 0.
\end{align*}
Consequently, at all times $t \geq t_\star$ we must necessarily have $\smash{\epsilon^\phi < \delta}$.
Because $\smash{\delta < 1}$, given eq. (\ref{app. dS-conjecture analysis: epsilon-phi}), we also have
\begin{equation}
    \epsilon <  1 - (1 - \delta) \Bigl[ 1 + \dfrac{k}{a^2 H^2} \Bigr] < 1,
\end{equation}
which, since it strictly holds at all times $t \geq t_\star$, implies the existence of a future cosmic horizon, contradicting the no-horizon theorem.
Therefore, we conclude that the universal condition $\smash{\gamma < \gamma_d}$, which allows this chain of implication, must be forbidden.
More specifically, this means that all $d$-dimensional EFTs descending from the reduction of a SSEC-fulfilling higher-dimensional theory must necessarily be such that the potential relative slope in Planck units cannot universally remain below the critical value $\gamma_d=2/\sqrt{d-2}$ throughout the future evolution, i.e.
\begin{equation} \label{app. dS-conjecture analysis: potential-gradient bound}
    \sup_{t \geq t_\star} \, \dfrac{1}{\kappa_d^2 V^2} \Bigl( \dfrac{\der V}{\der\phi^a} \dfrac{\der V}{\der\phi_a} \Bigr) \geq \dfrac{4}{d-2}.
\end{equation}
This is weaker than the strong dS conjecture: it constrains the supremum along the trajectory, without establishing a pointwise lower bound in an asymptotic field-space region.

We emphasize that the statement of ineq. (\ref{app. dS-conjecture analysis: potential-gradient bound}) is of a dynamical nature.
It is not a constraint on the off-shell shape of the scalar potential, but rather a prediction on the evolution of the potential along any on-shell field-space trajectories $\smash{\phi^a = \phi^a(t)}$.
Along a similar line of thought, it has already been observed that, whenever a string-theoretic reduction results in a multi-field multi-exponential EFT with a positive potential, the $d$-dimensional dilaton alone constrains the dynamical trajectory such that $\smash{\sqrt{(\der V)^2/\kappa_d^2 V^2} \geq 2/\sqrt{d-2}}$ at late-enough times \cite{Shiu:2023fhb}.

\end{document}